\documentclass[letterpaper,english,reprint,aps,pra,superscriptaddress,notitlepage,onecolumn,nofootinbib]{revtex4-2}
\usepackage[utf8]{inputenc}
\usepackage[T1]{fontenc}
\usepackage{amsmath,amssymb,amsfonts,amsthm}
\usepackage{physics}
\usepackage{graphicx}
\usepackage{hyperref}
\usepackage[margin=2.5cm]{geometry}
\usepackage{bbm}
\usepackage{subcaption}
\usepackage{xcolor}
\usepackage{rotating}
\usepackage{floatpag}
\rotfloatpagestyle{empty}

\usepackage{dsfont}
\usepackage{xspace,enumerate,epsfig}
\usepackage{graphicx}
\graphicspath{{.}{./figures/}}
\usepackage{marvosym}

\usepackage{tikzfig}
\usepackage{stmaryrd}
\usepackage{docmute}
\usepackage{keycommand}

\usetikzlibrary{shapes.multipart}

\tikzstyle{bwSpider}=[
       rectangle split,
       rectangle split parts=2,
       rectangle split part fill={black,white},
 minimum size=3.6 mm, inner sep=-2mm, draw=black,scale=0.5,rounded corners=0.8 mm
       ]
 \tikzstyle{wbSpider}=[
       rectangle split,
       rectangle split parts=2,
       rectangle split part fill={white,black},
 minimum size=3.6 mm, inner sep=-2mm, draw=black,scale=0.5,rounded corners=0.8 mm
       ]
\tikzstyle{qWire}=[line width = .5pt, color=black, double distance=1pt]
\tikzstyle{hyperbit}=[line width = 2pt, color=black]
\tikzstyle{pWire}=[line width = 1pt, color=red!60!black]
\tikzstyle{gWire}=[line width = 1.5pt, color=green!60!black]
\tikzstyle{cWire}=[color=blue,thin]
\tikzstyle{env}=[copoint,regular polygon rotate=0,minimum width=0.2cm, fill=black]

\tikzstyle{probs}=[shape=semicircle,fill=white,draw=black,shape border rotate=180,minimum width=1.2cm]

\tikzstyle{every picture}=[baseline=-0.25em,scale=0.5]
\tikzstyle{dotpic}=[] 
\tikzstyle{diredges}=[every to/.style={diredge}]
\tikzstyle{math matrix}=[matrix of math nodes,left delimiter=(,right delimiter=),inner sep=2pt,column sep=1em,row sep=0.5em,nodes={inner sep=0pt},text height=1.5ex, text depth=0.25ex]

\tikzstyle{inline text}=[text height=1.5ex, text depth=0.25ex,yshift=0.5mm]
\tikzstyle{label}=[font=\footnotesize,text height=1.5ex, text depth=0.25ex,yshift=0.5mm]
\tikzstyle{left label}=[label,anchor=east,xshift=1.5mm]
\tikzstyle{right label}=[label,anchor=west,xshift=-1mm]

\tikzstyle{braceedge}=[decorate,decoration={brace,amplitude=2mm,raise=-1mm}]
\tikzstyle{small braceedge}=[decorate,decoration={brace,amplitude=1mm,raise=-1mm}]

\tikzstyle{doubled}=[line width=1.6pt] 
\tikzstyle{boldedge}=[doubled,shorten <=-0.17mm,shorten >=-0.17mm]
\tikzstyle{boldedgegray}=[doubled,gray,shorten <=-0.17mm,shorten >=-0.17mm]
\tikzstyle{singleedgegray}=[gray]

\tikzstyle{semidoubled}=[line width=1.4pt] 
\tikzstyle{semiboldedgegray}=[semidoubled,gray,shorten <=-0.17mm,shorten >=-0.17mm]

\tikzstyle{boxedge}=[semiboldedgegray]

\tikzstyle{boldedgedashed}=[very thick,dashed,shorten <=-0.17mm,shorten >=-0.17mm]
\tikzstyle{vboldedgedashed}=[doubled,dashed,shorten <=-0.17mm,shorten >=-0.17mm]
\tikzstyle{left hook arrow}=[left hook-latex]
\tikzstyle{right hook arrow}=[right hook-latex]
\tikzstyle{sembracket}=[line width=0.5pt,shorten <=-0.07mm,shorten >=-0.07mm]

\tikzstyle{causal edge}=[->,thick,gray]
\tikzstyle{causal nondir}=[thick,gray]
\tikzstyle{timeline}=[thick,gray, dashed]

\tikzstyle{cedge}=[<->,thick,gray!70!white]

\tikzstyle{empty diagram}=[draw=gray!40!white,dashed,shape=rectangle,minimum width=1cm,minimum height=1cm]
\tikzstyle{empty diagram small}=[draw=gray!50!white,dashed,shape=rectangle,minimum width=0.6cm,minimum height=0.5cm]

\tikzstyle{dot}=[inner sep=0mm,minimum width=2mm,minimum height=2mm,draw,shape=circle]
\tikzstyle{leak}=[white dot, shape=regular polygon, minimum size=3.3 mm, regular polygon sides=3, outer sep=-0.2mm, regular polygon rotate=270]
\tikzstyle{proj}=[draw,fill=white,chamfered rectangle,chamfered rectangle angle=30, minimum width=2mm, minimum height= 1mm,scale=0.5, outer sep=-0.2mm]
\tikzstyle{wide proj}=[draw,fill=white,chamfered rectangle,chamfered rectangle angle=30, minimum width=15mm,minimum height=1mm,scale=0.5, outer sep=-0.2mm]
\tikzstyle{very wide proj}=[draw,fill=white,chamfered rectangle,chamfered rectangle angle=30, minimum width=25mm,minimum height=1mm,scale=0.5, outer sep=-0.2mm]
\tikzstyle{very very wide proj}=[draw,fill=white,chamfered rectangle,chamfered rectangle angle=30, minimum width=35mm,minimum height=1mm,scale=0.5, outer sep=-0.2mm]
\tikzstyle{preleak}=[proj]

\tikzstyle{split proj out}=[regular polygon,regular polygon sides=3,draw,scale=0.75,inner sep=-0.5pt,minimum width=3.3mm,fill=white,regular polygon rotate=180]
\tikzstyle{split proj in}=[regular polygon,regular polygon sides=3,draw,scale=0.75,inner sep=-0.5pt,minimum width=3.3mm,fill=white]
\tikzstyle{Vleak}=[white dot, shape=regular polygon, minimum size=3.3 mm, regular polygon sides=3, outer sep=-0.2mm, regular polygon rotate=90]
\tikzstyle{dleak}=[white dot, line width=1.6pt, shape=regular polygon, minimum size=3.3 mm, regular polygon sides=3, outer sep=-0.2mm, regular polygon rotate=270]

\tikzstyle{Wsquare}=[white dot, shape=regular polygon, rounded corners=0.8 mm, minimum size=3.3 mm, regular polygon sides=3, outer sep=-0.2mm]
\tikzstyle{Wsquareadj}=[white dot, shape=regular polygon, rounded corners=0.8 mm, minimum size=3.3 mm, regular polygon sides=3, outer sep=-0.2mm, regular polygon rotate=180]
\tikzstyle{ddot}=[inner sep=0mm, doubled, minimum width=2.5mm,minimum height=2.5mm,draw,shape=circle]

\tikzstyle{black dot}=[dot,fill=black]
\tikzstyle{white dot}=[dot,fill=white,,text depth=-0.2mm]
\tikzstyle{white Wsquare}=[Wsquare,fill=gray,,text depth=-0.2mm]
\tikzstyle{white Wsquareadj}=[Wsquareadj,fill=white,,text depth=-0.2mm]
\tikzstyle{green dot}=[white dot] 
\tikzstyle{gray dot}=[dot,fill=gray!40!white,,text depth=-0.2mm]
\tikzstyle{red dot}=[gray dot] 

\tikzstyle{black ddot}=[ddot,fill=black]
\tikzstyle{white ddot}=[ddot,fill=white]
\tikzstyle{gray ddot}=[ddot,fill=gray!40!white]

\tikzstyle{gray edge}=[gray!60!white]

\tikzstyle{small dot}=[inner sep=0.5mm,minimum width=0pt,minimum height=0pt,draw,shape=circle]

\tikzstyle{small black dot}=[small dot,fill=black]
\tikzstyle{small white dot}=[small dot,fill=white]
\tikzstyle{small gray dot}=[small dot,fill=gray!40!white]

\tikzstyle{causal dot}=[inner sep=0.4mm,minimum width=0pt,minimum height=0pt,draw=white,shape=circle,fill=gray!40!white]

\tikzstyle{phase dimensions}=[minimum size=5mm,font=\footnotesize,rectangle,rounded corners=2.5mm,inner sep=0.2mm,outer sep=-2mm]
\tikzstyle{dphase dimensions}=[minimum size=5mm,font=\footnotesize,rectangle,rounded corners=2.5mm,inner sep=0.2mm,outer sep=-2mm]

\tikzstyle{white phase dot}=[dot,fill=white,phase dimensions]
\tikzstyle{white phase ddot}=[ddot,fill=white,dphase dimensions]

\tikzstyle{white rect ddot}=[draw=black,fill=white,doubled,minimum size=5mm,font=\footnotesize,rectangle,rounded corners=2.5mm,inner sep=0.2mm]
\tikzstyle{gray rect ddot}=[draw=black,fill=gray!40!white,doubled,minimum size=6mm,font=\footnotesize,rectangle,rounded corners=3mm]

\tikzstyle{gray phase dot}=[dot,fill=gray!40!white,phase dimensions]
\tikzstyle{gray phase ddot}=[ddot,fill=gray!40!white,dphase dimensions]
\tikzstyle{grey phase dot}=[gray phase dot]
\tikzstyle{grey phase ddot}=[gray phase ddot]

\tikzstyle{small phase dimensions}=[minimum size=4mm,font=\tiny,rectangle,rounded corners=2mm,inner sep=0.2mm,outer sep=-2mm]
\tikzstyle{small dphase dimensions}=[minimum size=4mm,font=\tiny,rectangle,rounded corners=2mm,inner sep=0.2mm,outer sep=-2mm]

\tikzstyle{small gray phase dot}=[dot,fill=gray!40!white,small phase dimensions]
\tikzstyle{small gray phase ddot}=[ddot,fill=gray!40!white,small dphase dimensions]

\tikzstyle{small map}=[draw,shape=rectangle,minimum height=4mm,minimum width=4mm,fill=white]

\tikzstyle{cnot}=[fill=white,shape=circle,inner sep=-1.4pt]

\tikzstyle{asym hadamard}=[fill=white,draw,shape=NEbox,inner sep=0.6mm,font=\footnotesize,minimum height=4mm]
\tikzstyle{asym hadamard conj}=[fill=white,draw,shape=NWbox,inner sep=0.6mm,font=\footnotesize,minimum height=4mm]
\tikzstyle{asym hadamard dag}=[fill=white,draw,shape=SEbox,inner sep=0.6mm,font=\footnotesize,minimum height=4mm]

\tikzstyle{hadamard}=[fill=white,draw,inner sep=0.6mm,font=\footnotesize,minimum height=4mm,minimum width=4mm]
\tikzstyle{small hadamard}=[fill=white,draw,inner sep=0.6mm,minimum height=1.5mm,minimum width=1.5mm]
\tikzstyle{small hadamard rotate}=[small hadamard,rotate=45]
\tikzstyle{dhadamard}=[hadamard,doubled]
\tikzstyle{small dhadamard}=[small hadamard,doubled]
\tikzstyle{small dhadamard rotate}=[small hadamard rotate,doubled]
\tikzstyle{antipode}=[white dot,inner sep=0.3mm,font=\footnotesize]

\tikzstyle{scalar}=[diamond,draw,inner sep=0.5pt,font=\small]
\tikzstyle{dscalar}=[diamond,doubled, draw,inner sep=0.5pt,font=\small]

\tikzstyle{small box}=[rectangle,inline text,fill=white,draw,minimum height=5mm,yshift=-0.5mm,minimum width=5mm,font=\small]
\tikzstyle{small gray box}=[small box,fill=gray!30]
\tikzstyle{medium box}=[rectangle,inline text,fill=white,draw,minimum height=5mm,yshift=-0.5mm,minimum width=10mm,font=\small]
\tikzstyle{square box}=[small box] 
\tikzstyle{medium gray box}=[small box,fill=gray!30]
\tikzstyle{semilarge box}=[rectangle,inline text,fill=white,draw,minimum height=5mm,yshift=-0.5mm,minimum width=12.5mm,font=\small]
\tikzstyle{large box}=[rectangle,inline text,fill=white,draw,minimum height=5mm,yshift=-0.5mm,minimum width=15mm,font=\small]
\tikzstyle{large gray box}=[small box,fill=gray!30]

\tikzstyle{Bayes box}=[rectangle,fill=black,draw, minimum height=3mm, minimum width=3mm]

\tikzstyle{gray square point}=[small box,fill=gray!50]

\tikzstyle{dphase box white}=[dhadamard]
\tikzstyle{dphase box gray}=[dhadamard,fill=gray!50!white]
\tikzstyle{phase box white}=[hadamard]
\tikzstyle{phase box gray}=[hadamard,fill=gray!50!white]

\tikzstyle{point}=[regular polygon,regular polygon sides=3,draw,scale=0.75,inner sep=-0.5pt,minimum width=9mm,fill=white,regular polygon rotate=180]
\tikzstyle{point nosep}=[regular polygon,regular polygon sides=3,draw,scale=0.75,inner sep=-2pt,minimum width=9mm,fill=white,regular polygon rotate=180]
\tikzstyle{copoint}=[regular polygon,regular polygon sides=3,draw,scale=0.75,inner sep=-0.5pt,minimum width=9mm,fill=white]
\tikzstyle{dpoint}=[point,doubled]
\tikzstyle{dcopoint}=[copoint,doubled]

\tikzstyle{pointgrow}=[shape=cornerpoint,kpoint common,scale=0.75,inner sep=3pt]
\tikzstyle{pointgrow dag}=[shape=cornercopoint,kpoint common,scale=0.75,inner sep=3pt]

\tikzstyle{wide copoint}=[fill=white,draw,shape=isosceles triangle,shape border rotate=90,isosceles triangle stretches=true,inner sep=0pt,minimum width=1.5cm,minimum height=6.12mm]
\tikzstyle{wide point}=[fill=white,draw,shape=isosceles triangle,shape border rotate=-90,isosceles triangle stretches=true,inner sep=0pt,minimum width=1.5cm,minimum height=6.12mm,yshift=-0.0mm]
\tikzstyle{wide point plus}=[fill=white,draw,shape=isosceles triangle,shape border rotate=-90,isosceles triangle stretches=true,inner sep=0pt,minimum width=1.74cm,minimum height=7mm,yshift=-0.0mm]

\tikzstyle{wide dpoint}=[fill=white,doubled,draw,shape=isosceles triangle,shape border rotate=-90,isosceles triangle stretches=true,inner sep=0pt,minimum width=1.5cm,minimum height=6.12mm,yshift=-0.0mm]

\tikzstyle{tinypoint}=[regular polygon,regular polygon sides=3,draw,scale=0.55,inner sep=-0.15pt,minimum width=6mm,fill=white,regular polygon rotate=180]

\tikzstyle{white point}=[point]
\tikzstyle{white dpoint}=[dpoint]
\tikzstyle{green point}=[white point] 
\tikzstyle{white copoint}=[copoint]
\tikzstyle{gray point}=[point,fill=gray!40!white]
\tikzstyle{gray dpoint}=[gray point,doubled]
\tikzstyle{red point}=[gray point] 
\tikzstyle{gray copoint}=[copoint,fill=gray!40!white]
\tikzstyle{gray dcopoint}=[gray copoint,doubled]

\tikzstyle{white point guide}=[regular polygon,regular polygon sides=3,font=\scriptsize,draw,scale=0.65,inner sep=-0.5pt,minimum width=9mm,fill=white,regular polygon rotate=180]

\tikzstyle{black point}=[point,fill=black,font=\color{white}]
\tikzstyle{black copoint}=[copoint,fill=black,font=\color{white}]

\tikzstyle{tiny gray point}=[tinypoint,fill=gray!40!white]

\tikzstyle{diredge}=[->]
\tikzstyle{ddiredge}=[<->]
\tikzstyle{rdiredge}=[<-]
\tikzstyle{thickdiredge}=[->, very thick]
\tikzstyle{pointer edge}=[->,very thick,gray]
\tikzstyle{pointer edge part}=[very thick,gray]
\tikzstyle{dashed edge}=[dashed]
\tikzstyle{thick dashed edge}=[very thick,dashed]
\tikzstyle{thick gray dashed edge}=[thick dashed edge,gray!40]
\tikzstyle{thick map edge}=[very thick,|->]

\makeatletter
\newcommand{\boxshape}[3]{%
\pgfdeclareshape{#1}{
\inheritsavedanchors[from=rectangle] 
\inheritanchorborder[from=rectangle]
\inheritanchor[from=rectangle]{center}
\inheritanchor[from=rectangle]{north}
\inheritanchor[from=rectangle]{south}
\inheritanchor[from=rectangle]{west}
\inheritanchor[from=rectangle]{east}
\backgroundpath{
\southwest \pgf@xa=\pgf@x \pgf@ya=\pgf@y
\northeast \pgf@xb=\pgf@x \pgf@yb=\pgf@y

\@tempdima=#2
\@tempdimb=#3

\pgfpathmoveto{\pgfpoint{\pgf@xa - 5pt + \@tempdima}{\pgf@ya}}
\pgfpathlineto{\pgfpoint{\pgf@xa - 5pt - \@tempdima}{\pgf@yb}}
\pgfpathlineto{\pgfpoint{\pgf@xb + 5pt + \@tempdimb}{\pgf@yb}}
\pgfpathlineto{\pgfpoint{\pgf@xb + 5pt - \@tempdimb}{\pgf@ya}}
\pgfpathlineto{\pgfpoint{\pgf@xa - 5pt + \@tempdima}{\pgf@ya}}
\pgfpathclose
}
}}

\boxshape{NEbox}{0pt}{3pt}
\boxshape{SEbox}{0pt}{-3pt}
\boxshape{NWbox}{5pt}{0pt}
\boxshape{SWbox}{-5pt}{0pt}
\boxshape{EBox}{-3pt}{3pt}
\boxshape{WBox}{3pt}{-3pt}
\makeatother

\tikzstyle{cloud}=[shape=cloud,draw,minimum width=1.5cm,minimum height=1.5cm]

\tikzstyle{map}=[draw,shape=NEbox,inner sep=1pt,minimum height=4mm,fill=white]
\tikzstyle{dashedmap}=[draw,dashed,shape=NEbox,inner sep=2pt,minimum height=6mm,fill=white]
\tikzstyle{mapdag}=[draw,shape=SEbox,inner sep=1pt,minimum height=4mm,fill=white]
\tikzstyle{mapadj}=[draw,shape=SEbox,inner sep=2pt,minimum height=6mm,fill=white]
\tikzstyle{maptrans}=[draw,shape=SWbox,inner sep=2pt,minimum height=6mm,fill=white]
\tikzstyle{mapconj}=[draw,shape=NWbox,inner sep=2pt,minimum height=6mm,fill=white]

\tikzstyle{medium map}=[draw,shape=NEbox,inner sep=2pt,minimum height=6mm,fill=white,minimum width=7mm]
\tikzstyle{medium map dag}=[draw,shape=SEbox,inner sep=2pt,minimum height=6mm,fill=white,minimum width=7mm]
\tikzstyle{medium map adj}=[draw,shape=SEbox,inner sep=2pt,minimum height=6mm,fill=white,minimum width=7mm]
\tikzstyle{medium map trans}=[draw,shape=SWbox,inner sep=2pt,minimum height=6mm,fill=white,minimum width=7mm]
\tikzstyle{medium map conj}=[draw,shape=NWbox,inner sep=2pt,minimum height=6mm,fill=white,minimum width=7mm]
\tikzstyle{semilarge map}=[draw,shape=NEbox,inner sep=2pt,minimum height=6mm,fill=white,minimum width=9.5mm]
\tikzstyle{semilarge map trans}=[draw,shape=SWbox,inner sep=2pt,minimum height=6mm,fill=white,minimum width=9.5mm]
\tikzstyle{semilarge map adj}=[draw,shape=SEbox,inner sep=2pt,minimum height=6mm,fill=white,minimum width=9.5mm]
\tikzstyle{semilarge map dag}=[draw,shape=SEbox,inner sep=2pt,minimum height=6mm,fill=white,minimum width=9.5mm]
\tikzstyle{semilarge map conj}=[draw,shape=NWbox,inner sep=2pt,minimum height=6mm,fill=white,minimum width=9.5mm]
\tikzstyle{large map}=[draw,shape=NEbox,inner sep=2pt,minimum height=6mm,fill=white,minimum width=12mm]
\tikzstyle{large map conj}=[draw,shape=NWbox,inner sep=2pt,minimum height=6mm,fill=white,minimum width=12mm]
\tikzstyle{very large map}=[draw,shape=NEbox,inner sep=2pt,minimum height=6mm,fill=white,minimum width=17mm]

\tikzstyle{medium dmap}=[draw,doubled,shape=NEbox,inner sep=2pt,minimum height=6mm,fill=white,minimum width=7mm]
\tikzstyle{medium dmap dag}=[draw,doubled,shape=SEbox,inner sep=2pt,minimum height=6mm,fill=white,minimum width=7mm]
\tikzstyle{medium dmap adj}=[draw,doubled,shape=SEbox,inner sep=2pt,minimum height=6mm,fill=white,minimum width=7mm]
\tikzstyle{medium dmap trans}=[draw,doubled,shape=SWbox,inner sep=2pt,minimum height=6mm,fill=white,minimum width=7mm]
\tikzstyle{medium dmap conj}=[draw,doubled,shape=NWbox,inner sep=2pt,minimum height=6mm,fill=white,minimum width=7mm]
\tikzstyle{semilarge dmap}=[draw,doubled,shape=NEbox,inner sep=2pt,minimum height=6mm,fill=white,minimum width=9.5mm]
\tikzstyle{semilarge dmap trans}=[draw,doubled,shape=SWbox,inner sep=2pt,minimum height=6mm,fill=white,minimum width=9.5mm]
\tikzstyle{semilarge dmap adj}=[draw,doubled,shape=SEbox,inner sep=2pt,minimum height=6mm,fill=white,minimum width=9.5mm]
\tikzstyle{semilarge dmap dag}=[draw,doubled,shape=SEbox,inner sep=2pt,minimum height=6mm,fill=white,minimum width=9.5mm]
\tikzstyle{semilarge dmap conj}=[draw,doubled,shape=NWbox,inner sep=2pt,minimum height=6mm,fill=white,minimum width=9.5mm]
\tikzstyle{large dmap}=[draw,doubled,shape=NEbox,inner sep=2pt,minimum height=6mm,fill=white,minimum width=12mm]
\tikzstyle{large dmap conj}=[draw,doubled,shape=NWbox,inner sep=2pt,minimum height=6mm,fill=white,minimum width=12mm]
\tikzstyle{large dmap trans}=[draw,doubled,shape=SWbox,inner sep=2pt,minimum height=6mm,fill=white,minimum width=12mm]
\tikzstyle{large dmap adj}=[draw,doubled,shape=SEbox,inner sep=2pt,minimum height=6mm,fill=white,minimum width=12mm]
\tikzstyle{large dmap dag}=[draw,doubled,shape=SEbox,inner sep=2pt,minimum height=6mm,fill=white,minimum width=12mm]
\tikzstyle{very large dmap}=[draw,doubled,shape=NEbox,inner sep=2pt,minimum height=6mm,fill=white,minimum width=19.5mm]

\tikzstyle{muxbox}=[draw,shape=rectangle,minimum height=3mm,minimum width=3mm,fill=white]
\tikzstyle{dmuxbox}=[muxbox,doubled]

\tikzstyle{box}=[draw,shape=rectangle,inner sep=2pt,minimum height=6mm,minimum width=6mm,fill=white]
\tikzstyle{dbox}=[draw,doubled,shape=rectangle,inner sep=2pt,minimum height=6mm,minimum width=6mm,fill=white]
\tikzstyle{dmap}=[draw,doubled,shape=NEbox,inner sep=2pt,minimum height=6mm,fill=white]
\tikzstyle{dmapdag}=[draw,doubled,shape=SEbox,inner sep=2pt,minimum height=6mm,fill=white]
\tikzstyle{dmapadj}=[draw,doubled,shape=SEbox,inner sep=2pt,minimum height=6mm,fill=white]
\tikzstyle{dmaptrans}=[draw,doubled,shape=SWbox,inner sep=2pt,minimum height=6mm,fill=white]
\tikzstyle{dmapconj}=[draw,doubled,shape=NWbox,inner sep=2pt,minimum height=6mm,fill=white]

\tikzstyle{ddmap}=[draw,doubled,dashed,shape=NEbox,inner sep=2pt,minimum height=6mm,fill=white]
\tikzstyle{ddmapdag}=[draw,doubled,dashed,shape=SEbox,inner sep=2pt,minimum height=6mm,fill=white]
\tikzstyle{ddmapadj}=[draw,doubled,dashed,shape=SEbox,inner sep=2pt,minimum height=6mm,fill=white]
\tikzstyle{ddmaptrans}=[draw,doubled,dashed,shape=SWbox,inner sep=2pt,minimum height=6mm,fill=white]
\tikzstyle{ddmapconj}=[draw,doubled,dashed,shape=NWbox,inner sep=2pt,minimum height=6mm,fill=white]

\boxshape{sNEbox}{0pt}{3pt}
\boxshape{sSEbox}{0pt}{-3pt}
\boxshape{sNWbox}{3pt}{0pt}
\boxshape{sSWbox}{-3pt}{0pt}
\tikzstyle{smap}=[draw,shape=sNEbox,fill=white]
\tikzstyle{smapdag}=[draw,shape=sSEbox,fill=white]
\tikzstyle{smapadj}=[draw,shape=sSEbox,fill=white]
\tikzstyle{smaptrans}=[draw,shape=sSWbox,fill=white]
\tikzstyle{smapconj}=[draw,shape=sNWbox,fill=white]

\tikzstyle{dsmap}=[draw,dashed,shape=sNEbox,fill=white]
\tikzstyle{dsmapdag}=[draw,dashed,shape=sSEbox,fill=white]
\tikzstyle{dsmaptrans}=[draw,dashed,shape=sSWbox,fill=white]
\tikzstyle{dsmapconj}=[draw,dashed,shape=sNWbox,fill=white]

\boxshape{mNEbox}{0pt}{10pt}
\boxshape{mSEbox}{0pt}{-10pt}
\boxshape{mNWbox}{10pt}{0pt}
\boxshape{mSWbox}{-10pt}{0pt}
\tikzstyle{mmap}=[draw,shape=mNEbox]
\tikzstyle{mmapdag}=[draw,shape=mSEbox]
\tikzstyle{mmaptrans}=[draw,shape=mSWbox]
\tikzstyle{mmapconj}=[draw,shape=mNWbox]

\tikzstyle{mmapgray}=[draw,fill=gray!40!white,shape=mNEbox]
\tikzstyle{smapgray}=[draw,fill=gray!40!white,shape=sNEbox]

\makeatletter

\pgfdeclareshape{cornerpoint}{
\inheritsavedanchors[from=rectangle] 
\inheritanchorborder[from=rectangle]
\inheritanchor[from=rectangle]{center}
\inheritanchor[from=rectangle]{north}
\inheritanchor[from=rectangle]{south}
\inheritanchor[from=rectangle]{west}
\inheritanchor[from=rectangle]{east}
\backgroundpath{
\southwest \pgf@xa=\pgf@x \pgf@ya=\pgf@y
\northeast \pgf@xb=\pgf@x \pgf@yb=\pgf@y

\pgfmathsetmacro{\pgf@shorten@left}{\pgfkeysvalueof{/tikz/shorten left}}
\pgfmathsetmacro{\pgf@shorten@right}{\pgfkeysvalueof{/tikz/shorten right}}

\pgfpathmoveto{\pgfpoint{0.5 * (\pgf@xa + \pgf@xb)}{\pgf@ya - 5pt}}
\pgfpathlineto{\pgfpoint{\pgf@xa - 8pt + \pgf@shorten@left}{\pgf@yb - 1.5 * \pgf@shorten@left}}
\pgfpathlineto{\pgfpoint{\pgf@xa - 8pt + \pgf@shorten@left}{\pgf@yb}}
\pgfpathlineto{\pgfpoint{\pgf@xb + 8pt - \pgf@shorten@right}{\pgf@yb}}
\pgfpathlineto{\pgfpoint{\pgf@xb + 8pt - \pgf@shorten@right}{\pgf@yb - 1.5 * \pgf@shorten@right}}
\pgfpathclose
}
}

\pgfdeclareshape{cornercopoint}{
\inheritsavedanchors[from=rectangle] 
\inheritanchorborder[from=rectangle]
\inheritanchor[from=rectangle]{center}
\inheritanchor[from=rectangle]{north}
\inheritanchor[from=rectangle]{south}
\inheritanchor[from=rectangle]{west}
\inheritanchor[from=rectangle]{east}
\backgroundpath{
\southwest \pgf@xa=\pgf@x \pgf@ya=\pgf@y
\northeast \pgf@xb=\pgf@x \pgf@yb=\pgf@y

\pgfmathsetmacro{\pgf@shorten@left}{\pgfkeysvalueof{/tikz/shorten left}}
\pgfmathsetmacro{\pgf@shorten@right}{\pgfkeysvalueof{/tikz/shorten right}}

\pgfpathmoveto{\pgfpoint{0.5 * (\pgf@xa + \pgf@xb)}{\pgf@yb + 5pt}}
\pgfpathlineto{\pgfpoint{\pgf@xa - 8pt + \pgf@shorten@left}{\pgf@ya + 1.5 * \pgf@shorten@left}}
\pgfpathlineto{\pgfpoint{\pgf@xa - 8pt + \pgf@shorten@left}{\pgf@ya}}
\pgfpathlineto{\pgfpoint{\pgf@xb + 8pt - \pgf@shorten@right}{\pgf@ya}}
\pgfpathlineto{\pgfpoint{\pgf@xb + 8pt - \pgf@shorten@right}{\pgf@ya + 1.5 * \pgf@shorten@right}}
\pgfpathclose
}
}

\makeatother

\pgfkeyssetvalue{/tikz/shorten left}{0pt}
\pgfkeyssetvalue{/tikz/shorten right}{0pt}

\tikzstyle{kpoint common}=[draw,fill=white,inner sep=1pt,minimum height=4mm]
\tikzstyle{kpoint sc}=[shape=cornerpoint,kpoint common]
\tikzstyle{kpoint adjoint sc}=[shape=cornercopoint,kpoint common]
\tikzstyle{kpoint}=[shape=cornerpoint,shorten left=5pt,kpoint common]
\tikzstyle{kpoint adjoint}=[shape=cornercopoint,shorten left=5pt,kpoint common]
\tikzstyle{kpoint conjugate}=[shape=cornerpoint,shorten right=5pt,kpoint common]
\tikzstyle{kpoint transpose}=[shape=cornercopoint,shorten right=5pt,kpoint common]
\tikzstyle{kpoint symm}=[shape=cornerpoint,shorten left=5pt,shorten right=5pt,kpoint common]

\tikzstyle{wide kpoint sc}=[shape=cornerpoint,kpoint common, minimum width=1 cm]
\tikzstyle{wide kpointdag sc}=[shape=cornercopoint,kpoint common, minimum width=1 cm]

\tikzstyle{black kpoint}=[shape=cornerpoint,shorten left=5pt,kpoint common,fill=black,font=\color{white}]

\tikzstyle{black kpoint sm}=[shape=cornerpoint,shorten left=5pt,kpoint common,fill=black,font=\color{white},scale=0.75]

\tikzstyle{black kpoint adjoint}=[shape=cornercopoint,shorten left=5pt,kpoint common,fill=black,font=\color{white}]
\tikzstyle{black kpointadj}=[shape=cornercopoint,shorten left=5pt,kpoint common,fill=black,font=\color{white}]

\tikzstyle{black kpointadj sm}=[shape=cornercopoint,shorten left=5pt,kpoint common,fill=black,font=\color{white},scale=0.75]

\tikzstyle{black dkpoint}=[shape=cornerpoint,shorten left=5pt,kpoint common,fill=black, doubled,font=\color{white}]
\tikzstyle{black dkpoint adjoint}=[shape=cornercopoint,shorten left=5pt,kpoint common,fill=black, doubled,font=\color{white}]
\tikzstyle{black dkpointadj}=[shape=cornercopoint,shorten left=5pt,kpoint common,fill=black, doubled,font=\color{white}]

\tikzstyle{black dkpoint sm}=[shape=cornerpoint,shorten left=5pt,kpoint common,fill=black, doubled,font=\color{white},scale=0.75]
\tikzstyle{black dkpointadj sm}=[shape=cornercopoint,shorten left=5pt,kpoint common,fill=black, doubled,font=\color{white},scale=0.75]

\tikzstyle{kpointdag}=[kpoint adjoint]
\tikzstyle{kpointadj}=[kpoint adjoint]
\tikzstyle{kpointconj}=[kpoint conjugate]
\tikzstyle{kpointtrans}=[kpoint transpose]

\tikzstyle{big kpoint}=[kpoint, minimum width=1.2 cm, minimum height=8mm, inner sep=4pt, text depth=3mm]

\tikzstyle{wide kpoint}=[kpoint, minimum width=1 cm, inner sep=2pt]
\tikzstyle{wide kpointdag}=[kpointdag, minimum width=1 cm, inner sep=2pt]
\tikzstyle{wide kpointconj}=[kpointconj, minimum width=1 cm, inner sep=2pt]
\tikzstyle{wide kpointtrans}=[kpointtrans, minimum width=1 cm, inner sep=2pt]

\tikzstyle{wider kpoint}=[kpoint, minimum width=1.25 cm, inner sep=2pt]
\tikzstyle{wider kpointdag}=[kpointdag, minimum width=1.25 cm, inner sep=2pt]
\tikzstyle{wider kpointconj}=[kpointconj, minimum width=1.25 cm, inner sep=2pt]
\tikzstyle{wider kpointtrans}=[kpointtrans, minimum width=1.25 cm, inner sep=2pt]

\tikzstyle{gray kpoint}=[kpoint,fill=gray!50!white]
\tikzstyle{gray kpointdag}=[kpointdag,fill=gray!50!white]
\tikzstyle{gray kpointadj}=[kpointadj,fill=gray!50!white]
\tikzstyle{gray kpointconj}=[kpointconj,fill=gray!50!white]
\tikzstyle{gray kpointtrans}=[kpointtrans,fill=gray!50!white]

\tikzstyle{gray dkpoint}=[kpoint,fill=gray!50!white,doubled]
\tikzstyle{gray dkpointdag}=[kpointdag,fill=gray!50!white,doubled]
\tikzstyle{gray dkpointadj}=[kpointadj,fill=gray!50!white,doubled]
\tikzstyle{gray dkpointconj}=[kpointconj,fill=gray!50!white,doubled]
\tikzstyle{gray dkpointtrans}=[kpointtrans,fill=gray!50!white,doubled]

\tikzstyle{white label}=[draw,fill=white,rectangle,inner sep=0.7 mm]
\tikzstyle{gray label}=[draw,fill=gray!50!white,rectangle,inner sep=0.7 mm]
\tikzstyle{black label}=[draw,fill=black,rectangle,inner sep=0.7 mm]

\tikzstyle{dkpoint}=[kpoint,doubled]
\tikzstyle{wide dkpoint}=[wide kpoint,doubled]
\tikzstyle{dkpointdag}=[kpoint adjoint,doubled]
\tikzstyle{wide dkpointdag}=[wide kpointdag,doubled]
\tikzstyle{dkcopoint}=[kpoint adjoint,doubled]
\tikzstyle{dkpointadj}=[kpoint adjoint,doubled]
\tikzstyle{dkpointconj}=[kpoint conjugate,doubled]
\tikzstyle{dkpointtrans}=[kpoint transpose,doubled]

\tikzstyle{kscalar}=[kpoint common, shape=EBox, inner xsep=-1pt, inner ysep=3pt,font=\small]
\tikzstyle{kscalarconj}=[kpoint common, shape=WBox, inner xsep=-1pt, inner ysep=3pt,font=\small]

\tikzstyle{spekpoint}=[kpoint sc,minimum height=5mm,inner sep=3pt]
\tikzstyle{spekcopoint}=[kpoint adjoint sc,minimum height=5mm,inner sep=3pt]

\tikzstyle{dspekpoint}=[spekpoint,doubled]
\tikzstyle{dspekcopoint}=[spekcopoint,doubled]

 \tikzstyle{upground}=[circuit ee IEC,thick,ground,rotate=90,scale=2.5]
 \tikzstyle{downground}=[circuit ee IEC,thick,ground,rotate=-90,scale=2.5]
 \tikzstyle{bigground}=[regular polygon,regular polygon sides=3,draw=gray,scale=0.50,inner sep=-0.5pt,minimum width=10mm,fill=gray]

\tikzstyle{arrs}=[-latex,font=\small,auto]
\tikzstyle{arrow plain}=[arrs]
\tikzstyle{arrow dashed}=[dashed,arrs]
\tikzstyle{arrow bold}=[very thick,arrs]
\tikzstyle{arrow hide}=[draw=white!0,-]
\tikzstyle{arrow reverse}=[latex-]
\tikzstyle{cdnode}=[]

\usepackage{color}
\def\bR{\begin{color}{red}}
\def\bB{\begin{color}{blue}}
\def\bM{\begin{color}{magenta}}
\def\bC{\begin{color}{cyan}}
\def\bW{\begin{color}{white}}
\def\bBl{\begin{color}{black}}
\def\bG{\begin{color}{green}}
\def\bY{\begin{color}{yellow}}
\def\e{\end{color}\xspace}
\newcommand{\bit}{\begin{itemize}}
\newcommand{\eit}{\end{itemize}\par\noindent}
\newcommand{\ben}{\begin{enumerate}}
\newcommand{\een}{\end{enumerate}\par\noindent}
\newcommand{\beq}{\begin{equation}}
\newcommand{\eeq}{\end{equation}\par\noindent}
\newcommand{\beqa}{\begin{eqnarray*}}
\newcommand{\eeqa}{\end{eqnarray*}\par\noindent}
\newcommand{\beqn}{\begin{eqnarray}}
\newcommand{\eeqn}{\end{eqnarray}\par\noindent}

\def\jR{\begin{color}{DarkRed}}
\def\jB{\begin{color}{MidnightBlue}}
\def\jM{\begin{color}{magenta}}
\def\jC{\begin{color}{cyan}}
\def\jW{\begin{color}{white}}
\def\jBl{\begin{color}{black}}
\def\jG{\begin{color}{green}}
\def\jY{\begin{color}{yellow}}

\def\cR{\begin{color}{Crimson}}
\def\cB{\begin{color}{cyan}}
\def\cM{\begin{color}{magenta}}
\def\cC{\begin{color}{cyan}}
\def\cW{\begin{color}{white}}
\def\cBl{\begin{color}{black}}
\def\cG{\begin{color}{green}}
\def\cY{\begin{color}{yellow}}

\usepackage{tikz}
\usetikzlibrary{patterns}
\usepackage{pgfplots}
\usetikzlibrary{positioning}
\usetikzlibrary{decorations.markings}
\usetikzlibrary{arrows,decorations.text}
\usetikzlibrary{backgrounds,fit,decorations.pathreplacing}

 \newcommand{\id}{\openone}
\usepackage{scalerel,graphicx}
\newcommand{\bigtimes}{\mathop{\scalerel*{\times}{\sum}}}
\newcommand{\gs}{\color{red}}\newcommand{\blk}{\color{black}}
\begin{document}
	
\title{Entanglement detection via third-order local invariants from randomized measurements}	
	\author{Giovanni Scala}
	\affiliation{Dipartimento Interateneo di Fisica, Politecnico di Bari, 70126 Bari, Italy}
\affiliation{INFN, Sezione di Bari, 70126 Bari, Italy}
\email{giovanni.scala@poliba.it}
	
	\author{Anindita Bera}
	
	\affiliation{Department of Mathematics, Birla Institute of Technology Mesra, Jharkhand 835215, India}
	
	\author{Gniewomir Sarbicki}
	
	\affiliation{Institute of Physics, Faculty of Physics, Astronomy and Informatics,
		Nicolaus Copernicus University, Grudziadzka 5/7, 87-100 Toru\'{n},
		Poland}
	
	
	


\begin{abstract}
We compute all third-order local invariants accessible via randomised measurements and employ them to derive separability criteria.
The reconstruction of the invariants, yields experimentally accessible entanglement criteria 
for multipartite states with arbitrary local dimensions.
The results show that third-order invariants capture inter-subsystem correlations beyond second-order spectral criteria within more feasible entanglement detection protocols than full tomography. As an example, for Werner states in $d=3$, the entanglement is detected for $p>\frac 12$ for the second--order correlations and it is improved to $p>\frac 1{\sqrt[3]{10}}$ at the third-order.
\end{abstract}
\maketitle

\section{Introduction}
Testing whether a shared quantum state is entangled is a central question in quantum information theory and experiment \cite{HHHH,guhne2009entanglement,chruscinski2014entanglement}.  Standard approaches either require (i)~~\emph{full state tomography}\cite{Aaronson2020,Elben2020}, which scales exponentially with the system size, or (ii)~~\emph{non‑local measurements}, such as Bell‑state measurement (BSM)\cite{Zhou2020,Yu2021,Ghoreishi2025}, that are experimentally demanding when the parties are physically separated~\cite{Huang2020,Elben2022}. Once the entries of the density matrices are known, it is still challenging (NP hard~\cite{Gurvits2003}) to determine whether such a preparation, i.e. \textit{density matrix}, is entangled~\cite{Werner1989}. A way to tackle this open challenge is to build state-dependent entanglement witnesses (EWs) or mathematical separability criteria to address the central question~\cite{Doherty2004,RomeroPalleja2025,rico2025mixed}.

A complementary line of research exploits the fact that physicists are interested in using entanglement as a resource for quantum applications~\cite{nielsen2010quantum}
. In this case, we would like to answer the central question without knowing all the entries of the quantum state motivating the construction of separability criteria or entanglement witnesses. In this work, we derive two separability criteria starting from local invariant accessible by averaging twirling measurements. This experimental procedure would explore beyond the \emph{spectral} properties of the global and local density operators that, indeed, already impose constraints on separability.
 The most prominent example of spectral condition is \emph{majorisation criterion} \cite{Nielsen2001,Hiroshima2003,Partovi2012},
 which states that, for separable $\rho_{AB}$, the spectrum of $\rho_{AB}$ is majorised by those of its marginals.
This criterion follows from \emph{reduction criterion} \cite{Horodecki1999} obtained from the positivity of the map $R(X)=\Tr(X)\,\id-X$.
Since spectra are unitarily invariant, \emph{all} such criteria can be expressed in terms of the moments $\Tr \rho^k$~\cite{Yu2021,aggarwal2024entanglement}.

Van Enk and Beenakker \cite{vanEnk2012PRL} pointed out that each moment can, in principle, be estimated from \emph{single‑copy}\footnote{Here by single-copy we mean that at each run the degree of freedom of the experiment refers to $\rho$ and not to $ \rho^{\otimes n}$.} experiments by averaging measurement outcomes over a random unitary. 

Zoller et al. in Ref~\cite{brydges2019probing} have obtained a similar result showing that parties sharing a quantum state can estimate purities of all its marginals, averaging their measurements over local unitaries.  

In the present paper, we 
generalise the proof of  Ref~\cite{brydges2019probing} to the case of different dimensions of subsystems and then extend it to third-order correlations for bipartite systems, showing 10 local invariants which can be calculated from experimental data. Finally, we provide separability criteria expressible in terms of these invariants. 

\section{Invariants from randomised measurements}
\label{sec:moments}
In this section, we fix the notation. Let a single system $\rho\in\mathcal{B}(\mathbb{C}^{d_A})$ be an unknown density operator, 
$U^A\in \mathcal U(d_A)$ unitary picked from a Haar measure~\cite{Mezzadri2006,Cavalcanti2023,zyczkowski2011} to perform the projective measurement $\{ U^A\ket{i^A}\bra{i^A}U^{^\dagger}\}_{i=1}^{d_A}$.  The resulting probability distribution is
\begin{equation}
\,p(i^{A}_1|U_1)=\bra{i^A_1}U^{A\dagger}\rho U^{A}\ket{i^{A}_1}=\sum_{p_1^{A},qi^{A}_1=1}^{d_A} 
u_{p_1^{A}i_1^{A}}^*\,\,\rho_{p_1^{A}q_1^{A}}\,\, u_{q_1^{A}i^{A}_1}
\,,\qquad
i_1^{A}\in \{1,\dots,d_A\}=[d_A].
\end{equation}
For a sequence of indices $\{i_1^A, \dots, i_n^A\}$, a product of $n$ corresponding probabilities $p(i_1^{A}|U^A_1)\cdots p(i_n^{A}|U^A_1)$ is
a homogeneous polynomial of $n$-th degree in $U$ and in $U^\dagger$, which averaged over $U^A$ due to the Haar measure yields
$
 \langle p(i^A_1|U^A)\cdots p(i_n^A|U^A)\rangle_{\mathcal U(d_A)}
$
turning the monomials
into Weingarten integrals \cite{Collins2006,Puchala2017}
and $\langle p(i^A_1|U^A)\cdots p(i_n^A|U^A)\rangle_{\mathcal U(d_A)}$ into a unitarily invariant polynomial of order $n$ in the entries of $\rho$

In case  of $N$--partite system the index $i^A$ becomes a multi--index $I = \{i^{(1)}, \dots, i^{(N)} \}$ denoting results of measurements obtained by subsequent parties. In general, there are two sorts of correlations: the one among the outcomes in subsequent subsystems
in $p(I|\otimes_{\ell=1^N}U^{(\ell)})$ at the fixed measurement basis and the correlations between subsequent probabilities in the averaged product $\left\langle \prod _{k=1}^np(I_k|\otimes_{\ell=1^N} U^{(\ell)})\right\rangle_{\otimes_{\ell=1}^N \mathcal U(d^{(\ell)})}$ 
over local unitaries. 
For example, selecting only the subsystem $(1)$ we have
 \begin{equation}
     \left\langle \prod_{k=1}^n p(i_k^{(1)}|U)\right \rangle_{\mathcal U(d^{(1)})}=
     \sum_{\substack{p^{(1)}_1\dots p_n^{(1)}=1 \\ q_1^{(1)}\dots q_n^{(1)}=1}}^d\rho_{p_1^{(1)}q_1^{(1)}}\cdots\rho_{p_n^{(1)}q_n^{(1)}}
     \int_{ \mathcal U(d^{(1)})} u_{p_1^{(1)} i_1^{(1)}}^* \dots u_{p_n^{(1)} i_n^{(1)}}^* u_{q_1^{(1)} i_1^{(1)}}\dots u_{q_n^{(1)} i_n^{(1)}} \mathrm{d} U
 \end{equation}
 where
\begin{equation}\label{eq:wein}
    \int_{ \mathcal U(d)} u_{p_1 i_1}^* \dots u_{p_n i_n}^* u_{q_1 i_1}\dots u_{q_n i_n} \mathrm{d} U
    = \sum_{\sigma,\tau \in S_n}
    \delta_{i_1 i_{\sigma(1)}} \dots \delta_{i_n i_{\sigma(n)}}
     \mathrm{Wg}(\sigma \tau^{-1}, d)
     \delta_{p_1 q_{\tau(1)}} \dots \delta_{p_n q_{\tau(n)}}\,,
\end{equation}
where $S_n$ is the permutation group and 
\begin{equation}
    \mathrm{Wg}(\sigma,d)=\frac{1}{q!}\sum_\lambda
\frac{\chi^\lambda(1)^2\chi^\lambda(\sigma)}{s_{\lambda,d}(1)}.
\end{equation}
The sum is over all partitions $\lambda$ of $q$ \cite{Collins2006}. Here $\chi^\lambda$ is the character of $S_q$ corresponding to the partition $\lambda$ and $s$ is the Schur polynomial of $\lambda$, so that $s_{\lambda,d}$ is the dimension of the representation of $U(d)$ corresponding to $\lambda$. 
The Weingarten functions are rational functions in d. 
In this paper, we explicitly write the first Weingarten functions that we are going to use in the next sections,
\begin{align}\label{eq:Wg}
    \mathrm{Wg}(1,d_K)
    &= 
    \frac {1}{d_K},\qquad \text{(identity)}\\
    \left(\,\mathrm{Wg}(2,d_K),\,\mathrm{Wg}(1^2,d_K)\right)
    &=
    \frac{\left(\,-1,\,\,d_K\,\right)}{d_K(d_K^2-1)} 
    ,\qquad \text{(transposition, identity)}\label{Wg_2} \\
    \left(\,\mathrm{Wg}(3,d_K),\,\mathrm{Wg}(21,d_K)\,,\mathrm{Wg}(1^3,d_K)\,\right)
    &=
    \frac{\left(\,2,\,\,-d_K,\,\,d_K^2-2\,\right)}{d_K(d_K^2-1)(d_K^2-4)}\equiv (c^K,t^K,i^K), \label{Wg_3} \\
    & \text{(3-cycle, transposition, identity)} \nonumber  
\end{align}
In Eq. \eqref{Wg_2} we gather the Weingarten function for the two elements of $S_2=\{(2),(1^2)\}$, which corresponds to the transposition 
and the identity
.
In Eq. \eqref{Wg_3} the Weingarten function is defined only on the conjugacy classes $\{(1^3),(2),(3)\}$ that groups the six elements of $S_3=\{(1)(2)(3),(12),(23),(31),(312),(231)\}$
: the identity (1-cycle), the transpositions (2-cycle) and the 3-cycle. The symbols $i^K,t^K,c^K$ avoid cumbersome notation in Sec. \ref{sec:3order}. 
The simplest case is
\begin{equation}
    \langle p(i^{(1)}_1|U^1)\rangle_{\mathcal U(d^{(1)})}=\sum_{p^{(1)}_1q^{(1)}_1}
    \rho_{p_n^{(1)}q_n^{(1)}}\sum_{\sigma,\tau\in\{S_1\}}\delta_{i_1^{(1)}i^{(1)}_{\sigma(1)}}\mathrm{Wg}(\sigma\tau^{-1},d^{(1)})\delta_{p_1^{(1)}q_{\tau(1)}^{(1)}}=\frac{\mathrm{Tr}\rho}{d^{(1)}}
\end{equation}
and is the same for all $i^{(1)}_1\in \{ 1, \dots, d^{(1)}\}$.

In bipartite systems, the simplest case is obtained as follows. An inner loop, at fixed local $U^{(1)}\in \mathcal U(d^{(1)})$ and $U^{(2)}\in \mathcal U(d^{(2)})$ correlates the two subsystems

\begin{align}
    p(i^{(1)}_1,i^{(2)}_1|U^{(1)}\otimes U^{(2)})=&\bra{i^{(1)}_1,i^{(2)}_1}(U^{(1)\dagger}\otimes U^{(2)\dagger})\rho_{AB}(U^{(1)}\otimes U^{(2)})\ket{i^{(1)}_1,i^{(2)}_1}\nonumber\\
    =&\sum_{p^{(1)}_1 q_1^{(1)}=1}^{d^{(1)}}\sum_{p^{(2)}_1 q_1^{(2)}=1}^{d^{(2)}}
    u^{(1)*}_{p_1^{(1)}i_1^{(1)}}\,
    u^{(2)*}_{p_1^{(2)}i_1^{(2)}}\,\,
    \rho_{p_1^{(1)}p_1^{(2)},q_1^{(1)}q_1^{(2)}} \,\,
    u^{(1)}_{q_1^{(1)}i_1^{(1)}}\,
    u^{(2)}_{q_1^{(2)}i_1^{(2)}}\,
    .
\end{align}
Then averaging on an outer loop on local unitaries
\begin{align}\label{eq:nonloc}
    \langle p(i^{(1)}_1i^{(2)}_1|U^{(1)}\otimes U^{(2)})\rangle_{\mathcal U(d^{(1)})\otimes \mathcal U(d^{(2)})}=\nonumber\\
    =
    \sum_{p^{(1)}_1 q_1^{(1)}=1}^{d^{(1)}}\sum_{p^{(2)}_1 q_1^{(2)}=1}^{d^{(2)}}\,
    \rho_{p_1^{(1)}p_1^{(2)},q_1^{(1)}q_1^{(2)}}& 
    \,\int
    u^{(1)*}_{p_1^{(1)}i_1^{(1)}}\,
    u^{(1)}_{q_1^{(1)}i_1^{(1)}}\,\mathrm{d}U^{(1)}
    \,\,
    \int
    u^{(2)*}_{p_1^{(2)}i_1^{(2)}}\,\,
    u^{(2)}_{q_1^{(2)}i_1^{(2)}}\,\mathrm{d}U^{(2)}
    \nonumber\\
    =
    \sum_{p^{(1)}_1 q_1^{(1)}=1}^{d^{(1)}}\sum_{p^{(2)}_1 q_1^{(2)}=1}^{d^{(2)}}\,
    \rho_{p_1^{(1)}p_1^{(2)},q_1^{(1)}q_1^{(2)}}& 
    \,\,\, \delta_{i_1^{(1)},i_1^{(1)}}\mathrm{Wg}(1,d^{(1)})\delta_{p_1^{(1)},q_1^{(1)}}
    \,\,\, \delta_{i_1^{(2)},i_1^{(2)}}\mathrm{Wg}(1,d^{(2)})\delta_{p_1^{(2)},q_1^{(2)}}
    \nonumber\\
    =\frac{\mathrm{Tr}\rho}{d^{(1)}d^{(2)}}\qquad\qquad\qquad\qquad\quad\,&
\end{align}
In this case, there is only $1$ copy of probability of result specified by $I_1=(i_1^{(1)},i_1^{(2)})$ in the average over $\mathcal U(d^{(1)}) \otimes U(d^{(2)})$ 
and the same invariant $\Tr\rho$ is obtained. 

In the next section, we will study the less non-trivial bipartite scenario where the average over local unitaries it correlates the measurement results $I_1$ and $I_2$, and new local invariants become accessible. Before that, let us discuss the experimental procedure of the inner and outer loops~\cite{cieslinski2024analysing}.
\begin{itemize}
    \item Inner-loop --- fixing $\bigotimes_{\ell=1}^N U^{(\ell)}$ on the $N$--partite system we estimate the $n$--outcomes correlator
\begin{equation}
    \prod_{k=1}^n
    p\left(i^{(1)}_k\dots i^{(N)}_k\Big\vert\bigotimes_{\ell=1}^N U^{(\ell)}\right)
    =
    \prod_{k=1}^n 
    \bra{i^{(1)}_k\dots i^{(N)}_k}
    \left(\bigotimes_{\ell=1}^N U^{(\ell)}\right)^\dagger
    \rho
    \left(\bigotimes_{\ell=1}^N U^{(\ell)}\right)
    \ket{i^{(1)}_k\dots i^{(N)}_k},
\end{equation}
preparing the state $\rho\in\mathcal{B}(\bigotimes_{\ell=1}^N\mathbb{C}^{d^{(\ell)}})$ 
many times, applying 
\textit{the same} $\bigotimes_\ell U^{(\ell)}$, 
and recording outcomes $I_1,\dots,I_n$ 
one obtains the statistics 
$p(I_1|U^{(1)}\otimes\cdots U^{(n)})\cdots p(I_n|U^{(1)}\otimes\cdots U^{(n)})$. 
Each shot is only ``state-preparation $\to$ local gates $\to$ measurement''.
\item Outer-loop --- statistics sampling over $\{\bigotimes_\ell U^{(\ell)}\}$ needs to target the Haar average
\begin{equation}
    \left\langle \prod_{k=1}^n
    p\left(i^{(1)}_k\dots i^{(N)}_k\Big\vert\bigotimes_{\ell=1}^N U^{(\ell)}\right)\right\rangle_{\bigotimes_{\ell=1}^N \mathcal U(d^{(\ell)})}=
    \int \prod_{k=1}^n
    p\left(I_k|\bigotimes_{\ell=1}^N U^{(\ell)}\right)\,\,\mathrm{d}U^{(1)}\cdots\mathrm{d}U^{(N)}.
\end{equation}
Then, repeat the inner loop for independent choices of $\bigotimes_\ell U^{(\ell)}$; compile a fresh circuit each time. Sampling over $\{\bigotimes_\ell U^{(\ell)}\}$ can be simplified to sampling over an appropriate unitary design.
\end{itemize}
In the next section we analyze how to obtain a relation $\vec y^{(N)}_n=\mathbf M \,\,\vec x^{(N)}_n$, that relates the information of the experimental measurements carried by $\vec{y}^{(N)}_n$ with the information of the local invariants of $\rho$ carried by $\vec{x}^{(N)}_n$. Therefore one accesses to the local invariants of $\rho$ by solving $\vec{x}^{(N)}_n=\mathbf{M}^{-1}\,\,\vec{y}^{(N)}_n$
.
Although two statistical layers are involved in experimentally measuring $\vec{y}^{(N)}_n$, both can be drastically reduced compared to full tomographic access to $x^{(N)}_n$, where the cost increases exponentially~\cite{Elben2022}.

Specifically, in the next section, we obtain the relation $x^{(N)}_n=\mathbf{M}^{-1}\vec{y}^{(2)}_2$ where the vector of invariants $\vec{x}_2^{(2)}$
carries information about mixed moments such as $\Tr\rho_{AB}^2$, $\Tr\rho_A^2$ and $\Tr\rho_B^2$. 

\section{Second--order correlations in bipartite systems via local unitaries}
Let us start from
\begin{align}
    \langle p(i_1^A,i_1^B|U^A\otimes U^B)p(i_2^A,i_2^B|U^A\otimes U^B) \rangle_{\mathcal{U}(d_A) \otimes \mathcal{U}(d_B)}
        =&  \sum_{\substack{p_1^A,p_1^B,q_1^A,q_1^B \\ p_2^A,p_2^B,q_2^A,p_2^B}}
        \rho_{p_1^Ap_1^B,q_1^Aq_1^B}\,\, \rho_{p_2^Ap_2^B,q_2^Aq_2^B}
        \nonumber\\
        \times\int_{\mathcal U(d_A)} u_{p_1^Ai_1^A}^{A*} u_{p_2^Ai_2^A}^{A*} u_{q_1^Ai_1^A}^A &u_{q_2^Ai_2^A}^A \mathrm{d} U^A
        \int_{\mathcal U(d_B)} u_{p_1^Bi_1^B}^{B*} u_{p_2^Bi_2^B}^{B*} u_{q_1^Bi_1^B}^{B} u_{q_2^Bi_2^B}^{B} \mathrm{d} U^B.
\end{align}
LHS is measured experimentally by estimating the $(d_A d_B)^2$ possible outcomes for the inner--loop and for each of the possible outcomes, the outer-loop averages over $\mathcal U(d_A)\otimes \mathcal U(d_B)$
\begin{align}\label{eq:2outcomes}
    \{\langle p(i_1^A,i_1^B|U^A\otimes U^B)p(i_2^A,i_2^B|U^A\otimes U^B) \rangle_{\mathcal{U}(d_A)\otimes \mathcal{U}(d_B)}\}_{(i_1^{A},i_2^{A})\in [d_A]^2,(i_1^{B},i_2^{(B)})\in[d_B]^{2}}.
\end{align}
At fixed 
$i_1^{A},i_2^{A},i_1^{B},i_2^{B}$ using Eq. \eqref{eq:wein}
\begin{align}
\langle p(i_1^A,i_1^B|U^A\otimes U^B)p(i_2^A,i_2^B|U^A\otimes U^B) \rangle_{\mathcal{U}(d_A)\otimes \mathcal{U}(d_B)}&=\sum_{\stackrel{p_1^A,p_1^B;q_1^A q_1^B}{p_2^A,p_2^B;q_2^A q_2^B}}\rho_{p_1^Ap_1^B,q_1^Aq_1^B} \,\,\rho_{p_2^Ap_2^B,q_2^Aq_2^B}\nonumber\\
    \times\sum_{\sigma_{A},\tau_{A}\in S_{2}}&\delta_{i_{1}^Ai^A_{\sigma_{A}(1)}}\delta_{i_{2}^Ai^A_{\sigma_{A}(2)}}
    W\left(\sigma_{A}\tau_{A}^{-1},d_A\right)
    \delta_{p_{1}^Aq^A_{\tau_{A}(1)}}\delta_{p_{2}^Aq^A_{\tau_{A}(2)}}\nonumber\\
    \times\sum_{\sigma_{B},\tau_{B}\in S_{2}}&\delta_{i^B_{1}i^B_{\sigma_{B}(1)}}\delta_{i^B_{2}i^B_{\sigma_{B}(2)}}
    W\left(\sigma_{B}\tau_{B}^{-1},d_B\right)
    \delta_{p^B_{1}q^B_{\tau_{B}(1)}}\delta_{p^B_{2}q^B_{\tau_{B}(2)}}
\end{align}
where elements of $S_2$ are the identity $1^2:i\to i$ (cycle type: $1^2$) and the transposition $(2):i \mapsto i\oplus_2 1$ (cycle type: $2$). From Weingarten calculus,
      \begin{align}
        \langle p(i_1^A,i_1^B|U^A\otimes U^B)p(i_2^A,i_2^B|U^A\otimes U^B) \rangle_{\mathcal{U}(d_A)\otimes \mathcal{U}(d_B)}&
        = \sum_{\substack{p_1,p_1^B,q_1^A,q_1^B \\ p_2^A,p_2^B,q_2^A,q_2^B}}
        \rho_{p_1p_1^B,q_1^Aq_1^B} \rho_{p_2^Ap_2^B,q_2^Aq_2^B}\nonumber\\
        \times\Big( 
            (\delta_{i_1^Ai_1^A} \delta_{i^A_2i^A_2} \delta_{p_1^Aq_1^A}\delta_{p_2^Aq_2^A}
            + \delta_{i_1^Ai_2^A} \delta_{i_2^Ai_1^A}& \delta_{p_2^Aq_1^A}\delta_{p_1^Aq_2^A}) \mathrm{Wg} (1^2,d_A)\nonumber \\
            \,\,+  
            (\delta_{i_1^Ai_2^A} \delta_{i_2^Ai_1^A} \delta_{p_1^Aq_1^A}\delta_{p_2^Aq_2^A}
            + \delta_{i_1^Ai_1^A} \delta_{i_2^Ai_2^A}& \delta_{p_2^Aq_1^A}\delta_{p_1^Aq_2^A}) \mathrm{Wg} (2,d_A)
        \Big) \Big( \text{the same for }B \Big)
      \end{align}
Using Eq. \eqref{Wg_2},
\begin{align}\label{eq:curly2}
    \langle p(i_1^A,i_1^B|U^A\otimes U^B)\,p(i_2^A,i_2^B|U^A\otimes U^B)& \rangle_{\mathcal{U}(d_A)\otimes \mathcal{U}(d_B)}= \sum_{\substack{p_1,p_1^B,q_1^A,q_1^B \\ p_2^A,p_2^B,q_2^A,q_2^B}}
        \rho_{p_1p_1^B,q_1^Aq_1^B}\,\, \rho_{p_2^Ap_2^B,q_2^Aq_2^B}
         \nonumber\\
        \times\frac{1}{d_A(d_A^{2} - 1)}&\left\{ \begin{array}{cc} 
            (d_A-1) (\delta_{p_1^Aq_1^A}\,\delta_{p_2^Aq_2^A} + \delta_{p_2^Aq_1^A} \,\delta_{p_1^Aq_2^A}) & \text{if} \ i_1^A = i_2^A \\
            d_A\,\delta_{p_1^Aq_1^A}\,\delta_{p_2^A q_2^A} - \delta_{p_2^Aq_1^A}\,\delta_{p_1^Aq_2^A} & \text{if} \ i_1^A \ne i_2^A
        \end{array} \right.
        \nonumber\\
        \times\frac{1}{d_B(d_B^{2} - 1)}&\left\{ \begin{array}{cc} 
            (d_B-1) (\delta_{p_1^Bq_1^B}\,\delta_{p_2^Bq_2^B} + \delta_{p_2^Bq_1^B} \,\delta_{p_1^Bq_2^B}) & \text{if} \ i_1^B = i_2^B \\
            d_B\,\delta_{p_1^Bq_1^B}\,\delta_{p_2^B q_2^B} - \delta_{p_2^Bq_1^B}\,\delta_{p_1^Bq_2^B} & \text{if} \ i_1^B \ne i_2^B
        \end{array} \right.
\end{align}
Since now we will often represent $\rho_{p_1^Ap_1^B,q_1^Aq_1^B}$via \emph{Penrose's graphical notation} \cite{coecke2018picturing,penrose1971applications} as
\begin{equation}
\begin{tikzpicture}
	\begin{pgfonlayer}{nodelayer}
		\node (12) at (-0.25, 0.25) {};
		\node [style=none] (13) at (0, 0.5) {};
		\node [style=none] (14) at (0.5, 0.25) {};
		\node [style=none] (15) at (0.25, 0.5) {};
		\node [style=none] (16) at (0.5, -0.25) {};
		\node [style=none] (17) at (0.25, -0.5) {};
		\node [style=none] (18) at (-0.25, -0.25) {};
		\node [style=none] (19) at (0, -0.5) {};
		\node [style=none] (20) at (0.5, -0.25) {};
		\node [style=none] (21) at (-0.25, -0.25) {};
		\node [style=none] (22) at (-0.25, -0.25) {};
		\node [style=none] (24) at (-0.25, 0.25) {};
		\node [style=none] (27) at (0.5, -0.25) {};
		\node [style=none] (29) at (0.5, 0.25) {};
		\node [style=none] (30) at (-0.75, 0.25) {};
		\node [style=none] (31) at (-0.75, -0.25) {};
		\node [style=none] (32) at (1, 0.25) {};
		\node [style=none] (33) at (1, -0.25) {};
		\node [style=none] (36) at (1.5, -0.5) {$q_1^B$};
		\node [style=none] (37) at (1.5, 0.5) {$q_1^A$};
		\node [style=none] (38) at (-1.25, 0.5) {$p_1^A$};
		\node [style=none] (39) at (-1.25, -0.5) {$p_1^B$};
		\node [style=none] (41) at (4, 0) {bipartite,};
		\node [style=none] (42) at (8.75, 1) {};
		\node [style=none] (43) at (9, 1.25) {};
		\node [style=none] (44) at (9.5, 1) {};
		\node [style=none] (45) at (9.25, 1.25) {};
		\node [style=none] (46) at (9.5, -1.5) {};
		\node [style=none] (47) at (9.25, -1.75) {};
		\node [style=none] (48) at (8.75, -1.5) {};
		\node [style=none] (49) at (9, -1.75) {};
		\node [style=none] (50) at (9.5, -1.5) {};
		\node [style=none] (51) at (8.75, -1.5) {};
		\node [style=none] (52) at (8.75, -1.5) {};
		\node [style=none] (53) at (8.75, 1) {};
		\node [style=none] (54) at (9.5, -1.5) {};
		\node [style=none] (55) at (9.5, 1) {};
		\node [style=none] (56) at (8.25, 1) {};
		\node [style=none] (57) at (8.25, -1.5) {};
		\node [style=none] (58) at (10, 1) {};
		\node [style=none] (59) at (10, -1.5) {};
		\node [style=none] (60) at (10.5, -1.75) {$q_1^{(N)}$};
		\node [style=none] (61) at (10.5, 1.25) {$q_1^{(1)}$};
		\node [style=none] (62) at (7.75, 1.25) {$p_1^{(1)}$};
		\node [style=none] (63) at (7.75, -1.75) {$p_1^{(N)}$};
		\node [style=none] (65) at (20, 0.25) {$m$--partite};
		\node [style=none] (66) at (8.75, 0.5) {};
		\node [style=none] (67) at (8.25, 0.5) {};
		\node [style=none] (68) at (7.5, 0.25) {$p_1^{(2)}$};
		\node [style=none] (69) at (8.25, -0.5) {};
		\node [style=none] (70) at (8.25, -0.5) {$\vdots$};
		\node [style=none] (71) at (10, -0.5) {$\vdots$};
		\node [style=none] (72) at (9.5, 0.5) {};
		\node [style=none] (73) at (10, 0.5) {};
		\node [style=none] (74) at (10.5, 0.25) {$q_1^{(2)}$};
		\node [style=none] (75) at (15.25, 1) {};
		\node [style=none] (76) at (15.5, 1.25) {};
		\node [style=none] (77) at (16, 1) {};
		\node [style=none] (78) at (15.75, 1.25) {};
		\node [style=none] (79) at (16, -1.5) {};
		\node [style=none] (80) at (15.75, -1.75) {};
		\node [style=none] (81) at (15.25, -1.5) {};
		\node [style=none] (82) at (15.5, -1.75) {};
		\node [style=none] (83) at (16, -1.5) {};
		\node [style=none] (84) at (15.25, -1.5) {};
		\node [style=none] (85) at (15.25, -1.5) {};
		\node [style=none] (86) at (15.25, 1) {};
		\node [style=none] (87) at (16, -1.5) {};
		\node [style=none] (88) at (16, 1) {};
		\node [style=none] (89) at (14.75, 1) {};
		\node [style=none] (90) at (14.75, -1.5) {};
		\node [style=none] (91) at (16.5, 1) {};
		\node [style=none] (92) at (16.5, -1.5) {};
		\node [style=none] (93) at (17, -1.75) {$q_n^{(N)}$};
		\node [style=none] (94) at (17, 1.25) {$q_n^{(1)}$};
		\node [style=none] (95) at (14.25, 1.25) {$p_n^{(1)}$};
		\node [style=none] (96) at (14.25, -1.75) {$p_n^{(N)}$};
		\node [style=none] (97) at (15.25, 0.5) {};
		\node [style=none] (98) at (14.75, 0.5) {};
		\node [style=none] (99) at (14.75, -0.5) {};
		\node [style=none] (100) at (14.75, -0.5) {$\vdots$};
		\node [style=none] (101) at (16.5, -0.5) {$\vdots$};
		\node [style=none] (102) at (16, 0.5) {};
		\node [style=none] (103) at (16.5, 0.5) {};
		\node [style=none] (104) at (17, 0.25) {$q_n^{(2)}$};
		\node [style=none] (105) at (21.25, -0.75) {$n$--order correlation};
		\node [style=none] (106) at (12.25, -0.5) {$\cdots$};
		\node [style=none] (107) at (14.25, 0.25) {$q_n^{(2)}$};
	\end{pgfonlayer}
	\begin{pgfonlayer}{edgelayer}
		\draw [ bend right=45, looseness=1.75] (13.center) to (12.center);
		\draw [bend left=45, looseness=1.75] (15.center) to (14.center);
		\draw [bend right=45, looseness=1.75] (17.center) to (16.center);
		\draw [bend left=45, looseness=1.75] (19.center) to (18.center);
		\draw (12.center) to (18.center);
		\draw (14.center) to (16.center);
		\draw (19.center) to (17.center);
		\draw (13.center) to (15.center);
		\draw (32.center) to (29.center);
		\draw (33.center) to (27.center);
		\draw (22.center) to (31.center);
		\draw (24.center) to (30.center);
		\draw [bend right=45, looseness=1.75] (43.center) to (42.center);
		\draw [bend left=45, looseness=1.75] (45.center) to (44.center);
		\draw [bend right=45, looseness=1.75] (47.center) to (46.center);
		\draw [bend left=45, looseness=1.75] (49.center) to (48.center);
		\draw (42.center) to (48.center);
		\draw (44.center) to (46.center);
		\draw (49.center) to (47.center);
		\draw (43.center) to (45.center);
		\draw (58.center) to (55.center);
		\draw (59.center) to (54.center);
		\draw (52.center) to (57.center);
		\draw (53.center) to (56.center);
		\draw (66.center) to (67.center);
		\draw (73.center) to (72.center);
		\draw [bend right=45, looseness=1.75] (76.center) to (75.center);
		\draw [bend left=45, looseness=1.75] (78.center) to (77.center);
		\draw [bend right=45, looseness=1.75] (80.center) to (79.center);
		\draw [bend left=45, looseness=1.75] (82.center) to (81.center);
		\draw (75.center) to (81.center);
		\draw (77.center) to (79.center);
		\draw (82.center) to (80.center);
		\draw (76.center) to (78.center);
		\draw (91.center) to (88.center);
		\draw (92.center) to (87.center);
		\draw (85.center) to (90.center);
		\draw (86.center) to (89.center);
		\draw (97.center) to (98.center);
		\draw (103.center) to (102.center);
	\end{pgfonlayer}
\end{tikzpicture}}.
\end{equation}
Note that the cycles of $\tau_A$ correspond to copies of $\rho$ that are multiplied and traced over the subsystem $A$, while the cycles of $\tau_B$ correspond to copies of $\rho$ multiplied and traced over the subsystem $B$. All such $\tau$ permutations can thus be mapped to invariants represented using \textit{Penrose diagrams}, where all free pins of copies of $\rho$ are connected according to Kronecker deltas. All diagrams corresponding to all $\tau$'s are present in equation for any choice of $i_1^A,i_2^A,i_1^B,i_2^B$, while the permutations $\sigma_A$ and $\sigma_B$ determine the coefficients multiplying each invariant (diagram), being combinations of the Weingarten functions $\mathrm{Wg}$ in Eq.~\eqref{Wg_2}.

Each $\tau$ defines a map acting in the corresponding subsystem:
\begin{equation}\label{eq:Paction}
    \mathcal{P}_{\tau^A}:
    \rho^{\otimes 2}
    \mapsto 
    \sum_{p_1^A,p_2^A;q_1^Aq_2^A} 
    \rho_{p_1^A,\_;q_1^A,\_}\,\rho_{p_2^A,\_;q_2^A,\_}\,
    \delta_{p_{1}^Aq^A_{\tau_{A}(1)}}\delta_{p_{2}^Aq^A_{\tau_{A}(2)}}
\end{equation}
Let us denote $\mathcal{P}_{\tau_A}$ as $I_A$ if $\tau_A$ is identity and $T_A$ if $\tau_A$ is transposition.
Let us arrange $I_A$ and $I_B$ in a vector $r_A$. Its action on $\rho^{\otimes 2}$ is:
\begin{equation}\label{eq:r2}
    r_A \,\rho^{\otimes 2}
    =
    \left[\begin{array}{c}
        I_A \\ T_A
    \end{array}\right]
    \rho^{\otimes 2}
    =
    \left[
    \begin{array}{c}\begin{tikzpicture}[scale=.6]
                \draw [fill=green] (-.5,-.5) -- (-.5,.5) -- (.5,.5) -- (.5,-.5) -- cycle;
                \draw [fill=green] (1.5,-.5) -- (1.5,.5) -- (2.5,.5) -- (2.5,-.5) -- cycle;
                \node  (LU0) at (-0.5, 0.25) {};
                \node  (RU0) at (0.5, 0.25) {};
                \node  (LU1) at (1.5, 0.25) {};
                \node  (RU1) at (2.5, 0.25) {};
                \node  (U0) at (0, 1) {};
                \node  (U1) at (2, 1) {};
                \draw (LU0.center) to[out=180,in=180,looseness=1.5] (U0.center) to[out=0,in=0,looseness=1.5] (RU0.center);
                \draw (LU1.center) to[out=180,in=180,looseness=1.5] (U1.center) to[out=0,in=0,looseness=1.5] (RU1.center);
            \end{tikzpicture}
            \\
            \begin{tikzpicture}[scale=.6]
                \draw [fill=green] (-.5,-.5) -- (-.5,.5) -- (.5,.5) -- (.5,-.5) -- cycle;
                \draw [fill=green] (2-.5,-.5) -- (2-.5,.5) -- (2+.5,.5) -- (2+.5,-.5) -- cycle;
                \node  (LU0) at (-0.5, 0.25) {};
                \node  (RU0) at (0.5, 0.25) {};
                \node  (LU1) at (1.5, 0.25) {};
                \node  (RU1) at (2.5, 0.25) {};
                \node  (U0) at (0, 1) {};
                \node  (U1) at (2, 1) {};
                \draw (LU0.center) to[out=180,in=180,looseness=1.5] (U0.center) -- (U1.center) to[out=0,in=0,looseness=1.5] (RU1.center);
                \draw (RU0.center) -- (LU1.center);
            \end{tikzpicture}
    \end{array}\right]
\end{equation}
(similar for $B$). We can rewrite Eq.~\eqref{eq:curly2} as
\begin{align}\label{eq:sigmatau}
    \langle p(i_1^A,i_1^B|U^A\otimes U^B)\,p(i_2^A,i_2^B|U^A\otimes U^B) \rangle_{\mathcal{U}(d_A)\otimes \mathcal{U}(d_B)}
    &
    =
    \left[\begin{array}{cc}
         1& \delta_{i_1^A,i_2^A}
     \end{array}\right]
     \left[\begin{array}{cc}
         \mathrm{Wg}(1,d_A)&\mathrm{Wg}(2,d_A)\\ \mathrm{Wg}(2,d_A)&\mathrm{Wg}(1,d_A)
     \end{array}\right]
     \left[\begin{array}{cc}
         I_A\\T_A
    \end{array}\right]
    \nonumber\\
    &
    \otimes
    \left[\begin{array}{cc}
         1& \delta_{i_1^B,i_2^B}
     \end{array}\right]
     \left[\begin{array}{cc}
         \mathrm{Wg}(1,d_B)&\mathrm{Wg}(2,d_B)\\ \mathrm{Wg}(2,d_B)&\mathrm{Wg}(1,d_B)
     \end{array}\right]
      \left[\begin{array}{cc}
         I_B\\T_B
    \end{array}\right]
    \rho \otimes \rho
\end{align}
The row vectors $[1,\, \delta_{i_1^A,i_2^A}]$ and $[1,\, \delta_{i_1^B,i_2^B}]$ are due to the contribution of $\sigma_A$ and $\sigma_B$ respectively. 

Observe that by performing a permutation of basis vectors in each subsystem does not change the value of the LHS, hence we obtain only four different equations determined by equality / non-equality of indices in subsequent subsystems. 

For this reason, we define a matrix of equalities (inequalities), let us say $S$, to take into account the cases $i_1^A=i_2^A$ and $i_1^A\neq i_2^A$. This allows us to write Eq.~\eqref{eq:curly2} as follows 
\begin{equation}\label{eq:S2}
 \vec{y}^{(2,2)}= \mathbf M \,\vec x^{(2,2)},\text{ with }
 \mathbf M=SW\otimes SW,\qquad
 S=\left[
 \begin{array}{cc}
     1 & 1\\ 1&0
 \end{array}\right],\qquad
 W_A=\left[
 \begin{array}{cc}
    \mathrm{Wg}(1,d_A)&\mathrm{Wg}(2,d_A)\\ \mathrm{Wg}(2,d_A)&\mathrm{Wg}(1,d_A)
 \end{array}\right].
\end{equation}
where $\vec{y}^{(2,2)}$ are the experimentally accessible quantities, a.k.a. $(2,2)$--\textit{twirling} for two parties and two points measurement $I_1=(i_1^A,i_1^B)$ and $I_2=(i_2^A,i_2^B)$ and $\vec x^{(2,2)}=(1,\Tr\rho_B^2,\Tr\rho_A^2,\Tr\rho^2)$ are the invariants obtained by local unitaries represented in vector of diagrams as follows:
\begin{align}\label{eq:y22}
\vec{y}^{(2,2)}=&
\left\langle p\left[
\begin{pmatrix}
    i_1^A
    =
    i_2^A
    \\
    i_1^A
    \neq 
    i_2^A
\end{pmatrix}
\otimes
\begin{pmatrix}
    i_1^B
    =
    i_2^B
    \\
    i_1^B
    \neq 
    i_2^B
\end{pmatrix}
\Big\vert U\otimes V\right]
\right\rangle_{\mathcal{U}(d_A)\otimes \mathcal{U}(d_B)}\nonumber\\
=&
\frac{1}{d_Ad_B(d_A^{2}-1)(d_B^{2}-1)}\left[
\begin{array}{cc}
    d_A-1&d_A-1\\
    d_A&-1
\end{array}\right]
\otimes
\left[
\begin{array}{cc}
    d_B-1&d_B-1\\
    d_B&-1
\end{array}
\right]
\left[\begin{array}{c}
        I_A \\ T_A
    \end{array}\right]
\otimes
\left[\begin{array}{c}
        I_B \\ T_B
    \end{array}\right]
\rho^{\otimes 2}
\nonumber\\
=&
\frac{1}{d_Ad_B(d_A^{2}-1)(d_B^{2}-1)}\left[
\begin{array}{cc}
    d_A-1&d_A-1\\
    d_A&-1
\end{array}\right]
\otimes
\left[
\begin{array}{cc}
    d_B-1&d_B-1\\
    d_B&-1
\end{array}
\right]
\left[\begin{array}{c}
\begin{tikzpicture}[scale=.4]
                \draw [fill=green] (-.5,-.5) -- (-.5,.5) -- (.5,.5) -- (.5,-.5) -- cycle;
                \draw [fill=green] (1.5,-.5) -- (1.5,.5) -- (2.5,.5) -- (2.5,-.5) -- cycle;
                \node  (LU0) at (-0.5, 0.25) {};
                \node  (RU0) at (0.5, 0.25) {};
                \node  (LD0) at (-0.5, -0.25) {};
                \node  (RD0) at (0.5, -0.25) {};
                \node  (LU1) at (1.5, 0.25) {};
                \node  (RU1) at (2.5, 0.25) {};
                \node  (LD1) at (1.5, -0.25) {};
                \node  (RD1) at (2.5, -0.25) {};
                \node  (U0) at (0, 1) {};
                \node  (D0) at (0, -1) {};
                \node  (U1) at (2, 1) {};
                \node  (D1) at (2, -1) {};
                \draw (LU0.center) to[out=180,in=180,looseness=1.5] (U0.center) to[out=0,in=0,looseness=1.5] (RU0.center);
                \draw (LU1.center) to[out=180,in=180,looseness=1.5] (U1.center) to[out=0,in=0,looseness=1.5] (RU1.center);
                \draw (LD0.center) to[out=180,in=180,looseness=1.5] (D0.center) to[out=0,in=0,looseness=1.5] (RD0.center);
                \draw (LD1.center) to[out=180,in=180,looseness=1.5] (D1.center) to[out=0,in=0,looseness=1.5] (RD1.center);
            \end{tikzpicture}
            \\
            \begin{tikzpicture}[scale=.4]
                \draw [fill=green] (-.5,-.5) -- (-.5,.5) -- (.5,.5) -- (.5,-.5) -- cycle;
                \draw [fill=green] (2-.5,-.5) -- (2-.5,.5) -- (2+.5,.5) -- (2+.5,-.5) -- cycle;
                \node  (LU0) at (-0.5, 0.25) {};
                \node  (RU0) at (0.5, 0.25) {};
                \node  (LD0) at (-0.5, -0.25) {};
                \node  (RD0) at (0.5, -0.25) {};
                \node  (LU1) at (1.5, 0.25) {};
                \node  (RU1) at (2.5, 0.25) {};
                \node  (LD1) at (1.5, -0.25) {};
                \node  (RD1) at (2.5, -0.25) {};
                \node  (U0) at (0, 1) {};
                \node  (D0) at (0, -1) {};
                \node  (U1) at (2, 1) {};
                \node  (D1) at (2, -1) {};
                \draw (LU0.center) to[out=180,in=180,looseness=1.5] (U0.center) -- (U1.center) to[out=0,in=0,looseness=1.5] (RU1.center);
                \draw (RU0.center) -- (LU1.center);
                \draw (LD0.center) to[out=180,in=180,looseness=1.5] (D0.center) to[out=0,in=0,looseness=1.5] (RD0.center);
                \draw (LD1.center) to[out=180,in=180,looseness=1.5] (D1.center) to[out=0,in=0,looseness=1.5] (RD1.center);
            \end{tikzpicture}
            \\
            \begin{tikzpicture}[scale=.4]
                \draw [fill=green] (-.5,-.5) -- (-.5,.5) -- (.5,.5) -- (.5,-.5) -- cycle;
                \draw [fill=green] (1.5,-.5) -- (1.5,.5) -- (2.5,.5) -- (2.5,-.5) -- cycle;
                \node  (LU0) at (-0.5, 0.25) {};
                \node  (RU0) at (0.5, 0.25) {};
                \node  (LD0) at (-0.5, -0.25) {};
                \node  (RD0) at (0.5, -0.25) {};
                \node  (LU1) at (1.5, 0.25) {};
                \node  (RU1) at (2.5, 0.25) {};
                \node  (LD1) at (1.5, -0.25) {};
                \node  (RD1) at (2.5, -0.25) {};
                \node  (U0) at (0, 1) {};
                \node  (D0) at (0, -1) {};
                \node  (U1) at (2, 1) {};
                \node  (D1) at (2, -1) {};
                \draw (LD0.center) to[out=180,in=180,looseness=1.5] (D0.center) -- (D1.center) to[out=0,in=0,looseness=1.5] (RD1.center);
                \draw (RD0.center) -- (LD1.center);
                \draw (LU0.center) to[out=180,in=180,looseness=1.5] (U0.center) to[out=0,in=0,looseness=1.5] (RU0.center);
                \draw (LU1.center) to[out=180,in=180,looseness=1.5] (U1.center) to[out=0,in=0,looseness=1.5] (RU1.center);
            \end{tikzpicture}
            \\
            \begin{tikzpicture}[scale=.4]
                \draw [fill=green] (-.5,-.5) -- (-.5,.5) -- (.5,.5) -- (.5,-.5) -- cycle;
                \draw [fill=green] (1.5,-.5) -- (1.5,.5) -- (2.5,.5) -- (2.5,-.5) -- cycle;
                \node  (LU0) at (-0.5, 0.25) {};
                \node  (RU0) at (0.5, 0.25) {};
                \node  (LD0) at (-0.5, -0.25) {};
                \node  (RD0) at (0.5, -0.25) {};
                \node  (LU1) at (1.5, 0.25) {};
                \node  (RU1) at (2.5, 0.25) {};
                \node  (LD1) at (1.5, -0.25) {};
                \node  (RD1) at (2.5, -0.25) {};
                \node  (U0) at (0, 1) {};
                \node  (D0) at (0, -1) {};
                \node  (U1) at (2, 1) {};
                \node  (D1) at (2, -1) {};
                \draw (LD0.center) to[out=180,in=180,looseness=1.5] (D0.center) -- (D1.center) to[out=0,in=0,looseness=1.5] (RD1.center);
                \draw (RD0.center) -- (LD1.center);
                \draw (LU0.center) to[out=180,in=180,looseness=1.5] (U0.center) -- (U1.center) to[out=0,in=0,looseness=1.5] (RU1.center);
                \draw (RU0.center) -- (LU1.center);
            \end{tikzpicture}       
\end{array}\right].
\end{align}
We obtain the vector of local invariants by solving the above equation:

\begin{align}\label{eq:x22}
     & = 
    \left[ \begin{array}{c}
                \mathrm{Tr}(\mathrm{Tr_{AB}} \rho )^2      \\
                \mathrm{Tr}(\mathrm{Tr_A} \rho)^2           \\
                \mathrm{Tr}(\mathrm{Tr_B} \rho)^2           \\\mathrm{Tr}(\mathrm{Tr}_{\emptyset} \rho)^2
            \end{array} \right]
    = d_A \left[ \begin{array}{cc} 1 & (d_A - 1) \\ d_A & -(d_A - 1) \end{array} \right]
    \otimes
    d_B \left[ \begin{array}{cc} 1 & (d_B - 1) \\ d_B & -(d_B - 1) \end{array} \right]
    \left[ \begin{array}{c}
                p_{i_1^A=i_2^A,i_1^B=i_2^B} \\ p_{i_1^A=i_2^A,i_1^B\ne i_2^B} \\ p_{i_1^A\ne i_2^A,i_1^B=i_2^B} \\ p_{i_1^A\ne i_2^A,i_1^B\ne i_2^B}
            \end{array} \right]
    \nonumber\\
    =& 
    \left[ \begin{array}{cc} 1 & 1 \\ d_A & -1 \end{array} \right]
    \otimes
    \left[ \begin{array}{cc} 1 & 1 \\ d_B & -1 \end{array} \right]
    \left[ \begin{array}{c}
                \sum_{i_1^A=i_2^A} \sum_{i_1^B=i_2^B} \langle p(i^A_1,i^B_1|U^A\otimes U^B)p(i^A_2,i^B_2|U^A\otimes U^B)\rangle_{\mathcal{U}(d_A)\otimes \mathcal{U}(d_B)} \\ 
                \sum_{i_1^A=i_2^A} \sum_{i_1^B\ne i_2^B} \langle p(i^A_1,i^B_1|U^A\otimes U^B)p(i^A_2,i^B_2|U^A\otimes U^B)\rangle_{\mathcal{U}(d_A)\otimes \mathcal{U}(d_B)}  \\ 
                \sum_{i_1^A\ne i_2^A} \sum_{i_1^B=i_2^B} \langle p(i^A_1,i^B_1|U^A\otimes U^B)p(i^A_2,i^B_2|U^A\otimes U^B)\rangle_{\mathcal{U}(d_A)\otimes \mathcal{U}(d_B)}  \\ 
                \sum_{i_1^A\ne i_2^A} \sum_{i_1^B\neq i_2^B}\langle p(i^A_1,i^B_1|U^A\otimes U^B)p(i^A_2,i^B_2|U^A\otimes U^B)\rangle_{\mathcal{U}(d_A)\otimes \mathcal{U}(d_B)}  
            \end{array} \right].
\end{align}

In the last equality, we 
average over all possible choices of $i_1^A, i_2^A, i_1^B, i_2^B$ satisfying equality/non-equality relations in subsystems. 
There are $d_Ad_B$, $d_Ad_B(d_B-1)$, $d_A(d_A-1)d_B$, and $d_A(d_A-1)d_B(d_B-1)$ possibilities that satisfy  $i_1^A=i_2^A,i_1^B=i_2^B$, $i_1^A=i_2^A,i_1^B\neq i_2^B$, $i_1^A\neq i_2^A,i_1^B=i_2^B$, and $i_1^A\neq i_2^A,i_1^B\neq i_2^B$  respectively.
All of them are obviously equal due to be results of averaging over all possible local bases. 

\subsection{Purity for $N$--partite systems with arbitrary local dimensions }
We rewrite Eq.~\eqref{eq:x22} one more time to facilitate its generalization for the $N$--partite system with
$I_1=(i^A_1,i^B_1)$ and
    $I_2=(i^A_2,i^B_2)$
\begin{align}\label{eq:sum}
    \left[ \begin{array}{c}
                \mathrm{Tr}(\rho_\emptyset )^2      \\
                \mathrm{Tr}(\rho_{\{B\}})^2           \\
                \mathrm{Tr}(\rho_{\{A\}})^2           \\\mathrm{Tr}(\rho_{\{AB\}})^2
            \end{array} \right]
    &
    =
    \left[ \begin{array}{cc} 1 & 1 \\ d_A & -1 \end{array} \right]
    \otimes
    \left[ \begin{array}{cc} 1 & 1 \\ d_B & -1 \end{array} \right]
    \left[ \begin{array}{c}
                \sum_{
                \stackrel{
                I_1,I_2\in [d_A] \times [d_B]
                }
                {
                I_1^{(\ell)} \ne I_2^{(\ell)} \Leftrightarrow \ell \in \emptyset}
                }
                \langle p(I_1|U^A\otimes U^B)p(I_2|U^A\otimes U^B)\rangle_{\mathcal{U}(d_A)\otimes \mathcal{U}(d_B)} \\ 
                \sum_{
                \stackrel{
                I_1,I_2\in [d_A] \times [d_B]
                }
                {
                I_1^{(\ell)} \ne I_2^{(\ell)} \Leftrightarrow \ell  \in \{B\}}
                }
                \langle p(I_1|U^A\otimes U^B)p(I_2|U^A\otimes U^B)\rangle_{\mathcal{U}(d_A)\otimes \mathcal{U}(d_B)}\\ 
                 \sum_{
                \stackrel{
                I_1,I_2\in [d_A] \times [d_B]
                }
                {
                I_1^{(\ell)} \ne I_2^{(\ell)} \Leftrightarrow \ell  \in \{A\}}
                }
                \langle p(I_1|U^A\otimes U^B)p(I_2|U^A\otimes U^B)\rangle_{\mathcal{U}(d_A)\otimes \mathcal{U}(d_B)} \\ 
                 \sum_{
                \stackrel{
                I_1,I_2\in [d_A] \times [d_B]
                }
                {
                I_1^{(\ell)} \ne I_2^{(\ell)} \Leftrightarrow \ell  \in \{A,B\}}
                }
                \langle p(I_1|U^A\otimes U^B)p(I_2|U^A\otimes U^B)\rangle_{\mathcal{U}(d_A)\otimes \mathcal{U}(d_B)}
            \end{array} \right],
\end{align}
In particular, $\Tr_{AB}\rho=\rho_\emptyset=1$ and $\Tr_\emptyset\rho=\rho$.
Now, Eq.~\eqref{eq:sum} can be easily generalised to an arbitrary number of $N$ subsystems. The LHS vector and the RHS vectors are characterized respectively by $P,Q$, the subsets of $\{1,\dots,N\}$ ordered from the empty set to the whole set binarly ($\emptyset$, $\{1\}$, $\{2\}$, $\{1,2\}, \{3\}, \dots$).
Subsystems in $Q$ are the ones where the corresponding entries in multi-indices have to be different and $P$ denotes the subsystems comming into the marginal on which the purity is calculated,
\begin{equation}
\left[ \mathrm{Tr} \rho_P^2 \right]_{P \subset \{1, \dots, N\}} =
\bigotimes_{\ell = 1}^{N}
\begin{bmatrix}
1 & 1 \\
d_\ell & -1
\end{bmatrix}
\left[
\sum_{\substack{
I_1, I_2 \in \bigtimes_{\ell \in \{1, \dots, N\}} [d_\ell]\\
I_1^{(\ell)} \ne I_2^{(\ell)} \Leftrightarrow \ell \in Q}}
\left\langle 
p\left(I_1 \Big| \bigotimes_{\ell=1}^N U_\ell \right)
p\left(I_2 \Big| \bigotimes_{\ell=1}^N U_\ell \right)
\right\rangle_{\bigotimes_{\ell=1}^N \mathcal{U}(d^{(\ell)})}
\right]_{Q \subset \{1, \dots, N\}}.
\end{equation}

Finally, observe that due to the first row vectors of ones in the tensor product of matrices, if $\ell \not \in P$, then the condition $I^{(\ell)}_1\neq I^{(\ell)}_2$ is not checked on the $\ell$-th subsystem and the formula takes the form:


\begin{align}
\mathrm{Tr} \rho_P^2 = 
& \bigotimes_{\ell \in P} \left[ \begin{array}{cc} d_\ell & -1 \end{array} \right] 
\left[ 
\sum_{\substack{I_1,I_2 \in \bigtimes_{\ell \in \{1, \dots, N\}} [d_\ell]\\ I_1^{(\ell)} \ne I_2^{(\ell)} \Leftrightarrow \ell  \in Q}} 
\left\langle 
p\left(I_1\Big|\bigotimes_{\ell=1}^N U_\ell\right)
\,p\left(I_2\Big|\bigotimes_{\ell=1}^N U_\ell\right)
\right\rangle_{\bigotimes_{\ell=1}^N \mathcal U(d^{(\ell)})}
\right]_{Q \subset \{ 1, \dots, N \}}.
\nonumber\\
= & \prod_{\ell \in P} d_\ell 
\sum_{I_1,I_2 \in \bigtimes_{\ell \in P} [d_\ell]}\,\,
\prod_{
\stackrel{\ell=1}{I_1^{(\ell)}\neq I_2^{(\ell)}}
}^N \left( - \frac 1{d_\ell} \right) \left\langle 
p\left(I_1\Big|\bigotimes_{\ell=1}^N U_\ell\right)
\,p\left(I_2\Big|\bigotimes_{\ell=1}^N U_\ell\right)
\right\rangle_{\bigotimes_{\ell=1}^N \mathcal{U}(d^{(\ell)})}
\end{align}
if all the subsystems have equal dimensions products become powers and the solution takes the form:
\begin{equation}
    \mathrm{Tr} \rho_P^2 = d^{\# P} \sum_{I_1,I_2 \in \{1,\dots,d\}^P} \left( -d \right)^{-D(I_1,I_2)}\left\langle 
p\left(I_1\Big|\bigotimes_{\ell=1}^N U_\ell\right)
\,p\left(I_2\Big|\bigotimes_{\ell=1}^N U_\ell\right)
\right\rangle_{\bigotimes_{\ell=1}^N \mathcal U(d^{(\ell)})}.
\end{equation}
where $D$ is the Hamming distance and $\#$ is the counting measure. In this special case, we obtain the expression by Zoller et al (see Supp. Mat. of Ref~\cite{brydges2019probing}).

\subsection{Entanglement criterion from second--order correlations in bipartite systems via local unitaries}
\begin{figure}[h!]
        \centering
        \includegraphics[width=.35\linewidth]{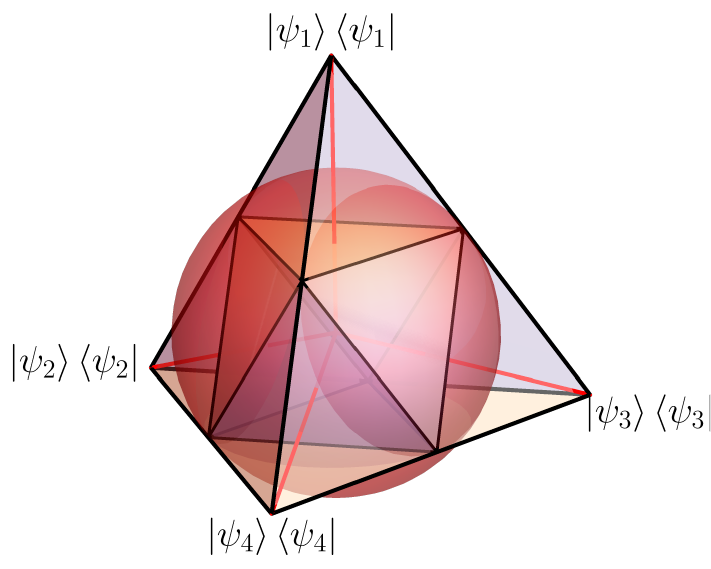}
    \caption{The tetrahedron is a set of all Bell diagonal states of two qubits, $\rho=\sum_i\lambda_i\ket{\psi_i}\bra{\psi_i}$. The PPT criterion cuts the octahedron of separable states spanned by the middle points of the edges of the tetrahedron. The separability condition \ref{sep_from_purities} detects states outside the ball circumscribed on the octahedron. Red lines defined the Werner states.}
    \label{fig:tetra}
\end{figure}

For every separable state, the following inequality holds (see, e.g. Ref. \cite{guhne2009entanglement} Sec. II.C.])
\begin{equation} \label{sep_from_purities}
\operatorname{Tr}\rho_{\text{sep}}^2\;\le\;
\min\!\Bigl\{\operatorname{Tr}\rho_A^2,\;
           \operatorname{Tr}\rho_B^2\Bigr\}.
\end{equation}
indeed,
$$
\operatorname{Tr}\rho_{\text{sep}}^2=\!\!
\sum_{k,\ell} p_k p_\ell\,
\underbrace{\operatorname{Tr}\bigl(\rho_A^{(k)}\rho_A^{(\ell)}\bigr)}_{\le 1}\,
\operatorname{Tr}\bigl(\rho_B^{(k)}\rho_B^{(\ell)}\bigr)
\le
\sum_{k,\ell} p_k p_\ell\,\operatorname{Tr}\bigl(\rho_B^{(k)}\rho_B^{(\ell)}\bigr)
=
\operatorname{Tr}\rho_B^2,
$$
and the same with $A\leftrightarrow B$. More generally, for marginals of an $N$-partite system in a separable state:
\begin{equation}
    \forall \ P,Q \subset \{ 1, \dots, N \} \ Q \subset P \Leftrightarrow \mathrm{Tr}\rho_P^2 \le \mathrm{Tr}\rho_Q^2
\end{equation}
These are quantities that we can measure experimentally in the randomised measurement scheme, as we explained in the previous section. Therefore, a separability criterion from the data is
\begin{equation}
    \forall \rho_{\mathrm{sep}} \Longrightarrow  \min\{x^{(2,2)}_1,x^{(2,2)}_2\}\ge x^{(2,2)}_3\,, \qquad
    \vec x^{(2,2)}=(1,\Tr_B\rho^2,\Tr_A\rho^2,\Tr\rho^2).
\end{equation}
For pure states $x_3=1$. If the state is entangled, the reduced states are mixed, so $x^{(2,2)}_{1}<1$ and $x^{(2,2)}_{2}<1$, therefore, it always gives a violation certifying
  entanglement. Given two--qubits states in the Bell basis $\{\psi_i\}_i$,
\begin{equation}
    \rho = \sum_{i=1}^4 \lambda_i \ketbra{\psi_i}{\psi_i} = \frac{1}{2}\left[ \begin{array}{cc|cc} \lambda_1 + \lambda_2 & \cdot & \cdot & \lambda_1 - \lambda_2 \\
    \cdot & \lambda_3+\lambda_4 & \lambda_3-\lambda_4 & \cdot \\ \hline 
    \cdot & \lambda_3-\lambda_4 & \lambda_3+\lambda_4 & \cdot \\
    \lambda_1 - \lambda_2 & \cdot & \cdot & \lambda_1 + \lambda_2 \\
    \end{array} \right], \qquad
    \lambda_i\in [0,1],\,\,\sum_i\lambda_i=1,
\end{equation}
$\rho$ is NPT iff $\max_i\lambda_i > \frac 12$ (see Fig. \ref{fig:tetra}). This condition comes from PPT criterion where the positivity condition on the eigenvalues of $\rho^\Gamma$ is $1-2\lambda_i\ge0$ for all $i=1,\dots,4$ and determines the middle points between two Bell states. Instead the criterion \eqref{sep_from_purities} gives that $\rho$ is separable iff $\sum_i \lambda_i^2\le 1/2$.
For Werner-state $\rho_W(p)=p\,|\Psi^-\rangle\!\langle\Psi^-|+(1-p)\,\bm 1/4$  one finds
$
x_3(p)=\tfrac14\!\bigl(1+3p^2\bigr)$ and $
x_1(p)=x_2(p)=\tfrac12 .
$
The criterion detects entanglement whenever $p>\sqrt{1/3}\approx0.577$, whereas the true entanglement threshold is $p>1/3$. 

\section{Third--order correlations in bipartite systems via local unitaries}\label{sec:3order}

In this section, we will calculate the third-order correlations for a bipartite system:
\begin{align}\label{eq:rho3}
\langle p(I_1|U^A\otimes U^B)&p(I_2|U^A\otimes U^B)p(I_3|U^A\otimes U^B)\rangle_{\mathcal{U}(d_A)\otimes \mathcal{U}(d_B)}
=
\int\prod_{k=1}^3
\left\langle I_{k}|U^{A\dagger}\otimes U^{B\dagger}\rho U^{A}\otimes U^{B}|I_k\right\rangle\,
\mathrm dU^A\mathrm dU^B\nonumber\\
=&\prod_{k=1}^3\rho_{p_{k}^Ap_{k}^B,q_{k}^Aq_{k}^B}
\int 
\prod_{n=1}^3
u_{p_{k}^Ai_{k}^A}^{A*}u_{q_{k}^Ai_{k}^A}^{A}\,\mathrm dU^A
\int \prod_{k'=1}^3 u^{B*}_{p_{k'}^Bi_{k'}^B}
u_{q_{k'}^B i_{k'}^{B}}^{B}
\mathrm d U^B \nonumber \\
=&
\rho_{p_{1}^Ap_{1}^B,q_{1}^Aq_{1}^B}\,
\dots
\rho_{p_{3}^Ap_{3}^B,q_{3}^Aq_{3}^B}\,
\sum_{\sigma_{A},\tau_{A}\in S_{3}}\delta_{i_{1}^A i^A_{\sigma_{A}(1)}}
\dots
\delta_{i^A_{3}i^A_{\sigma_{A}(3)}}
\mathrm{Wg}\left(\sigma_{A}\tau_{A}^{-1},d_A\right)
\delta_{p_{1}^A q^A_{\tau_{A}(1)}}\dots
\delta_{p_{3}^Aq^A_{\tau_{A}(3)}}\nonumber\\
\times&\sum_{\sigma_{B},\tau_{B}\in S_{3}}
\delta_{i^B_{1}i^B_{\sigma_{B}(1)}}
\dots
\delta_{i^B_{3}i^B_{\sigma_{B}(3)}}
\mathrm{Wg}\left(\sigma_B\tau_{B}^{-1},d_B\right)
\delta_{p^B_{1}q^B_{\tau_{B}(1)}}\dots\delta_{p^B_{3}q^B_{\tau_{B}(3)}}
\nonumber\\
=&\left[\begin{array}{ccc}
     1& \vec \delta_{(12)}^A& \vec \delta_{(3)}^A 
\end{array}\right]
W_A\,r_A \otimes 
\left[\begin{array}{ccc}
     1& \vec \delta_{(12)}^B& \vec \delta_{(3)}^B 
\end{array}\right]
W_B\,r_B 
\,\rho^{\otimes 3}
.
\end{align}
The equation is analogous to Eq. \eqref{eq:r2}-\eqref{eq:S2} where the equality/inequality of the outcomes is ruled by the row vector $[1,\,\delta_{i^A_1i^A_{\sigma_A(1)}}]$. 
In this case, the corresponding vector has 6 components:
$\vec\delta^A_{(12)}=(\delta_{i_1^Ai_2^A},\delta_{i_2^Ai_3^A},\delta_{i_3^Ai_1^A})$, $\delta_{(3)}^A=\delta_{i_1^Ai_2^A}\delta_{i_2^Ai_3^A}\delta_{i_3^Ai_1^A}(1,1)$. The vector 
$r_A=[I_A,\, \vec{T}_A,\, \vec{C}_A]^T$, where $\vec{T}_A$ contains the 3 maps corresponding to transpositions and $\vec{C}_A$ contains the two maps corresponding to the two 3-cycles. 
Observe that $\sigma\tau^{-1}$ is the identity if $\sigma = \tau$, two different transpositions or two different cycles result in a cycle, a cycle and a transposition result in a transposition.
Now one obtains the matrix $W$ using Eq.\eqref{eq:Wg}:
\begin{equation}\label{eq:wgd}
W = \frac{1}{d\left(d^{2}-4\right)\left(d^{2}-1\right)}
\left[ \begin{array}{c|ccc|cc} 
d^{2}-2 & -d & -d & -d & 2 & 2 \\
\hline
-d & d^{2}-2 & 2 & 2 & -d & -d \\
-d & 2 & d^{2}-2 & 2 & -d & -d \\
-d & 2 & 2 & d^{2}-2 & -d & -d \\
\hline 
2 & -d & -d & -d & d^{2}-2 & 2 \\
2 & -d & -d & -d & 2 & d^{2}-2
\end{array} \right]
\end{equation}
The  equality/inequality relations between the outcomes encoded in the vectors $[1,\,\vec{\delta}^A_{(12)},\,\vec{\delta}^A_{(3)}]$ form the rows of the $5\times 6$ 
equalities
matrix $S$ that multiply the matrices $W_A$ and $W_B$ like in Eq. \eqref{eq:S2} (the matrix $S$ is $5\times 6$ because the elements of $\vec{\delta}^A_{(3)}$ are the same).
We write the matrices $S$ and $W_K, K \in \{A,B\}$ in block form, grouping together permutations of the same conjugacy class.
\begin{equation}
(1)(2)(3) \in [1^3], \qquad
(12),(23),(31) \in [12],\qquad
(312),(231) \in [123]
\end{equation}
which yields
\begin{equation}
    S=\left[\begin{array}{c|c|c}
        1 & 0 & 0\\\hline
        \mathbbm 1_3& \mathbb{I}_3 & \mathbbm 1_3 \mathbbm 1_2^T\\\hline
        1& \mathbbm{1}_3^T & \mathbbm{1}_2^T
    \end{array}\right],
    \qquad
    W_K=\left[\begin{array}{c|c|c}
        i^K & t^K \mathbbm{1}_3^T & c^K  \mathbbm{1}_2^T \\\hline
        t^K \mathbbm{1}_3& (i^K-c^K) \mathbb{I}_3 + c_K \mathbbm{1}_3\mathbbm{1}_3^T & t^K \mathbbm{1}_3 \mathbbm{1}_2^T \\\hline
        c^K \mathbbm{1}_2& t^K \mathbbm{1}_2 \mathbbm{1}_3^T & (i^K-c^K) \mathbb{I}_2 + c^K \mathbbm{1}_2 \mathbbm{1}_2^T
    \end{array}\right],
\end{equation}
where $\mathbbm 1_n$ denotes the vector of $1$'s and $i^K, t^K, c^K$ are the values of the Weingarten function defined in (\ref{Wg_3}).
Hence, the $(3,2)$--\textit{twirling} for two parties and three points measurement $I_k=(i_k^A,i_k^B)$, $(k=1,2,3)$ is 
\begin{equation}\label{eq:yMx}
    \vec{y}^{(2,3)}=\bm M \vec{x}^{(2,3)},\quad
    \vec{y}^{(2,3)}=\left\{\left\langle \prod_{k=1}^3p(I_k|U^A\otimes U^B)\right\rangle_{\mathcal{U}(d_A)\otimes \mathcal{U}(d_B)}\right\}_{I^{A}_k\neq I^{A}_{k'}\Longleftrightarrow k\in Q}
    ,\quad Q=\{\emptyset,\{1\},\dots,\{1,2,3\}\}
\end{equation}
with $\vec{x}^{(2,3)}=r_A\otimes r_B\,\rho^{\otimes3}$, $\bm M=SW_A\otimes SW_B.$
The matrix $\bm M\in \mathcal{M}_{25,36}$ but since the vector $\vec{x}^{(2,3)}$ is determined by the diagrams resulting from the action of $r_A \otimes r_B$ on $\rho^{\otimes 3}$ represented in tab \ref{tab:Otau} and \ref{tab:Otau1}, the 
Eq. \eqref{eq:yMx} can be simplified in the following steps:
\begin{enumerate}
    \item Since vector $\vec x^{(2,3)}$ has only 11 different entries, related to diagrams in table \ref{tab:Otau} and table \ref{tab:Otau1}, we sum together columns related to the same $x_i$, resulting in $25 \times 11$ matrix.
    \item Among 25 equations there are only 10 different equations, roughly corresponding to pairs of conjugacy classes, except the pair $([21],[21])$ which is represented by two non-equivalent equality types: 
    $(i^A_j=i^A_k\neq i^A_\ell,i^B_j=i^B_k\neq i^B_\ell)$ and $(i^A_j=i^A_k\neq i^A_\ell,i^B_j \neq i^B_k = i^B_\ell)$.
    \item Observe, that the last two columns are the same. We reduce number of columns introducing variable $x_S = (x_9 + x_{10})/2$, resulting in $10 \times 10$ matrix:
\end{enumerate}

\begin{equation}\label{eq:big}
    \left[\begin{array}{c}
    y_0\\\vdots\\y_9
    \end{array}\right]
    =
    \left[\begin{array}{ccc}
    i_A\left[\begin{array}{ccc} 
i^B& 3t^B& 2c^B\\
    a^B & f^B & 2b^B\\
    f^B&3f^B& 2f^B
    \end{array}\right]
& 3t_A\left[\begin{array}{cccc} 
i^B& 2t^B&t^B& 2c^B\\
    a^B & 2f^B& f^B & 2b^B\\
    f^B&2f^B& f^B& 2f^B
    \end{array}\right]
& 2c_A\left[\begin{array}{ccc} 
i^B& 3t^B& 2c^B\\
    a^B & f^B & 2b^B\\
    f^B&3f^B& 2f^B
    \end{array}\right]
\\
    a^A\left[\begin{array}{ccc} 
i^B& 3t^B& 2c^B\\
    a^B & f^B & 2b^B\\
    a^B & f^B & 2b^B\\
    f^B&3f^B& 2f^B
    \end{array}\right]
 & f^A \left[\begin{array}{cccc} 
i^B& 2t^B& t^B& 2c^B\\
    a^B & \varepsilon_1+\eta& \varepsilon_2 & 2b^B\\
    a^B & \varepsilon_1& \varepsilon_2 +\eta& 2b^B\\
    a^B & f^B & 2b^B\\
    f^B&2f^B&f^B& 2f^B
    \end{array}\right]
& 2b^A\left[\begin{array}{ccc} 
i^B& 3t^B& 2c^B\\
    a^B & f^B & 2b^B\\
    a^B & f^B & 2b^B\\
    f^B&3f^B& 2f^B
    \end{array}\right]\\
    f^A\left[\begin{array}{ccc} 
i^B& 3t^B& 2c^B\\
    a^B & f^B & 2b^B\\
    f^B&3f^B& 2f^B
    \end{array}\right]&3f^A\left[\begin{array}{cccc} 
i^B& 2t^B&t^B& 2c^B\\
    a^B & 2f^B& f^B & 2b^B\\
    f^B&2f^B& f^B& 2f^B
    \end{array}\right]& 2f^A\left[\begin{array}{ccc} 
i^B& 3t^B& 2c^B\\
    a^B & f^B & 2b^B\\
    f^B&3f^B& 2f^B
    \end{array}\right]
    \end{array}\right]
    \left[\begin{array}{c}
    x_0\\\vdots\\x_8\\x_S
    \end{array}\right],
\end{equation}
where
\begin{align}
    &a^K=t^K+i^K, &b^K=c^K+t^K, \qquad&f^K=a^K+2b^K, \nonumber\\ 
    &\varepsilon_1=2b^Ab^B+(a^A+a^B)(b^A+b^B), 
    &\varepsilon_2=a^Aa^B+b^Ab^B,  
    \qquad&\eta=(b^A-a^A)(a^B-b^B)
\end{align}
for $K=A,B$.
We observe that
\begin{equation}
    y_4-y_5=\eta(x_4-x_5) \Longrightarrow x_4=\frac{y_4-y_5}{\eta}+x_5\Longrightarrow \vec{x}^{(2,3)}=[x_0,\dots,x_5,x_5,\dots]^T+\Delta[0,\dots,1,0,\dots]^T
\end{equation}
We call these two addends as $\vec{x}^{(2,3)}=\tilde{\vec{x}}^{(2,3)}+\Delta\,\hat{\vec{x}}^{(2,3)}$ with $\Delta=(y_4-y_5)/\eta$. We remove the redundancy of $x_5$ from $\tilde{\vec{x}}^{(2,3)}$ replacing the $4$th and $5$th column of M by their sum, and removing the $4$-th row, as it becomes equal to the $5$th, so that $\tilde{\vec{x}}^{(2,3)}$  becomes a vector with $9$ components and $\tilde{\bm M}$ a $9\times 9$ matrix factorizable as $\tilde{\mathbf{M}} = Q_A \otimes Q_B$. Therefore, let us call $\hat{\vec{y}}^{(2,3)}=\tilde{\mathbf{M}} \,\hat{\vec{x}}^{(2,3)}$, then Eq. \eqref{eq:big} becomes
\begin{equation}
    \tilde{\vec{y}}^{(2)}_3 -\Delta \,\hat{\vec{y}}^{(2)}_3=Q_A \otimes Q_B \tilde{\vec{x}}^{(2)}_3,
    \qquad
    Q_K=\frac{1}{d_k(d_k^2-1)(d_k^2-4)}\begin{pmatrix}
        d_k^2-1&-3d_k&4\\d_k+1&d_k-1&-2\\1&3&2
    \end{pmatrix}
\end{equation}
(see the details in Appendix \ref{sec:3rd}).
\begin{table}[h]
\centering{}%
\resizebox{\textwidth}{!}{
\begin{tabular}{|c||c||c|c|c||c|c|}
\hline 
 & $1^3$ & $(12)$ & $(23)$ & $(13)$ & $(123)$ & $(132)$\tabularnewline
\hline 
\hline 
$1^3$ & $%
\begin{tikzpicture}[scale = .75]
        \draw [fill=green] (0-.5,-.5) -- (0-.5,.5) -- (0+.5,.5) -- (0+.5,-.5) -- cycle;
        \node  (LU0) at (0-0.5, 0.25) {};
        \node  (RU0) at (0+0.5, 0.25) {};
        \node  (LD0) at (0-0.5, -0.25) {};
        \node  (RD0) at (0+0.5, -0.25) {};
        \node  (U0) at (0, 1.2) {};
        \node  (D0) at (0, -1.2) {};
        \draw [fill=green] (2-.5,-.5) -- (2-.5,.5) -- (2+.5,.5) -- (2+.5,-.5) -- cycle;
        \node  (LU1) at (2-0.5, 0.25) {};
        \node  (RU1) at (2+0.5, 0.25) {};
        \node  (LD1) at (2-0.5, -0.25) {};
        \node  (RD1) at (2+0.5, -0.25) {};
        \node  (U1) at (2, 1.2) {};
        \node  (D1) at (2, -1.2) {};
        \draw [fill=green] (4-.5,-.5) -- (4-.5,.5) -- (4+.5,.5) -- (4+.5,-.5) -- cycle;
        \node  (LU2) at (4-0.5, 0.25) {};
        \node  (RU2) at (4+0.5, 0.25) {};
        \node  (LD2) at (4-0.5, -0.25) {};
        \node  (RD2) at (4+0.5, -0.25) {};
        \node  (U2) at (4, 1.2) {};
        \node  (D2) at (4, -1.2) {};
        \draw (LU0.center) to[out=180,in=180,looseness=1.5]  (U0.center)  to[out=0,in=0,looseness=1.5] (RU0.center);
        \draw (LU1.center) to[out=180,in=180,looseness=1.5]  (U1.center)  to[out=0,in=0,looseness=1.5] (RU1.center);
        \draw (LU2.center) to[out=180,in=180,looseness=1.5]  (U2.center)  to[out=0,in=0,looseness=1.5] (RU2.center);\draw (LD0.center) to[out=180,in=180,looseness=1.5]  (D0.center)  to[out=0,in=0,looseness=1.5] (RD0.center);
        \draw (LD1.center) to[out=180,in=180,looseness=1.5]  (D1.center)  to[out=0,in=0,looseness=1.5] (RD1.center);
        \draw (LD2.center) to[out=180,in=180,looseness=1.5]  (D2.center)  to[out=0,in=0,looseness=1.5] (RD2.center);
\end{tikzpicture}}$ & $%
\begin{tikzpicture}[scale = .75]
        \draw [fill=green] (0-.5,-.5) -- (0-.5,.5) -- (0+.5,.5) -- (0+.5,-.5) -- cycle;
        \node  (LU0) at (0-0.5, 0.25) {};
        \node  (RU0) at (0+0.5, 0.25) {};
        \node  (LD0) at (0-0.5, -0.25) {};
        \node  (RD0) at (0+0.5, -0.25) {};
        \node  (U0) at (0, 1.2) {};
        \node  (D0) at (0, -1.2) {};
        \draw [fill=green] (2-.5,-.5) -- (2-.5,.5) -- (2+.5,.5) -- (2+.5,-.5) -- cycle;
        \node  (LU1) at (2-0.5, 0.25) {};
        \node  (RU1) at (2+0.5, 0.25) {};
        \node  (LD1) at (2-0.5, -0.25) {};
        \node  (RD1) at (2+0.5, -0.25) {};
        \node  (U1) at (2, 1.2) {};
        \node  (D1) at (2, -1.2) {};
        \draw [fill=green] (4-.5,-.5) -- (4-.5,.5) -- (4+.5,.5) -- (4+.5,-.5) -- cycle;
        \node  (LU2) at (4-0.5, 0.25) {};
        \node  (RU2) at (4+0.5, 0.25) {};
        \node  (LD2) at (4-0.5, -0.25) {};
        \node  (RD2) at (4+0.5, -0.25) {};
        \node  (U2) at (4, 1.2) {};
        \node  (D2) at (4, -1.2) {};
        \draw (LU0.center) to[out=180,in=180,looseness=1.5]  (U0.center)  to[out=0,in=0,looseness=1.5] (RU0.center);
        \draw (LU1.center) to[out=180,in=180,looseness=1.5]  (U1.center)  to[out=0,in=0,looseness=1.5] (RU1.center);
        \draw (LU2.center) to[out=180,in=180,looseness=1.5]  (U2.center)  to[out=0,in=0,looseness=1.5] (RU2.center);
        \draw (LD2.center) to[out=180,in=180,looseness=1.5]  (D2.center)  to[out=0,in=0,looseness=1.5] (RD2.center);
        \draw (LD0.center) to[out=180,in=180,looseness=1.5]  (D0.center) -- (D1.center)  to[out=0,in=0,looseness=1.5] (RD1.center);
        \draw (RD0.center) -- (LD1.center);
    \end{tikzpicture}}$ & 
$%
\begin{tikzpicture}[baseline=0,scale=.75] 
        \draw [fill=green] (0-.5,-.5) -- (0-.5,.5) -- (0+.5,.5) -- (0+.5,-.5) -- cycle;
        \node  (LU0) at (0-0.5, 0.25) {};
        \node  (RU0) at (0+0.5, 0.25) {};
        \node  (LD0) at (0-0.5, -0.25) {};
        \node  (RD0) at (0+0.5, -0.25) {};
        \node  (U0) at (0, 1.2) {};
        \node  (D0) at (0, -1.2) {};
        \draw [fill=green] (2-.5,-.5) -- (2-.5,.5) -- (2+.5,.5) -- (2+.5,-.5) -- cycle;
        \node  (LU1) at (2-0.5, 0.25) {};
        \node  (RU1) at (2+0.5, 0.25) {};
        \node  (LD1) at (2-0.5, -0.25) {};
        \node  (RD1) at (2+0.5, -0.25) {};
        \node  (U1) at (2, 1.2) {};
        \node  (D1) at (2, -1.2) {};
        \draw [fill=green] (4-.5,-.5) -- (4-.5,.5) -- (4+.5,.5) -- (4+.5,-.5) -- cycle;
        \node  (LU2) at (4-0.5, 0.25) {};
        \node  (RU2) at (4+0.5, 0.25) {};
        \node  (LD2) at (4-0.5, -0.25) {};
        \node  (RD2) at (4+0.5, -0.25) {};
        \node  (U2) at (4, 1.2) {};
        \node  (D2) at (4, -1.2) {};
        \draw (LU0.center) to[out=180,in=180,looseness=1.5]  (U0.center)  to[out=0,in=0,looseness=1.5] (RU0.center);
        \draw (LU1.center) to[out=180,in=180,looseness=1.5]  (U1.center)  to[out=0,in=0,looseness=1.5] (RU1.center);
        \draw (LU2.center) to[out=180,in=180,looseness=1.5]  (U2.center)  to[out=0,in=0,looseness=1.5] (RU2.center);
        \draw (LD0.center) to[out=180,in=180,looseness=1.5]  (D0.center)  to[out=0,in=0,looseness=1.5] (RD0.center);
        \draw (LD1.center) to[out=180,in=180,looseness=1.5]  (D1.center) -- (D2.center)  to[out=0,in=0,looseness=1.5] (RD2.center);
        \draw (RD1.center) -- (LD2.center);
    \end{tikzpicture}}$ & 
$%
\begin{tikzpicture}[scale = .75] 
\draw [fill=green] (0-.5,-.5) -- (0-.5,.5) -- (0+.5,.5) -- (0+.5,-.5) -- cycle;
        \node  (LU0) at (0-0.5, 0.25) {};
        \node  (RU0) at (0+0.5, 0.25) {};
        \node  (LD0) at (0-0.5, -0.25) {};
        \node  (RD0) at (0+0.5, -0.25) {};
        \node  (U0) at (0, 1.2) {};
        \node  (D0) at (0, -1.2) {};
        \draw [fill=green] (2-.5,-.5) -- (2-.5,.5) -- (2+.5,.5) -- (2+.5,-.5) -- cycle;
        \node  (LU1) at (2-0.5, 0.25) {};
        \node  (RU1) at (2+0.5, 0.25) {};
        \node  (LD1) at (2-0.5, -0.25) {};
        \node  (RD1) at (2+0.5, -0.25) {};
        \node  (U1) at (2, 1.2) {};
        \node  (D11) at (2, -1) {};
        \node  (D111) at (2, -1.2) {};
        \node  (D1) at (2, -.8) {};
        \draw [fill=green] (4-.5,-.5) -- (4-.5,.5) -- (4+.5,.5) -- (4+.5,-.5) -- cycle;
        \node  (LU2) at (4-0.5, 0.25) {};
        \node  (RU2) at (4+0.5, 0.25) {};
        \node  (LD2) at (4-0.5, -0.25) {};
        \node  (RD2) at (4+0.5, -0.25) {};
        \node  (U2) at (4, 1.2) {};
        \node  (D2) at (4, -1.2) {};

        
        \draw (LD0.center) to[out=180,in=180,looseness=1.5]  (D111.center)
        to[out=0,in=0,looseness=1.5] (RD2.center);
        \draw (RD0.center) to[out=0,in=180,looseness=1.5]  (D11.center) to[out=0,in=180,looseness=1.5]  (LD2.center);
       \draw (LD1.center) to[out=180,in=180,looseness=1.5]  (D1.center)  to[out=0,in=0,looseness=1.5] (RD1.center);
        \draw (LU0.center) to[out=180,in=180,looseness=1.5]  (U0.center)  to[out=0,in=0,looseness=1.5] (RU0.center);
        \draw (LU1.center) to[out=180,in=180,looseness=1.5]  (U1.center)  to[out=0,in=0,looseness=1.5] (RU1.center);
        \draw (LU2.center) to[out=180,in=180,looseness=1.5]  (U2.center)  to[out=0,in=0,looseness=1.5] (RU2.center);
        \end{tikzpicture}}$ & 
$%
 \begin{tikzpicture}[scale = .75] 
        \draw [fill=green] (0-.5,-.5) -- (0-.5,.5) -- (0+.5,.5) -- (0+.5,-.5) -- cycle;
        \node  (LU0) at (0-0.5, 0.25) {};
        \node  (RU0) at (0+0.5, 0.25) {};
        \node  (LD0) at (0-0.5, -0.25) {};
        \node  (RD0) at (0+0.5, -0.25) {};
        \node  (U0) at (0, 1.2) {};
        \node  (D0) at (0, -1.2) {};
        \draw [fill=green] (2-.5,-.5) -- (2-.5,.5) -- (2+.5,.5) -- (2+.5,-.5) -- cycle;
        \node  (LU1) at (2-0.5, 0.25) {};
        \node  (RU1) at (2+0.5, 0.25) {};
        \node  (LD1) at (2-0.5, -0.25) {};
        \node  (RD1) at (2+0.5, -0.25) {};
        \node  (U1) at (2, 1.2) {};
        \node  (D1) at (2, -1.2) {};
        \draw [fill=green] (4-.5,-.5) -- (4-.5,.5) -- (4+.5,.5) -- (4+.5,-.5) -- cycle;
        \node  (LU2) at (4-0.5, 0.25) {};
        \node  (RU2) at (4+0.5, 0.25) {};
        \node  (LD2) at (4-0.5, -0.25) {};
        \node  (RD2) at (4+0.5, -0.25) {};
        \node  (U2) at (4, 1.2) {};
        \node  (D2) at (4, -1.2) {};
        \draw (LU0.center) to[out=180,in=180,looseness=1.5]  (U0.center)  to[out=0,in=0,looseness=1.5] (RU0.center);
        \draw (LU1.center) to[out=180,in=180,looseness=1.5]  (U1.center)  to[out=0,in=0,looseness=1.5] (RU1.center);
        \draw (LU2.center) to[out=180,in=180,looseness=1.5]  (U2.center)  to[out=0,in=0,looseness=1.5] (RU2.center);
        \draw (LD0.center) to[out=180,in=180,looseness=1.5]  (D0.center) -- (D1.center) -- (D2.center)  to[out=0,in=0,looseness=1.5] (RD2.center);
        \draw (RD0.center) -- (LD1.center);
        \draw (RD1.center) -- (LD2.center);
    \end{tikzpicture}}$ & $%
 \begin{tikzpicture}[scale = .75] 
        \draw [fill=green] (0-.5,-.5) -- (0-.5,.5) -- (0+.5,.5) -- (0+.5,-.5) -- cycle;
        \node  (LU0) at (0-0.5, 0.25) {};
        \node  (RU0) at (0+0.5, 0.25) {};
        \node  (LD0) at (0-0.5, -0.25) {};
        \node  (RD0) at (0+0.5, -0.25) {};
        \node  (U0) at (0, 1.2) {};
        \node  (D0) at (0, -1.2) {};
        \draw [fill=green] (2-.5,-.5) -- (2-.5,.5) -- (2+.5,.5) -- (2+.5,-.5) -- cycle;
        \node  (LU1) at (2-0.5, 0.25) {};
        \node  (RU1) at (2+0.5, 0.25) {};
        \node  (LD1) at (2-0.5, -0.25) {};
        \node  (RD1) at (2+0.5, -0.25) {};
        \node  (U1) at (2, 1.2) {};
        \node  (D1) at (2, -1.2) {};
        \draw [fill=green] (4-.5,-.5) -- (4-.5,.5) -- (4+.5,.5) -- (4+.5,-.5) -- cycle;
        \node  (LU2) at (4-0.5, 0.25) {};
        \node  (RU2) at (4+0.5, 0.25) {};
        \node  (LD2) at (4-0.5, -0.25) {};
        \node  (RD2) at (4+0.5, -0.25) {};
        \node  (U2) at (4, 1.2) {};
        \node  (D2) at (4, -1.2) {};
        \draw (LU0.center) to[out=180,in=180,looseness=1.5]  (U0.center)  to[out=0,in=0,looseness=1.5] (RU0.center);
        \draw (LU1.center) to[out=180,in=180,looseness=1.5]  (U1.center)  to[out=0,in=0,looseness=1.5] (RU1.center);
        \draw (LU2.center) to[out=180,in=180,looseness=1.5]  (U2.center)  to[out=0,in=0,looseness=1.5] (RU2.center);
        \draw (LD0.center) to[out=180,in=180,looseness=2]  (D0.center) to[out=0,in=0,looseness=1.2] (RD1.center);
        \draw (LD1.center) to[out=180,in=180,looseness=1.2] (D2.center) to[out=0,in=0,looseness=2] (RD2.center);
        \draw (RD0.center) to[out=0,in=180,looseness=1] (D1.center) to[out=0,in=180,looseness=1] (LD2.center);
    \end{tikzpicture}}$\tabularnewline
\hline 
\hline 
$(12)$ & $%
 \begin{tikzpicture}[scale = .75] 
        \draw [fill=green] (0-.5,-.5) -- (0-.5,.5) -- (0+.5,.5) -- (0+.5,-.5) -- cycle;
        \node  (LU0) at (0-0.5, 0.25) {};
        \node  (RU0) at (0+0.5, 0.25) {};
        \node  (LD0) at (0-0.5, -0.25) {};
        \node  (RD0) at (0+0.5, -0.25) {};
        \node  (U0) at (0, 1.2) {};
        \node  (D0) at (0, -1.2) {};
        \draw [fill=green] (2-.5,-.5) -- (2-.5,.5) -- (2+.5,.5) -- (2+.5,-.5) -- cycle;
        \node  (LU1) at (2-0.5, 0.25) {};
        \node  (RU1) at (2+0.5, 0.25) {};
        \node  (LD1) at (2-0.5, -0.25) {};
        \node  (RD1) at (2+0.5, -0.25) {};
        \node  (U1) at (2, 1.2) {};
        \node  (D1) at (2, -1.2) {};
        \draw [fill=green] (4-.5,-.5) -- (4-.5,.5) -- (4+.5,.5) -- (4+.5,-.5) -- cycle;
        \node  (LU2) at (4-0.5, 0.25) {};
        \node  (RU2) at (4+0.5, 0.25) {};
        \node  (LD2) at (4-0.5, -0.25) {};
        \node  (RD2) at (4+0.5, -0.25) {};
        \node  (U2) at (4, 1.2) {};
        \node  (D2) at (4, -1.2) {};
        \draw (LU0.center) to[out=180,in=180,looseness=1.5]  (U0.center) -- (U1.center)  to[out=0,in=0,looseness=1.5] (RU1.center);
        \draw (LU2.center) to[out=180,in=180,looseness=1.5]  (U2.center)  to[out=0,in=0,looseness=1.5] (RU2.center);
        \draw (RU0.center) -- (LU1.center);
        \draw (LD0.center) to[out=180,in=180,looseness=1.5]  (D0.center) to[out=0,in=0,looseness=1.5] (RD0.center);
         \draw (LD1.center) to[out=180,in=180,looseness=1.5]  (D1.center)  to[out=0,in=0,looseness=1.5] (RD1.center);
        \draw (LD2.center) to[out=180,in=180,looseness=1.5]  (D2.center)  to[out=0,in=0,looseness=1.5] (RD2.center);
    \end{tikzpicture}

      }$ & $%
 \begin{tikzpicture}[scale = .75] 
        \draw [fill=green] (0-.5,-.5) -- (0-.5,.5) -- (0+.5,.5) -- (0+.5,-.5) -- cycle;
        \node  (LU0) at (0-0.5, 0.25) {};
        \node  (RU0) at (0+0.5, 0.25) {};
        \node  (LD0) at (0-0.5, -0.25) {};
        \node  (RD0) at (0+0.5, -0.25) {};
        \node  (U0) at (0, 1.2) {};
        \node  (D0) at (0, -1.2) {};
        \draw [fill=green] (2-.5,-.5) -- (2-.5,.5) -- (2+.5,.5) -- (2+.5,-.5) -- cycle;
        \node  (LU1) at (2-0.5, 0.25) {};
        \node  (RU1) at (2+0.5, 0.25) {};
        \node  (LD1) at (2-0.5, -0.25) {};
        \node  (RD1) at (2+0.5, -0.25) {};
        \node  (U1) at (2, 1.2) {};
        \node  (D1) at (2, -1.2) {};
        \draw [fill=green] (4-.5,-.5) -- (4-.5,.5) -- (4+.5,.5) -- (4+.5,-.5) -- cycle;
        \node  (LU2) at (4-0.5, 0.25) {};
        \node  (RU2) at (4+0.5, 0.25) {};
        \node  (LD2) at (4-0.5, -0.25) {};
        \node  (RD2) at (4+0.5, -0.25) {};
        \node  (U2) at (4, 1.2) {};
        \node  (D2) at (4, -1.2) {};
        \draw (LU0.center) to[out=180,in=180,looseness=1.5]  (U0.center) -- (U1.center)  to[out=0,in=0,looseness=1.5] (RU1.center);
        \draw (LU2.center) to[out=180,in=180,looseness=1.5]  (U2.center)  to[out=0,in=0,looseness=1.5] (RU2.center);
        \draw (RU0.center) -- (LU1.center);
        \draw (LD2.center) to[out=180,in=180,looseness=1.5]  (D2.center)  to[out=0,in=0,looseness=1.5] (RD2.center);
        \draw (LD0.center) to[out=180,in=180,looseness=1.5]  (D0.center) -- (D1.center)  to[out=0,in=0,looseness=1.5] (RD1.center);
        \draw (RD0.center) -- (LD1.center);
    \end{tikzpicture}

      }$ &
$%
 \begin{tikzpicture}[scale = .75] 
        \draw [fill=green] (0-.5,-.5) -- (0-.5,.5) -- (0+.5,.5) -- (0+.5,-.5) -- cycle;
        \node  (LU0) at (0-0.5, 0.25) {};
        \node  (RU0) at (0+0.5, 0.25) {};
        \node  (LD0) at (0-0.5, -0.25) {};
        \node  (RD0) at (0+0.5, -0.25) {};
        \node  (U0) at (0, 1.2) {};
        \node  (D0) at (0, -1.2) {};
        \draw [fill=green] (2-.5,-.5) -- (2-.5,.5) -- (2+.5,.5) -- (2+.5,-.5) -- cycle;
        \node  (LU1) at (2-0.5, 0.25) {};
        \node  (RU1) at (2+0.5, 0.25) {};
        \node  (LD1) at (2-0.5, -0.25) {};
        \node  (RD1) at (2+0.5, -0.25) {};
        \node  (U1) at (2, 1.2) {};
        \node  (D1) at (2, -1.2) {};
        \draw [fill=green] (4-.5,-.5) -- (4-.5,.5) -- (4+.5,.5) -- (4+.5,-.5) -- cycle;
        \node  (LU2) at (4-0.5, 0.25) {};
        \node  (RU2) at (4+0.5, 0.25) {};
        \node  (LD2) at (4-0.5, -0.25) {};
        \node  (RD2) at (4+0.5, -0.25) {};
        \node  (U2) at (4, 1.2) {};
        \node  (D2) at (4, -1.2) {};
        \draw (LU0.center) to[out=180,in=180,looseness=1.5]  (U0.center) -- (U1.center)  to[out=0,in=0,looseness=1.5] (RU1.center);
        \draw (LU2.center) to[out=180,in=180,looseness=1.5]  (U2.center)  to[out=0,in=0,looseness=1.5] (RU2.center);
        \draw (RU0.center) -- (LU1.center);
        \draw (LD0.center) to[out=180,in=180,looseness=1.5]  (D0.center)  to[out=0,in=0,looseness=1.5] (RD0.center);
        \draw (LD1.center) to[out=180,in=180,looseness=1.5]  (D1.center) -- (D2.center)  to[out=0,in=0,looseness=1.5] (RD2.center);
        \draw (RD1.center) -- (LD2.center);
    \end{tikzpicture}

      }$ &
$%
 \begin{tikzpicture}[scale = .75] 
        \draw [fill=green] (0-.5,-.5) -- (0-.5,.5) -- (0+.5,.5) -- (0+.5,-.5) -- cycle;
        \node  (LU0) at (0-0.5, 0.25) {};
        \node  (RU0) at (0+0.5, 0.25) {};
        \node  (LD0) at (0-0.5, -0.25) {};
        \node  (RD0) at (0+0.5, -0.25) {};
        \node  (U0) at (0, 1.2) {};
        \node  (D0) at (0, -1.2) {};
        \draw [fill=green] (2-.5,-.5) -- (2-.5,.5) -- (2+.5,.5) -- (2+.5,-.5) -- cycle;
        \node  (LU1) at (2-0.5, 0.25) {};
        \node  (RU1) at (2+0.5, 0.25) {};
        \node  (LD1) at (2-0.5, -0.25) {};
        \node  (RD1) at (2+0.5, -0.25) {};
        \node  (U1) at (2, 1.2) {};
        \node  (D1) at (2, -1.2) {};
        \draw [fill=green] (4-.5,-.5) -- (4-.5,.5) -- (4+.5,.5) -- (4+.5,-.5) -- cycle;
        \node  (LU2) at (4-0.5, 0.25) {};
        \node  (RU2) at (4+0.5, 0.25) {};
        \node  (LD2) at (4-0.5, -0.25) {};
        \node  (RD2) at (4+0.5, -0.25) {};
        \node  (U2) at (4, 1.2) {};
        \node  (D2) at (4, -1.2) {};
        \draw (LU0.center) to[out=180,in=180,looseness=1.5]  (U0.center) -- (U1.center)  to[out=0,in=0,looseness=1.5] (RU1.center);
        \draw (LU2.center) to[out=180,in=180,looseness=1.5]  (U2.center)  to[out=0,in=0,looseness=1.5] (RU2.center);
        \draw (RU0.center) -- (LU1.center);
        \draw (LD0.center) to[out=180,in=180,looseness=1.5]  (D111.center)
        to[out=0,in=0,looseness=1.5] (RD2.center);
        \draw (RD0.center) to[out=0,in=180,looseness=1.5]  (D11.center) to[out=0,in=180,looseness=1.5]  (LD2.center);
       \draw (LD1.center) to[out=180,in=180,looseness=1.5]  (D1.center)  to[out=0,in=0,looseness=1.5] (RD1.center);
    \end{tikzpicture}

      }$ &
$%
 \begin{tikzpicture}[scale = .75] 
        \draw [fill=green] (0-.5,-.5) -- (0-.5,.5) -- (0+.5,.5) -- (0+.5,-.5) -- cycle;
        \node  (LU0) at (0-0.5, 0.25) {};
        \node  (RU0) at (0+0.5, 0.25) {};
        \node  (LD0) at (0-0.5, -0.25) {};
        \node  (RD0) at (0+0.5, -0.25) {};
        \node  (U0) at (0, 1.2) {};
        \node  (D0) at (0, -1.2) {};
        \draw [fill=green] (2-.5,-.5) -- (2-.5,.5) -- (2+.5,.5) -- (2+.5,-.5) -- cycle;
        \node  (LU1) at (2-0.5, 0.25) {};
        \node  (RU1) at (2+0.5, 0.25) {};
        \node  (LD1) at (2-0.5, -0.25) {};
        \node  (RD1) at (2+0.5, -0.25) {};
        \node  (U1) at (2, 1.2) {};
        \node  (D1) at (2, -1.2) {};
        \draw [fill=green] (4-.5,-.5) -- (4-.5,.5) -- (4+.5,.5) -- (4+.5,-.5) -- cycle;
        \node  (LU2) at (4-0.5, 0.25) {};
        \node  (RU2) at (4+0.5, 0.25) {};
        \node  (LD2) at (4-0.5, -0.25) {};
        \node  (RD2) at (4+0.5, -0.25) {};
        \node  (U2) at (4, 1.2) {};
        \node  (D2) at (4, -1.2) {};
        \draw (LU0.center) to[out=180,in=180,looseness=1.5]  (U0.center) -- (U1.center)  to[out=0,in=0,looseness=1.5] (RU1.center);
        \draw (LU2.center) to[out=180,in=180,looseness=1.5]  (U2.center)  to[out=0,in=0,looseness=1.5] (RU2.center);
        \draw (RU0.center) -- (LU1.center);
       \draw (LD0.center) to[out=180,in=180,looseness=1.5]  (D0.center) -- (D1.center) -- (D2.center)  to[out=0,in=0,looseness=1.5] (RD2.center);
        \draw (RD0.center) -- (LD1.center);
        \draw (RD1.center) -- (LD2.center);
    \end{tikzpicture}

      }$&
$%
 \begin{tikzpicture}[scale = .75] 
        \draw [fill=green] (0-.5,-.5) -- (0-.5,.5) -- (0+.5,.5) -- (0+.5,-.5) -- cycle;
        \node  (LU0) at (0-0.5, 0.25) {};
        \node  (RU0) at (0+0.5, 0.25) {};
        \node  (LD0) at (0-0.5, -0.25) {};
        \node  (RD0) at (0+0.5, -0.25) {};
        \node  (U0) at (0, 1.2) {};
        \node  (D0) at (0, -1.2) {};
        \draw [fill=green] (2-.5,-.5) -- (2-.5,.5) -- (2+.5,.5) -- (2+.5,-.5) -- cycle;
        \node  (LU1) at (2-0.5, 0.25) {};
        \node  (RU1) at (2+0.5, 0.25) {};
        \node  (LD1) at (2-0.5, -0.25) {};
        \node  (RD1) at (2+0.5, -0.25) {};
        \node  (U1) at (2, 1.2) {};
        \node  (D1) at (2, -1.2) {};
        \draw [fill=green] (4-.5,-.5) -- (4-.5,.5) -- (4+.5,.5) -- (4+.5,-.5) -- cycle;
        \node  (LU2) at (4-0.5, 0.25) {};
        \node  (RU2) at (4+0.5, 0.25) {};
        \node  (LD2) at (4-0.5, -0.25) {};
        \node  (RD2) at (4+0.5, -0.25) {};
        \node  (U2) at (4, 1.2) {};
        \node  (D2) at (4, -1.2) {};
        \draw (LU0.center) to[out=180,in=180,looseness=1.5]  (U0.center) -- (U1.center)  to[out=0,in=0,looseness=1.5] (RU1.center);
        \draw (LU2.center) to[out=180,in=180,looseness=1.5]  (U2.center)  to[out=0,in=0,looseness=1.5] (RU2.center);
        \draw (RU0.center) -- (LU1.center);
        \draw (LD0.center) to[out=180,in=180,looseness=2]  (D0.center) to[out=0,in=0,looseness=1.2] (RD1.center);
        \draw (LD1.center) to[out=180,in=180,looseness=1.2] (D2.center) to[out=0,in=0,looseness=2] (RD2.center);
        \draw (RD0.center) to[out=0,in=180,looseness=1] (D1.center) to[out=0,in=180,looseness=1] (LD2.center);
    \end{tikzpicture}

      }$\tabularnewline
\hline 
$(23)$ & $%
 \begin{tikzpicture}[scale = .75] 
        \draw [fill=green] (0-.5,-.5) -- (0-.5,.5) -- (0+.5,.5) -- (0+.5,-.5) -- cycle;
        \node  (LU0) at (0-0.5, 0.25) {};
        \node  (RU0) at (0+0.5, 0.25) {};
        \node  (LD0) at (0-0.5, -0.25) {};
        \node  (RD0) at (0+0.5, -0.25) {};
        \node  (U0) at (0, 1.2) {};
        \node  (D0) at (0, -1.2) {};
        \draw [fill=green] (2-.5,-.5) -- (2-.5,.5) -- (2+.5,.5) -- (2+.5,-.5) -- cycle;
        \node  (LU1) at (2-0.5, 0.25) {};
        \node  (RU1) at (2+0.5, 0.25) {};
        \node  (LD1) at (2-0.5, -0.25) {};
        \node  (RD1) at (2+0.5, -0.25) {};
        \node  (U1) at (2, 1.2) {};
        \node  (D1) at (2, -1.2) {};
        \draw [fill=green] (4-.5,-.5) -- (4-.5,.5) -- (4+.5,.5) -- (4+.5,-.5) -- cycle;
        \node  (LU2) at (4-0.5, 0.25) {};
        \node  (RU2) at (4+0.5, 0.25) {};
        \node  (LD2) at (4-0.5, -0.25) {};
        \node  (RD2) at (4+0.5, -0.25) {};
        \node  (U2) at (4, 1.2) {};
        \node  (D2) at (4, -1.2) {}; 
        \draw (LU0.center) to[out=180,in=180,looseness=1.5]  (U0.center)  to[out=0,in=0,looseness=1.5] (RU0.center);
        \draw (LU1.center) to[out=180,in=180,looseness=1.5]  (U1.center)-- (U2.center)  to[out=0,in=0,looseness=1.5] (RU2.center);
\draw (LD0.center) to[out=180,in=180,looseness=1.5]  (D0.center)  to[out=0,in=0,looseness=1.5] (RD0.center);
        
        \draw (RU1.center) -- (LU2.center);
       \draw (LD1.center) to[out=180,in=180,looseness=1.5]  (D1.center)  to[out=0,in=0,looseness=1.5] (RD1.center);
        \draw (LD2.center) to[out=180,in=180,looseness=1.5]  (D2.center)  to[out=0,in=0,looseness=1.5] (RD2.center);
    \end{tikzpicture}

   }$ &
$%
 \begin{tikzpicture}[scale = .75] 
        \draw [fill=green] (0-.5,-.5) -- (0-.5,.5) -- (0+.5,.5) -- (0+.5,-.5) -- cycle;
        \node  (LU0) at (0-0.5, 0.25) {};
        \node  (RU0) at (0+0.5, 0.25) {};
        \node  (LD0) at (0-0.5, -0.25) {};
        \node  (RD0) at (0+0.5, -0.25) {};
        \node  (U0) at (0, 1.2) {};
        \node  (D0) at (0, -1.2) {};
        \draw [fill=green] (2-.5,-.5) -- (2-.5,.5) -- (2+.5,.5) -- (2+.5,-.5) -- cycle;
        \node  (LU1) at (2-0.5, 0.25) {};
        \node  (RU1) at (2+0.5, 0.25) {};
        \node  (LD1) at (2-0.5, -0.25) {};
        \node  (RD1) at (2+0.5, -0.25) {};
        \node  (U1) at (2, 1.2) {};
        \node  (D1) at (2, -1.2) {};
        \draw [fill=green] (4-.5,-.5) -- (4-.5,.5) -- (4+.5,.5) -- (4+.5,-.5) -- cycle;
        \node  (LU2) at (4-0.5, 0.25) {};
        \node  (RU2) at (4+0.5, 0.25) {};
        \node  (LD2) at (4-0.5, -0.25) {};
        \node  (RD2) at (4+0.5, -0.25) {};
        \node  (U2) at (4, 1.2) {};
        \node  (D2) at (4, -1.2) {}; 
        \draw (LU0.center) to[out=180,in=180,looseness=1.5]  (U0.center)  to[out=0,in=0,looseness=1.5] (RU0.center);
        \draw (LU1.center) to[out=180,in=180,looseness=1.5]  (U1.center)-- (U2.center)  to[out=0,in=0,looseness=1.5] (RU2.center);

        \draw (LD2.center) to[out=180,in=180,looseness=1.5]  (D2.center)  to[out=0,in=0,looseness=1.5] (RD2.center);
        \draw (LD0.center) to[out=180,in=180,looseness=1.5]  (D0.center) -- (D1.center)  to[out=0,in=0,looseness=1.5] (RD1.center);
        \draw (RD0.center) -- (LD1.center);
    \end{tikzpicture}

   }$&
$%
 \begin{tikzpicture}[scale = .75] 
        \draw [fill=green] (0-.5,-.5) -- (0-.5,.5) -- (0+.5,.5) -- (0+.5,-.5) -- cycle;
        \node  (LU0) at (0-0.5, 0.25) {};
        \node  (RU0) at (0+0.5, 0.25) {};
        \node  (LD0) at (0-0.5, -0.25) {};
        \node  (RD0) at (0+0.5, -0.25) {};
        \node  (U0) at (0, 1.2) {};
        \node  (D0) at (0, -1.2) {};
        \draw [fill=green] (2-.5,-.5) -- (2-.5,.5) -- (2+.5,.5) -- (2+.5,-.5) -- cycle;
        \node  (LU1) at (2-0.5, 0.25) {};
        \node  (RU1) at (2+0.5, 0.25) {};
        \node  (LD1) at (2-0.5, -0.25) {};
        \node  (RD1) at (2+0.5, -0.25) {};
        \node  (U1) at (2, 1.2) {};
        \node  (D1) at (2, -1.2) {};
        \draw [fill=green] (4-.5,-.5) -- (4-.5,.5) -- (4+.5,.5) -- (4+.5,-.5) -- cycle;
        \node  (LU2) at (4-0.5, 0.25) {};
        \node  (RU2) at (4+0.5, 0.25) {};
        \node  (LD2) at (4-0.5, -0.25) {};
        \node  (RD2) at (4+0.5, -0.25) {};
        \node  (U2) at (4, 1.2) {};
        \node  (D2) at (4, -1.2) {}; 
        \draw (LU0.center) to[out=180,in=180,looseness=1.5]  (U0.center)  to[out=0,in=0,looseness=1.5] (RU0.center);
        \draw (LU1.center) to[out=180,in=180,looseness=1.5]  (U1.center)-- (U2.center)  to[out=0,in=0,looseness=1.5] (RU2.center);
        
       \draw (LD0.center) to[out=180,in=180,looseness=1.5]  (D0.center)  to[out=0,in=0,looseness=1.5] (RD0.center);
        \draw (LD1.center) to[out=180,in=180,looseness=1.5]  (D1.center) -- (D2.center)  to[out=0,in=0,looseness=1.5] (RD2.center);
        \draw (RD1.center) -- (LD2.center);
    \end{tikzpicture}

   }$&
$%
 \begin{tikzpicture}[scale = .75] 
        \draw [fill=green] (0-.5,-.5) -- (0-.5,.5) -- (0+.5,.5) -- (0+.5,-.5) -- cycle;
        \node  (LU0) at (0-0.5, 0.25) {};
        \node  (RU0) at (0+0.5, 0.25) {};
        \node  (LD0) at (0-0.5, -0.25) {};
        \node  (RD0) at (0+0.5, -0.25) {};
        \node  (U0) at (0, 1.2) {};
        \node  (D0) at (0, -1.2) {};
        \draw [fill=green] (2-.5,-.5) -- (2-.5,.5) -- (2+.5,.5) -- (2+.5,-.5) -- cycle;
        \node  (LU1) at (2-0.5, 0.25) {};
        \node  (RU1) at (2+0.5, 0.25) {};
        \node  (LD1) at (2-0.5, -0.25) {};
        \node  (RD1) at (2+0.5, -0.25) {};
        \node  (U1) at (2, 1.2) {};
        \node  (D1) at (2, -1.2) {};
        \draw [fill=green] (4-.5,-.5) -- (4-.5,.5) -- (4+.5,.5) -- (4+.5,-.5) -- cycle;
        \node  (LU2) at (4-0.5, 0.25) {};
        \node  (RU2) at (4+0.5, 0.25) {};
        \node  (LD2) at (4-0.5, -0.25) {};
        \node  (RD2) at (4+0.5, -0.25) {};
        \node  (U2) at (4, 1.2) {};
        \node  (D2) at (4, -1.2) {}; 
        \draw (LU0.center) to[out=180,in=180,looseness=1.5]  (U0.center)  to[out=0,in=0,looseness=1.5] (RU0.center);
        \draw (LU1.center) to[out=180,in=180,looseness=1.5]  (U1.center)-- (U2.center)  to[out=0,in=0,looseness=1.5] (RU2.center);
  \draw (LD0.center) to[out=180,in=180,looseness=1.5]  (D111.center)
        to[out=0,in=0,looseness=1.5] (RD2.center);
        \draw (RD0.center) to[out=0,in=180,looseness=1.5]  (D11.center) to[out=0,in=180,looseness=1.5]  (LD2.center);
       \draw (LD1.center) to[out=180,in=180,looseness=1.5]  (D1.center)  to[out=0,in=0,looseness=1.5] (RD1.center);
    \end{tikzpicture}

   }$&
$%
 \begin{tikzpicture}[scale = .75] 
        \draw [fill=green] (0-.5,-.5) -- (0-.5,.5) -- (0+.5,.5) -- (0+.5,-.5) -- cycle;
        \node  (LU0) at (0-0.5, 0.25) {};
        \node  (RU0) at (0+0.5, 0.25) {};
        \node  (LD0) at (0-0.5, -0.25) {};
        \node  (RD0) at (0+0.5, -0.25) {};
        \node  (U0) at (0, 1.2) {};
        \node  (D0) at (0, -1.2) {};
        \draw [fill=green] (2-.5,-.5) -- (2-.5,.5) -- (2+.5,.5) -- (2+.5,-.5) -- cycle;
        \node  (LU1) at (2-0.5, 0.25) {};
        \node  (RU1) at (2+0.5, 0.25) {};
        \node  (LD1) at (2-0.5, -0.25) {};
        \node  (RD1) at (2+0.5, -0.25) {};
        \node  (U1) at (2, 1.2) {};
        \node  (D1) at (2, -1.2) {};
        \draw [fill=green] (4-.5,-.5) -- (4-.5,.5) -- (4+.5,.5) -- (4+.5,-.5) -- cycle;
        \node  (LU2) at (4-0.5, 0.25) {};
        \node  (RU2) at (4+0.5, 0.25) {};
        \node  (LD2) at (4-0.5, -0.25) {};
        \node  (RD2) at (4+0.5, -0.25) {};
        \node  (U2) at (4, 1.2) {};
        \node  (D2) at (4, -1.2) {}; 
        \draw (LU0.center) to[out=180,in=180,looseness=1.5]  (U0.center)  to[out=0,in=0,looseness=1.5] (RU0.center);
        \draw (LU1.center) to[out=180,in=180,looseness=1.5]  (U1.center)-- (U2.center)  to[out=0,in=0,looseness=1.5] (RU2.center);
\draw (LD0.center) to[out=180,in=180,looseness=1.5]  (D0.center) -- (D1.center) -- (D2.center)  to[out=0,in=0,looseness=1.5] (RD2.center);
        \draw (RD0.center) -- (LD1.center);
        \draw (RD1.center) -- (LD2.center);
    \end{tikzpicture}

   }$&
$%
 \begin{tikzpicture}[scale = .75] 
        \draw [fill=green] (0-.5,-.5) -- (0-.5,.5) -- (0+.5,.5) -- (0+.5,-.5) -- cycle;
        \node  (LU0) at (0-0.5, 0.25) {};
        \node  (RU0) at (0+0.5, 0.25) {};
        \node  (LD0) at (0-0.5, -0.25) {};
        \node  (RD0) at (0+0.5, -0.25) {};
        \node  (U0) at (0, 1.2) {};
        \node  (D0) at (0, -1.2) {};
        \draw [fill=green] (2-.5,-.5) -- (2-.5,.5) -- (2+.5,.5) -- (2+.5,-.5) -- cycle;
        \node  (LU1) at (2-0.5, 0.25) {};
        \node  (RU1) at (2+0.5, 0.25) {};
        \node  (LD1) at (2-0.5, -0.25) {};
        \node  (RD1) at (2+0.5, -0.25) {};
        \node  (U1) at (2, 1.2) {};
        \node  (D1) at (2, -1.2) {};
        \draw [fill=green] (4-.5,-.5) -- (4-.5,.5) -- (4+.5,.5) -- (4+.5,-.5) -- cycle;
        \node  (LU2) at (4-0.5, 0.25) {};
        \node  (RU2) at (4+0.5, 0.25) {};
        \node  (LD2) at (4-0.5, -0.25) {};
        \node  (RD2) at (4+0.5, -0.25) {};
        \node  (U2) at (4, 1.2) {};
        \node  (D2) at (4, -1.2) {}; 
        \draw (LU0.center) to[out=180,in=180,looseness=1.5]  (U0.center)  to[out=0,in=0,looseness=1.5] (RU0.center);
        \draw (LU1.center) to[out=180,in=180,looseness=1.5]  (U1.center)-- (U2.center)  to[out=0,in=0,looseness=1.5] (RU2.center);
\draw (LD0.center) to[out=180,in=180,looseness=2]  (D0.center) to[out=0,in=0,looseness=1.2] (RD1.center);
        \draw (LD1.center) to[out=180,in=180,looseness=1.2] (D2.center) to[out=0,in=0,looseness=2] (RD2.center);
        \draw (RD0.center) to[out=0,in=180,looseness=1] (D1.center) to[out=0,in=180,looseness=1] (LD2.center);
    \end{tikzpicture}

   }$\tabularnewline
\hline 
$(13)$ & 
$%
 \begin{tikzpicture}[scale = .75] 
\draw [fill=green] (0-.5,-.5) -- (0-.5,.5) -- (0+.5,.5) -- (0+.5,-.5) -- cycle;
        \node  (LU0) at (0-0.5, 0.25) {};
        \node  (RU0) at (0+0.5, 0.25) {};
        \node  (LD0) at (0-0.5, -0.25) {};
        \node  (RD0) at (0+0.5, -0.25) {};
        \node  (U0) at (0, 1.2) {};
        \node  (D0) at (0, -1.2) {};
        \draw [fill=green] (2-.5,-.5) -- (2-.5,.5) -- (2+.5,.5) -- (2+.5,-.5) -- cycle;
        \node  (LU1) at (2-0.5, 0.25) {};
        \node  (RU1) at (2+0.5, 0.25) {};
        \node  (LD1) at (2-0.5, -0.25) {};
        \node  (RD1) at (2+0.5, -0.25) {};
        \node  (U1) at (2, .8) {};
        \node  (U11) at (2, 1) {};
        \node  (U111) at (2, 1.2) {};
        \node  (D1) at (2, -1.2) {};
        \draw [fill=green] (4-.5,-.5) -- (4-.5,.5) -- (4+.5,.5) -- (4+.5,-.5) -- cycle;
        \node  (LU2) at (4-0.5, 0.25) {};
        \node  (RU2) at (4+0.5, 0.25) {};
        \node  (LD2) at (4-0.5, -0.25) {};
        \node  (RD2) at (4+0.5, -0.25) {};
        \node  (U2) at (4, 1,2) {};
        \node  (D2) at (4, -1.2) {};

        
        \draw (LU0.center) to[out=180,in=180,looseness=1.5]  (U111.center)
        to[out=0,in=0,looseness=1.5] (RU2.center);
        \draw (RU0.center) to[out=0,in=180,looseness=1.5]  (U11.center) to[out=0,in=180,looseness=1.5]  (LU2.center);
       \draw (LU1.center) to[out=180,in=180,looseness=1.5]  (U1.center)  to[out=0,in=0,looseness=1.5] (RU1.center);
        \draw (LD0.center) to[out=180,in=180,looseness=1.5]  (D0.center)  to[out=0,in=0,looseness=1.5] (RD0.center);
        \draw (LD1.center) to[out=180,in=180,looseness=1.5]  (D1.center)  to[out=0,in=0,looseness=1.5] (RD1.center);
        \draw (LD2.center) to[out=180,in=180,looseness=1.5]  (D2.center)  to[out=0,in=0,looseness=1.5] (RD2.center);
        \end{tikzpicture}}$ &
$%
 \begin{tikzpicture}[scale = .75] 
\draw [fill=green] (0-.5,-.5) -- (0-.5,.5) -- (0+.5,.5) -- (0+.5,-.5) -- cycle;
        \node  (LU0) at (0-0.5, 0.25) {};
        \node  (RU0) at (0+0.5, 0.25) {};
        \node  (LD0) at (0-0.5, -0.25) {};
        \node  (RD0) at (0+0.5, -0.25) {};
        \node  (U0) at (0, 1.2) {};
        \node  (D0) at (0, -1.2) {};
        \draw [fill=green] (2-.5,-.5) -- (2-.5,.5) -- (2+.5,.5) -- (2+.5,-.5) -- cycle;
        \node  (LU1) at (2-0.5, 0.25) {};
        \node  (RU1) at (2+0.5, 0.25) {};
        \node  (LD1) at (2-0.5, -0.25) {};
        \node  (RD1) at (2+0.5, -0.25) {};
        \node  (U1) at (2, .8) {};
        \node  (U11) at (2, 1) {};
        \node  (U111) at (2, 1.2) {};
        \node  (D1) at (2, -1.2) {};
        \draw [fill=green] (4-.5,-.5) -- (4-.5,.5) -- (4+.5,.5) -- (4+.5,-.5) -- cycle;
        \node  (LU2) at (4-0.5, 0.25) {};
        \node  (RU2) at (4+0.5, 0.25) {};
        \node  (LD2) at (4-0.5, -0.25) {};
        \node  (RD2) at (4+0.5, -0.25) {};
        \node  (U2) at (4, 1,2) {};
        \node  (D2) at (4, -1.2) {};

        
        \draw (LU0.center) to[out=180,in=180,looseness=1.5]  (U111.center)
        to[out=0,in=0,looseness=1.5] (RU2.center);
        \draw (RU0.center) to[out=0,in=180,looseness=1.5]  (U11.center) to[out=0,in=180,looseness=1.5]  (LU2.center);
       \draw (LU1.center) to[out=180,in=180,looseness=1.5]  (U1.center)  to[out=0,in=0,looseness=1.5] (RU1.center);
        \draw (LD2.center) to[out=180,in=180,looseness=1.5]  (D2.center)  to[out=0,in=0,looseness=1.5] (RD2.center);
        \draw (LD0.center) to[out=180,in=180,looseness=1.5]  (D0.center) -- (D1.center)  to[out=0,in=0,looseness=1.5] (RD1.center);
        \draw (RD0.center) -- (LD1.center);
        \end{tikzpicture}}$&
$%
 \begin{tikzpicture}[scale = .75] 
\draw [fill=green] (0-.5,-.5) -- (0-.5,.5) -- (0+.5,.5) -- (0+.5,-.5) -- cycle;
        \node  (LU0) at (0-0.5, 0.25) {};
        \node  (RU0) at (0+0.5, 0.25) {};
        \node  (LD0) at (0-0.5, -0.25) {};
        \node  (RD0) at (0+0.5, -0.25) {};
        \node  (U0) at (0, 1.2) {};
        \node  (D0) at (0, -1.2) {};
        \draw [fill=green] (2-.5,-.5) -- (2-.5,.5) -- (2+.5,.5) -- (2+.5,-.5) -- cycle;
        \node  (LU1) at (2-0.5, 0.25) {};
        \node  (RU1) at (2+0.5, 0.25) {};
        \node  (LD1) at (2-0.5, -0.25) {};
        \node  (RD1) at (2+0.5, -0.25) {};
        \node  (U1) at (2, .8) {};
        \node  (U11) at (2, 1) {};
        \node  (U111) at (2, 1.2) {};
        \node  (D1) at (2, -1.2) {};
        \draw [fill=green] (4-.5,-.5) -- (4-.5,.5) -- (4+.5,.5) -- (4+.5,-.5) -- cycle;
        \node  (LU2) at (4-0.5, 0.25) {};
        \node  (RU2) at (4+0.5, 0.25) {};
        \node  (LD2) at (4-0.5, -0.25) {};
        \node  (RD2) at (4+0.5, -0.25) {};
        \node  (U2) at (4, 1,2) {};
        \node  (D2) at (4, -1.2) {};

        
        \draw (LU0.center) to[out=180,in=180,looseness=1.5]  (U111.center)
        to[out=0,in=0,looseness=1.5] (RU2.center);
        \draw (RU0.center) to[out=0,in=180,looseness=1.5]  (U11.center) to[out=0,in=180,looseness=1.5]  (LU2.center);
       \draw (LU1.center) to[out=180,in=180,looseness=1.5]  (U1.center)  to[out=0,in=0,looseness=1.5] (RU1.center);
        \draw (LD0.center) to[out=180,in=180,looseness=1.5]  (D0.center)  to[out=0,in=0,looseness=1.5] (RD0.center);
        \draw (LD1.center) to[out=180,in=180,looseness=1.5]  (D1.center) -- (D2.center)  to[out=0,in=0,looseness=1.5] (RD2.center);
        \draw (RD1.center) -- (LD2.center);
        \end{tikzpicture}}$&
$%
 \begin{tikzpicture}[scale = .75] 
\draw [fill=green] (0-.5,-.5) -- (0-.5,.5) -- (0+.5,.5) -- (0+.5,-.5) -- cycle;
        \node  (LU0) at (0-0.5, 0.25) {};
        \node  (RU0) at (0+0.5, 0.25) {};
        \node  (LD0) at (0-0.5, -0.25) {};
        \node  (RD0) at (0+0.5, -0.25) {};
        \node  (U0) at (0, 1.2) {};
        \node  (D0) at (0, -1.2) {};
        \draw [fill=green] (2-.5,-.5) -- (2-.5,.5) -- (2+.5,.5) -- (2+.5,-.5) -- cycle;
        \node  (LU1) at (2-0.5, 0.25) {};
        \node  (RU1) at (2+0.5, 0.25) {};
        \node  (LD1) at (2-0.5, -0.25) {};
        \node  (RD1) at (2+0.5, -0.25) {};
        \node  (U1) at (2, .8) {};
        \node  (U11) at (2, 1) {};
        \node  (U111) at (2, 1.2) {};
        \node  (D1) at (2, -1.2) {};
        \draw [fill=green] (4-.5,-.5) -- (4-.5,.5) -- (4+.5,.5) -- (4+.5,-.5) -- cycle;
        \node  (LU2) at (4-0.5, 0.25) {};
        \node  (RU2) at (4+0.5, 0.25) {};
        \node  (LD2) at (4-0.5, -0.25) {};
        \node  (RD2) at (4+0.5, -0.25) {};
        \node  (U2) at (4, 1,2) {};
        \node  (D2) at (4, -1.2) {};

        
        \draw (LU0.center) to[out=180,in=180,looseness=1.5]  (U111.center)
        to[out=0,in=0,looseness=1.5] (RU2.center);
        \draw (RU0.center) to[out=0,in=180,looseness=1.5]  (U11.center) to[out=0,in=180,looseness=1.5]  (LU2.center);
       \draw (LU1.center) to[out=180,in=180,looseness=1.5]  (U1.center)  to[out=0,in=0,looseness=1.5] (RU1.center);
         \draw (LD0.center) to[out=180,in=180,looseness=1.5]  (D111.center)
        to[out=0,in=0,looseness=1.5] (RD2.center);
        \draw (RD0.center) to[out=0,in=180,looseness=1.5]  (D11.center) to[out=0,in=180,looseness=1.5]  (LD2.center);
       \draw (LD1.center) to[out=180,in=180,looseness=1.5]  (D1.center)  to[out=0,in=0,looseness=1.5] (RD1.center);
        \end{tikzpicture}}$&
$%
 \begin{tikzpicture}[scale = .75] 
\draw [fill=green] (0-.5,-.5) -- (0-.5,.5) -- (0+.5,.5) -- (0+.5,-.5) -- cycle;
        \node  (LU0) at (0-0.5, 0.25) {};
        \node  (RU0) at (0+0.5, 0.25) {};
        \node  (LD0) at (0-0.5, -0.25) {};
        \node  (RD0) at (0+0.5, -0.25) {};
        \node  (U0) at (0, 1.2) {};
        \node  (D0) at (0, -1.2) {};
        \draw [fill=green] (2-.5,-.5) -- (2-.5,.5) -- (2+.5,.5) -- (2+.5,-.5) -- cycle;
        \node  (LU1) at (2-0.5, 0.25) {};
        \node  (RU1) at (2+0.5, 0.25) {};
        \node  (LD1) at (2-0.5, -0.25) {};
        \node  (RD1) at (2+0.5, -0.25) {};
        \node  (U1) at (2, .8) {};
        \node  (U11) at (2, 1) {};
        \node  (U111) at (2, 1.2) {};
        \node  (D1) at (2, -1.2) {};
        \draw [fill=green] (4-.5,-.5) -- (4-.5,.5) -- (4+.5,.5) -- (4+.5,-.5) -- cycle;
        \node  (LU2) at (4-0.5, 0.25) {};
        \node  (RU2) at (4+0.5, 0.25) {};
        \node  (LD2) at (4-0.5, -0.25) {};
        \node  (RD2) at (4+0.5, -0.25) {};
        \node  (U2) at (4, 1,2) {};
        \node  (D2) at (4, -1.2) {};

        
        \draw (LU0.center) to[out=180,in=180,looseness=1.5]  (U111.center)
        to[out=0,in=0,looseness=1.5] (RU2.center);
        \draw (RU0.center) to[out=0,in=180,looseness=1.5]  (U11.center) to[out=0,in=180,looseness=1.5]  (LU2.center);
       \draw (LU1.center) to[out=180,in=180,looseness=1.5]  (U1.center)  to[out=0,in=0,looseness=1.5] (RU1.center);
      \draw (LD0.center) to[out=180,in=180,looseness=1.5]  (D0.center) -- (D1.center) -- (D2.center)  to[out=0,in=0,looseness=1.5] (RD2.center);
        \draw (RD0.center) -- (LD1.center);
        \draw (RD1.center) -- (LD2.center);
        \end{tikzpicture}}$&
$%
 \begin{tikzpicture}[scale = .75] 
\draw [fill=green] (0-.5,-.5) -- (0-.5,.5) -- (0+.5,.5) -- (0+.5,-.5) -- cycle;
        \node  (LU0) at (0-0.5, 0.25) {};
        \node  (RU0) at (0+0.5, 0.25) {};
        \node  (LD0) at (0-0.5, -0.25) {};
        \node  (RD0) at (0+0.5, -0.25) {};
        \node  (U0) at (0, 1.2) {};
        \node  (D0) at (0, -1.2) {};
        \draw [fill=green] (2-.5,-.5) -- (2-.5,.5) -- (2+.5,.5) -- (2+.5,-.5) -- cycle;
        \node  (LU1) at (2-0.5, 0.25) {};
        \node  (RU1) at (2+0.5, 0.25) {};
        \node  (LD1) at (2-0.5, -0.25) {};
        \node  (RD1) at (2+0.5, -0.25) {};
        \node  (U1) at (2, .8) {};
        \node  (U11) at (2, 1) {};
        \node  (U111) at (2, 1.2) {};
        \node  (D1) at (2, -1.2) {};
        \draw [fill=green] (4-.5,-.5) -- (4-.5,.5) -- (4+.5,.5) -- (4+.5,-.5) -- cycle;
        \node  (LU2) at (4-0.5, 0.25) {};
        \node  (RU2) at (4+0.5, 0.25) {};
        \node  (LD2) at (4-0.5, -0.25) {};
        \node  (RD2) at (4+0.5, -0.25) {};
        \node  (U2) at (4, 1,2) {};
        \node  (D2) at (4, -1.2) {};

        
        \draw (LU0.center) to[out=180,in=180,looseness=1.5]  (U111.center)
        to[out=0,in=0,looseness=1.5] (RU2.center);
        \draw (RU0.center) to[out=0,in=180,looseness=1.5]  (U11.center) to[out=0,in=180,looseness=1.5]  (LU2.center);
       \draw (LU1.center) to[out=180,in=180,looseness=1.5]  (U1.center)  to[out=0,in=0,looseness=1.5] (RU1.center);
        \draw (LD0.center) to[out=180,in=180,looseness=2]  (D0.center) to[out=0,in=0,looseness=1.2] (RD1.center);
        \draw (LD1.center) to[out=180,in=180,looseness=1.2] (D2.center) to[out=0,in=0,looseness=2] (RD2.center);
        \draw (RD0.center) to[out=0,in=180,looseness=1] (D1.center) to[out=0,in=180,looseness=1] (LD2.center);
        \end{tikzpicture}}$\tabularnewline
\hline 
\hline 
$(123)$ &
$%
\begin{tikzpicture}[scale=.75]
    \draw [fill=green] (0-.5,-.5) -- (0-.5,.5) -- (0+.5,.5) -- (0+.5,-.5) -- cycle;
        \node  (LU0) at (0-0.5, 0.25) {};
        \node  (RU0) at (0+0.5, 0.25) {};
        \node  (LD0) at (0-0.5, -0.25) {};
        \node  (RD0) at (0+0.5, -0.25) {};
        \node  (U0) at (0, 1.2) {};
        \node  (D0) at (0, -1.2) {};
        \draw [fill=green] (2-.5,-.5) -- (2-.5,.5) -- (2+.5,.5) -- (2+.5,-.5) -- cycle;
        \node  (LU1) at (2-0.5, 0.25) {};
        \node  (RU1) at (2+0.5, 0.25) {};
        \node  (LD1) at (2-0.5, -0.25) {};
        \node  (RD1) at (2+0.5, -0.25) {};
        \node  (U1) at (2, 1.2) {};
        \node  (D1) at (2, -1.2) {};
        \draw [fill=green] (4-.5,-.5) -- (4-.5,.5) -- (4+.5,.5) -- (4+.5,-.5) -- cycle;
        \node  (LU2) at (4-0.5, 0.25) {};
        \node  (RU2) at (4+0.5, 0.25) {};
        \node  (LD2) at (4-0.5, -0.25) {};
        \node  (RD2) at (4+0.5, -0.25) {};
        \node  (U2) at (4, 1.2) {};
        \node  (D2) at (4, -1.2) {};
        \draw (LU0.center) to[out=180,in=180,looseness=1.5]  (U0.center) -- (U1.center) -- (U2.center)  to[out=0,in=0,looseness=1.5] (RU2.center);
        \draw (RU0.center) -- (LU1.center);
        \draw (RU1.center) -- (LU2.center);
        \draw (LD0.center) to[out=180,in=180,looseness=1.5]  (D0.center)  to[out=0,in=0,looseness=1.5] (RD0.center);
        \draw (LD1.center) to[out=180,in=180,looseness=1.5]  (D1.center)  to[out=0,in=0,looseness=1.5] (RD1.center);
        \draw (LD2.center) to[out=180,in=180,looseness=1.5]  (D2.center)  to[out=0,in=0,looseness=1.5] (RD2.center);
\end{tikzpicture}}$ &
$%
\begin{tikzpicture}[scale=.75]
    \draw [fill=green] (0-.5,-.5) -- (0-.5,.5) -- (0+.5,.5) -- (0+.5,-.5) -- cycle;
        \node  (LU0) at (0-0.5, 0.25) {};
        \node  (RU0) at (0+0.5, 0.25) {};
        \node  (LD0) at (0-0.5, -0.25) {};
        \node  (RD0) at (0+0.5, -0.25) {};
        \node  (U0) at (0, 1.2) {};
        \node  (D0) at (0, -1.2) {};
        \draw [fill=green] (2-.5,-.5) -- (2-.5,.5) -- (2+.5,.5) -- (2+.5,-.5) -- cycle;
        \node  (LU1) at (2-0.5, 0.25) {};
        \node  (RU1) at (2+0.5, 0.25) {};
        \node  (LD1) at (2-0.5, -0.25) {};
        \node  (RD1) at (2+0.5, -0.25) {};
        \node  (U1) at (2, 1.2) {};
        \node  (D1) at (2, -1.2) {};
        \draw [fill=green] (4-.5,-.5) -- (4-.5,.5) -- (4+.5,.5) -- (4+.5,-.5) -- cycle;
        \node  (LU2) at (4-0.5, 0.25) {};
        \node  (RU2) at (4+0.5, 0.25) {};
        \node  (LD2) at (4-0.5, -0.25) {};
        \node  (RD2) at (4+0.5, -0.25) {};
        \node  (U2) at (4, 1.2) {};
        \node  (D2) at (4, -1.2) {};
        \draw (LU0.center) to[out=180,in=180,looseness=1.5]  (U0.center) -- (U1.center) -- (U2.center)  to[out=0,in=0,looseness=1.5] (RU2.center);
        \draw (RU0.center) -- (LU1.center);
        \draw (RU1.center) -- (LU2.center);
        \draw (LD2.center) to[out=180,in=180,looseness=1.5]  (D2.center)  to[out=0,in=0,looseness=1.5] (RD2.center);
        \draw (LD0.center) to[out=180,in=180,looseness=1.5]  (D0.center) -- (D1.center)  to[out=0,in=0,looseness=1.5] (RD1.center);
        \draw (RD0.center) -- (LD1.center);
\end{tikzpicture}}$&
$%
\begin{tikzpicture}[scale=.75]
    \draw [fill=green] (0-.5,-.5) -- (0-.5,.5) -- (0+.5,.5) -- (0+.5,-.5) -- cycle;
        \node  (LU0) at (0-0.5, 0.25) {};
        \node  (RU0) at (0+0.5, 0.25) {};
        \node  (LD0) at (0-0.5, -0.25) {};
        \node  (RD0) at (0+0.5, -0.25) {};
        \node  (U0) at (0, 1.2) {};
        \node  (D0) at (0, -1.2) {};
        \draw [fill=green] (2-.5,-.5) -- (2-.5,.5) -- (2+.5,.5) -- (2+.5,-.5) -- cycle;
        \node  (LU1) at (2-0.5, 0.25) {};
        \node  (RU1) at (2+0.5, 0.25) {};
        \node  (LD1) at (2-0.5, -0.25) {};
        \node  (RD1) at (2+0.5, -0.25) {};
        \node  (U1) at (2, 1.2) {};
        \node  (D1) at (2, -1.2) {};
        \draw [fill=green] (4-.5,-.5) -- (4-.5,.5) -- (4+.5,.5) -- (4+.5,-.5) -- cycle;
        \node  (LU2) at (4-0.5, 0.25) {};
        \node  (RU2) at (4+0.5, 0.25) {};
        \node  (LD2) at (4-0.5, -0.25) {};
        \node  (RD2) at (4+0.5, -0.25) {};
        \node  (U2) at (4, 1.2) {};
        \node  (D2) at (4, -1.2) {};
        \draw (LU0.center) to[out=180,in=180,looseness=1.5]  (U0.center) -- (U1.center) -- (U2.center)  to[out=0,in=0,looseness=1.5] (RU2.center);
        \draw (RU0.center) -- (LU1.center);
        \draw (RU1.center) -- (LU2.center);
        \draw (LD0.center) to[out=180,in=180,looseness=1.5]  (D0.center)  to[out=0,in=0,looseness=1.5] (RD0.center);
        \draw (LD1.center) to[out=180,in=180,looseness=1.5]  (D1.center) -- (D2.center)  to[out=0,in=0,looseness=1.5] (RD2.center);
        \draw (RD1.center) -- (LD2.center);
\end{tikzpicture}}$&
$%
\begin{tikzpicture}[scale=.75]
    \draw [fill=green] (0-.5,-.5) -- (0-.5,.5) -- (0+.5,.5) -- (0+.5,-.5) -- cycle;
        \node  (LU0) at (0-0.5, 0.25) {};
        \node  (RU0) at (0+0.5, 0.25) {};
        \node  (LD0) at (0-0.5, -0.25) {};
        \node  (RD0) at (0+0.5, -0.25) {};
        \node  (U0) at (0, 1.2) {};
        \node  (D0) at (0, -1.2) {};
        \draw [fill=green] (2-.5,-.5) -- (2-.5,.5) -- (2+.5,.5) -- (2+.5,-.5) -- cycle;
        \node  (LU1) at (2-0.5, 0.25) {};
        \node  (RU1) at (2+0.5, 0.25) {};
        \node  (LD1) at (2-0.5, -0.25) {};
        \node  (RD1) at (2+0.5, -0.25) {};
        \node  (U1) at (2, 1.2) {};
        \node  (D1) at (2, -1.2) {};
        \draw [fill=green] (4-.5,-.5) -- (4-.5,.5) -- (4+.5,.5) -- (4+.5,-.5) -- cycle;
        \node  (LU2) at (4-0.5, 0.25) {};
        \node  (RU2) at (4+0.5, 0.25) {};
        \node  (LD2) at (4-0.5, -0.25) {};
        \node  (RD2) at (4+0.5, -0.25) {};
        \node  (U2) at (4, 1.2) {};
        \node  (D2) at (4, -1.2) {};
        \draw (LU0.center) to[out=180,in=180,looseness=1.5]  (U0.center) -- (U1.center) -- (U2.center)  to[out=0,in=0,looseness=1.5] (RU2.center);
        \draw (RU0.center) -- (LU1.center);
        \draw (RU1.center) -- (LU2.center);
          \draw (LD0.center) to[out=180,in=180,looseness=1.5]  (D111.center)
        to[out=0,in=0,looseness=1.5] (RD2.center);
        \draw (RD0.center) to[out=0,in=180,looseness=1.5]  (D11.center) to[out=0,in=180,looseness=1.5]  (LD2.center);
       \draw (LD1.center) to[out=180,in=180,looseness=1.5]  (D1.center)  to[out=0,in=0,looseness=1.5] (RD1.center);
\end{tikzpicture}}$&
$%
\begin{tikzpicture}[scale=.75]
    \draw [fill=green] (0-.5,-.5) -- (0-.5,.5) -- (0+.5,.5) -- (0+.5,-.5) -- cycle;
        \node  (LU0) at (0-0.5, 0.25) {};
        \node  (RU0) at (0+0.5, 0.25) {};
        \node  (LD0) at (0-0.5, -0.25) {};
        \node  (RD0) at (0+0.5, -0.25) {};
        \node  (U0) at (0, 1.2) {};
        \node  (D0) at (0, -1.2) {};
        \draw [fill=green] (2-.5,-.5) -- (2-.5,.5) -- (2+.5,.5) -- (2+.5,-.5) -- cycle;
        \node  (LU1) at (2-0.5, 0.25) {};
        \node  (RU1) at (2+0.5, 0.25) {};
        \node  (LD1) at (2-0.5, -0.25) {};
        \node  (RD1) at (2+0.5, -0.25) {};
        \node  (U1) at (2, 1.2) {};
        \node  (D1) at (2, -1.2) {};
        \draw [fill=green] (4-.5,-.5) -- (4-.5,.5) -- (4+.5,.5) -- (4+.5,-.5) -- cycle;
        \node  (LU2) at (4-0.5, 0.25) {};
        \node  (RU2) at (4+0.5, 0.25) {};
        \node  (LD2) at (4-0.5, -0.25) {};
        \node  (RD2) at (4+0.5, -0.25) {};
        \node  (U2) at (4, 1.2) {};
        \node  (D2) at (4, -1.2) {};
        \draw (LU0.center) to[out=180,in=180,looseness=1.5]  (U0.center) -- (U1.center) -- (U2.center)  to[out=0,in=0,looseness=1.5] (RU2.center);
        \draw (RU0.center) -- (LU1.center);
        \draw (RU1.center) -- (LU2.center);
       \draw (LD0.center) to[out=180,in=180,looseness=1.5]  (D0.center) -- (D1.center) -- (D2.center)  to[out=0,in=0,looseness=1.5] (RD2.center);
        \draw (RD0.center) -- (LD1.center);
        \draw (RD1.center) -- (LD2.center);
\end{tikzpicture}}$&
$%
\begin{tikzpicture}[scale=.75]
    \draw [fill=green] (0-.5,-.5) -- (0-.5,.5) -- (0+.5,.5) -- (0+.5,-.5) -- cycle;
        \node  (LU0) at (0-0.5, 0.25) {};
        \node  (RU0) at (0+0.5, 0.25) {};
        \node  (LD0) at (0-0.5, -0.25) {};
        \node  (RD0) at (0+0.5, -0.25) {};
        \node  (U0) at (0, 1.2) {};
        \node  (D0) at (0, -1.2) {};
        \draw [fill=green] (2-.5,-.5) -- (2-.5,.5) -- (2+.5,.5) -- (2+.5,-.5) -- cycle;
        \node  (LU1) at (2-0.5, 0.25) {};
        \node  (RU1) at (2+0.5, 0.25) {};
        \node  (LD1) at (2-0.5, -0.25) {};
        \node  (RD1) at (2+0.5, -0.25) {};
        \node  (U1) at (2, 1.2) {};
        \node  (D1) at (2, -1.2) {};
        \draw [fill=green] (4-.5,-.5) -- (4-.5,.5) -- (4+.5,.5) -- (4+.5,-.5) -- cycle;
        \node  (LU2) at (4-0.5, 0.25) {};
        \node  (RU2) at (4+0.5, 0.25) {};
        \node  (LD2) at (4-0.5, -0.25) {};
        \node  (RD2) at (4+0.5, -0.25) {};
        \node  (U2) at (4, 1.2) {};
        \node  (D2) at (4, -1.2) {};
        \draw (LU0.center) to[out=180,in=180,looseness=1.5]  (U0.center) -- (U1.center) -- (U2.center)  to[out=0,in=0,looseness=1.5] (RU2.center);
        \draw (RU0.center) -- (LU1.center);
        \draw (RU1.center) -- (LU2.center);
       \draw (LD0.center) to[out=180,in=180,looseness=2]  (D0.center) to[out=0,in=0,looseness=1.2] (RD1.center);
        \draw (LD1.center) to[out=180,in=180,looseness=1.2] (D2.center) to[out=0,in=0,looseness=2] (RD2.center);
        \draw (RD0.center) to[out=0,in=180,looseness=1] (D1.center) to[out=0,in=180,looseness=1] (LD2.center);
\end{tikzpicture}}$\tabularnewline
\hline 
$(132)$ & 
$%
\begin{tikzpicture}[scale=.75]
    \draw [fill=green] (0-.5,-.5) -- (0-.5,.5) -- (0+.5,.5) -- (0+.5,-.5) -- cycle;
        \node  (LU0) at (0-0.5, 0.25) {};
        \node  (RU0) at (0+0.5, 0.25) {};
        \node  (LD0) at (0-0.5, -0.25) {};
        \node  (RD0) at (0+0.5, -0.25) {};
        \node  (U0) at (0, 1.2) {};
        \node  (D0) at (0, -1.2) {};
        \draw [fill=green] (2-.5,-.5) -- (2-.5,.5) -- (2+.5,.5) -- (2+.5,-.5) -- cycle;
        \node  (LU1) at (2-0.5, 0.25) {};
        \node  (RU1) at (2+0.5, 0.25) {};
        \node  (LD1) at (2-0.5, -0.25) {};
        \node  (RD1) at (2+0.5, -0.25) {};
        \node  (U1) at (2, 1.2) {};
        \node  (D1) at (2, -1.2) {};
        \draw [fill=green] (4-.5,-.5) -- (4-.5,.5) -- (4+.5,.5) -- (4+.5,-.5) -- cycle;
        \node  (LU2) at (4-0.5, 0.25) {};
        \node  (RU2) at (4+0.5, 0.25) {};
        \node  (LD2) at (4-0.5, -0.25) {};
        \node  (RD2) at (4+0.5, -0.25) {};
        \node  (U2) at (4, 1.2) {};
        \node  (D2) at (4, -1.2) {};
        \draw (LU0.center) to[out=180,in=180,looseness=2]  (U0.center) to[out=0,in=0,looseness=1.2] (RU1.center);
        \draw (LU1.center) to[out=180,in=180,looseness=1.2] (U2.center) to[out=0,in=0,looseness=2] (RU2.center);
        \draw (RU0.center) to[out=0,in=180,looseness=1] (U1.center) to[out=0,in=180,looseness=1] (LU2.center);
        \draw (LD0.center) to[out=180,in=180,looseness=1.5]  (D0.center)  to[out=0,in=0,looseness=1.5] (RD0.center);
        \draw (LD1.center) to[out=180,in=180,looseness=1.5]  (D1.center)  to[out=0,in=0,looseness=1.5] (RD1.center);
        \draw (LD2.center) to[out=180,in=180,looseness=1.5]  (D2.center)  to[out=0,in=0,looseness=1.5] (RD2.center);
\end{tikzpicture}}$ &
$%
\begin{tikzpicture}[scale=.75]
    \draw [fill=green] (0-.5,-.5) -- (0-.5,.5) -- (0+.5,.5) -- (0+.5,-.5) -- cycle;
        \node  (LU0) at (0-0.5, 0.25) {};
        \node  (RU0) at (0+0.5, 0.25) {};
        \node  (LD0) at (0-0.5, -0.25) {};
        \node  (RD0) at (0+0.5, -0.25) {};
        \node  (U0) at (0, 1.2) {};
        \node  (D0) at (0, -1.2) {};
        \draw [fill=green] (2-.5,-.5) -- (2-.5,.5) -- (2+.5,.5) -- (2+.5,-.5) -- cycle;
        \node  (LU1) at (2-0.5, 0.25) {};
        \node  (RU1) at (2+0.5, 0.25) {};
        \node  (LD1) at (2-0.5, -0.25) {};
        \node  (RD1) at (2+0.5, -0.25) {};
        \node  (U1) at (2, 1.2) {};
        \node  (D1) at (2, -1.2) {};
        \draw [fill=green] (4-.5,-.5) -- (4-.5,.5) -- (4+.5,.5) -- (4+.5,-.5) -- cycle;
        \node  (LU2) at (4-0.5, 0.25) {};
        \node  (RU2) at (4+0.5, 0.25) {};
        \node  (LD2) at (4-0.5, -0.25) {};
        \node  (RD2) at (4+0.5, -0.25) {};
        \node  (U2) at (4, 1.2) {};
        \node  (D2) at (4, -1.2) {};
        \draw (LU0.center) to[out=180,in=180,looseness=2]  (U0.center) to[out=0,in=0,looseness=1.2] (RU1.center);
        \draw (LU1.center) to[out=180,in=180,looseness=1.2] (U2.center) to[out=0,in=0,looseness=2] (RU2.center);
        \draw (RU0.center) to[out=0,in=180,looseness=1] (U1.center) to[out=0,in=180,looseness=1] (LU2.center);
        \draw (LD2.center) to[out=180,in=180,looseness=1.5]  (D2.center)  to[out=0,in=0,looseness=1.5] (RD2.center);
        \draw (LD0.center) to[out=180,in=180,looseness=1.5]  (D0.center) -- (D1.center)  to[out=0,in=0,looseness=1.5] (RD1.center);
        \draw (RD0.center) -- (LD1.center);
\end{tikzpicture}}$&
$%
\begin{tikzpicture}[scale=.75]
    \draw [fill=green] (0-.5,-.5) -- (0-.5,.5) -- (0+.5,.5) -- (0+.5,-.5) -- cycle;
        \node  (LU0) at (0-0.5, 0.25) {};
        \node  (RU0) at (0+0.5, 0.25) {};
        \node  (LD0) at (0-0.5, -0.25) {};
        \node  (RD0) at (0+0.5, -0.25) {};
        \node  (U0) at (0, 1.2) {};
        \node  (D0) at (0, -1.2) {};
        \draw [fill=green] (2-.5,-.5) -- (2-.5,.5) -- (2+.5,.5) -- (2+.5,-.5) -- cycle;
        \node  (LU1) at (2-0.5, 0.25) {};
        \node  (RU1) at (2+0.5, 0.25) {};
        \node  (LD1) at (2-0.5, -0.25) {};
        \node  (RD1) at (2+0.5, -0.25) {};
        \node  (U1) at (2, 1.2) {};
        \node  (D1) at (2, -1.2) {};
        \draw [fill=green] (4-.5,-.5) -- (4-.5,.5) -- (4+.5,.5) -- (4+.5,-.5) -- cycle;
        \node  (LU2) at (4-0.5, 0.25) {};
        \node  (RU2) at (4+0.5, 0.25) {};
        \node  (LD2) at (4-0.5, -0.25) {};
        \node  (RD2) at (4+0.5, -0.25) {};
        \node  (U2) at (4, 1.2) {};
        \node  (D2) at (4, -1.2) {};
        \draw (LU0.center) to[out=180,in=180,looseness=2]  (U0.center) to[out=0,in=0,looseness=1.2] (RU1.center);
        \draw (LU1.center) to[out=180,in=180,looseness=1.2] (U2.center) to[out=0,in=0,looseness=2] (RU2.center);
        \draw (RU0.center) to[out=0,in=180,looseness=1] (U1.center) to[out=0,in=180,looseness=1] (LU2.center);
         \draw (LD0.center) to[out=180,in=180,looseness=1.5]  (D0.center)  to[out=0,in=0,looseness=1.5] (RD0.center);
        \draw (LD1.center) to[out=180,in=180,looseness=1.5]  (D1.center) -- (D2.center)  to[out=0,in=0,looseness=1.5] (RD2.center);
        \draw (RD1.center) -- (LD2.center);
\end{tikzpicture}}$&
$%
\begin{tikzpicture}[scale=.75]
    \draw [fill=green] (0-.5,-.5) -- (0-.5,.5) -- (0+.5,.5) -- (0+.5,-.5) -- cycle;
        \node  (LU0) at (0-0.5, 0.25) {};
        \node  (RU0) at (0+0.5, 0.25) {};
        \node  (LD0) at (0-0.5, -0.25) {};
        \node  (RD0) at (0+0.5, -0.25) {};
        \node  (U0) at (0, 1.2) {};
        \node  (D0) at (0, -1.2) {};
        \draw [fill=green] (2-.5,-.5) -- (2-.5,.5) -- (2+.5,.5) -- (2+.5,-.5) -- cycle;
        \node  (LU1) at (2-0.5, 0.25) {};
        \node  (RU1) at (2+0.5, 0.25) {};
        \node  (LD1) at (2-0.5, -0.25) {};
        \node  (RD1) at (2+0.5, -0.25) {};
        \node  (U1) at (2, 1.2) {};
        \node  (D1) at (2, -1.2) {};
        \draw [fill=green] (4-.5,-.5) -- (4-.5,.5) -- (4+.5,.5) -- (4+.5,-.5) -- cycle;
        \node  (LU2) at (4-0.5, 0.25) {};
        \node  (RU2) at (4+0.5, 0.25) {};
        \node  (LD2) at (4-0.5, -0.25) {};
        \node  (RD2) at (4+0.5, -0.25) {};
        \node  (U2) at (4, 1.2) {};
        \node  (D2) at (4, -1.2) {};
        \draw (LU0.center) to[out=180,in=180,looseness=2]  (U0.center) to[out=0,in=0,looseness=1.2] (RU1.center);
        \draw (LU1.center) to[out=180,in=180,looseness=1.2] (U2.center) to[out=0,in=0,looseness=2] (RU2.center);
        \draw (RU0.center) to[out=0,in=180,looseness=1] (U1.center) to[out=0,in=180,looseness=1] (LU2.center);
          \draw (LD0.center) to[out=180,in=180,looseness=1.5]  (D111.center)
        to[out=0,in=0,looseness=1.5] (RD2.center);
        \draw (RD0.center) to[out=0,in=180,looseness=1.5]  (D11.center) to[out=0,in=180,looseness=1.5]  (LD2.center);
       \draw (LD1.center) to[out=180,in=180,looseness=1.5]  (D1.center)  to[out=0,in=0,looseness=1.5] (RD1.center);
\end{tikzpicture}}$&
$%
\begin{tikzpicture}[scale=.75]
    \draw [fill=green] (0-.5,-.5) -- (0-.5,.5) -- (0+.5,.5) -- (0+.5,-.5) -- cycle;
        \node  (LU0) at (0-0.5, 0.25) {};
        \node  (RU0) at (0+0.5, 0.25) {};
        \node  (LD0) at (0-0.5, -0.25) {};
        \node  (RD0) at (0+0.5, -0.25) {};
        \node  (U0) at (0, 1.2) {};
        \node  (D0) at (0, -1.2) {};
        \draw [fill=green] (2-.5,-.5) -- (2-.5,.5) -- (2+.5,.5) -- (2+.5,-.5) -- cycle;
        \node  (LU1) at (2-0.5, 0.25) {};
        \node  (RU1) at (2+0.5, 0.25) {};
        \node  (LD1) at (2-0.5, -0.25) {};
        \node  (RD1) at (2+0.5, -0.25) {};
        \node  (U1) at (2, 1.2) {};
        \node  (D1) at (2, -1.2) {};
        \draw [fill=green] (4-.5,-.5) -- (4-.5,.5) -- (4+.5,.5) -- (4+.5,-.5) -- cycle;
        \node  (LU2) at (4-0.5, 0.25) {};
        \node  (RU2) at (4+0.5, 0.25) {};
        \node  (LD2) at (4-0.5, -0.25) {};
        \node  (RD2) at (4+0.5, -0.25) {};
        \node  (U2) at (4, 1.2) {};
        \node  (D2) at (4, -1.2) {};
        \draw (LU0.center) to[out=180,in=180,looseness=2]  (U0.center) to[out=0,in=0,looseness=1.2] (RU1.center);
        \draw (LU1.center) to[out=180,in=180,looseness=1.2] (U2.center) to[out=0,in=0,looseness=2] (RU2.center);
        \draw (RU0.center) to[out=0,in=180,looseness=1] (U1.center) to[out=0,in=180,looseness=1] (LU2.center);
       \draw (LD0.center) to[out=180,in=180,looseness=1.5]  (D0.center) -- (D1.center) -- (D2.center)  to[out=0,in=0,looseness=1.5] (RD2.center);
        \draw (RD0.center) -- (LD1.center);
        \draw (RD1.center) -- (LD2.center);
\end{tikzpicture}}$&
$%
\begin{tikzpicture}[scale=.75]
    \draw [fill=green] (0-.5,-.5) -- (0-.5,.5) -- (0+.5,.5) -- (0+.5,-.5) -- cycle;
        \node  (LU0) at (0-0.5, 0.25) {};
        \node  (RU0) at (0+0.5, 0.25) {};
        \node  (LD0) at (0-0.5, -0.25) {};
        \node  (RD0) at (0+0.5, -0.25) {};
        \node  (U0) at (0, 1.2) {};
        \node  (D0) at (0, -1.2) {};
        \draw [fill=green] (2-.5,-.5) -- (2-.5,.5) -- (2+.5,.5) -- (2+.5,-.5) -- cycle;
        \node  (LU1) at (2-0.5, 0.25) {};
        \node  (RU1) at (2+0.5, 0.25) {};
        \node  (LD1) at (2-0.5, -0.25) {};
        \node  (RD1) at (2+0.5, -0.25) {};
        \node  (U1) at (2, 1.2) {};
        \node  (D1) at (2, -1.2) {};
        \draw [fill=green] (4-.5,-.5) -- (4-.5,.5) -- (4+.5,.5) -- (4+.5,-.5) -- cycle;
        \node  (LU2) at (4-0.5, 0.25) {};
        \node  (RU2) at (4+0.5, 0.25) {};
        \node  (LD2) at (4-0.5, -0.25) {};
        \node  (RD2) at (4+0.5, -0.25) {};
        \node  (U2) at (4, 1.2) {};
        \node  (D2) at (4, -1.2) {};
        \draw (LU0.center) to[out=180,in=180,looseness=2]  (U0.center) to[out=0,in=0,looseness=1.2] (RU1.center);
        \draw (LU1.center) to[out=180,in=180,looseness=1.2] (U2.center) to[out=0,in=0,looseness=2] (RU2.center);
        \draw (RU0.center) to[out=0,in=180,looseness=1] (U1.center) to[out=0,in=180,looseness=1] (LU2.center);
       \draw (LD0.center) to[out=180,in=180,looseness=2]  (D0.center) to[out=0,in=0,looseness=1.2] (RD1.center);
        \draw (LD1.center) to[out=180,in=180,looseness=1.2] (D2.center) to[out=0,in=0,looseness=2] (RD2.center);
        \draw (RD0.center) to[out=0,in=180,looseness=1] (D1.center) to[out=0,in=180,looseness=1] (LD2.center);
\end{tikzpicture}}$\tabularnewline
\hline 
\end{tabular}
}
\caption{The action of $r_A\otimes r_B\,\rho^{\otimes 3}$ similar to Eqs. \eqref{eq:Paction}-\eqref{eq:r2}.}
\label{tab:Otau}
\end{table}

The components of $\tilde{\vec{x}}^{(2,3)}$ are the non-equivalents diagrams of Tab. \ref{tab:Otau}-\ref{tab:Otau1} corresponds to the following quantities
\begin{table}[h]
\centering{}%
\begin{tabular}{|c||c||c|c|c||c|c|}
\hline 
 & $1^3$ & $(12)$ & $(23)$ & $(13)$ & $(312)$ & $(231)$\tabularnewline
\hline 
\hline 
$1^3$ & 
$x_0$ & 
$x_1$ & 
$x_1$ & 
$x_1$ & 
$x_2$ & $x_2$\tabularnewline
\hline
\hline
$(12)$ &
$x_3$ &
$x_5$ &
$x_4$ &
$x_4$ &
$x_6$  &
$x_6$\tabularnewline
\hline 
$(23)$ & $x_3$ &
$x_4$&
$x_5$&
$x_4$&
$x_6$&
$x_6$\tabularnewline
\hline 
$(13)$ & 
$x_3$&
$x_4$ &
$x_4$&
$x_5$&
$x_6$&
$x_6$\tabularnewline
\hline 
\hline
$(312)$ &
$x_7$ &
$x_8$&
$x_8$&
$x_8$&
$x_9$&
$x_{10}$
\tabularnewline
\hline 
$(231)$ & 
$x_7$ &
$x_8$&
$x_8$&
$x_8$&
$x_{10}$&
$x_9$\tabularnewline
\hline 
\end{tabular}\caption{The diagrams from Tab.\ref{tab:Otau} gives the invariants defined in Eq. \eqref{x_for_3rd_order}.}
\label{tab:Otau1}
\end{table}
    with 
\begin{align} \label{x_for_3rd_order}
    x_0=(\Tr \rho)^3,\quad&
    x_1= \Tr (\Tr_A\rho)^2\Tr \rho,&
    x_2=\Tr(\Tr_A\rho)^3\quad\nonumber\\
    x_3= \Tr(\Tr_B\rho)^2\Tr\rho,\quad&
    x_4= x_5 + \Delta = \Tr(\Tr_B\rho\otimes \Tr_A\rho\cdot \rho )&
    x_5=\Tr\rho^2\Tr\rho,\nonumber\\
    x_6=\Tr(\Tr_A\rho^2\Tr_A\rho),\quad&
    x_7=\Tr(\Tr_B\rho)^3,&
    x_8=\Tr(\Tr_B\rho^2\Tr_B\rho)\nonumber\\
    &2x_S = x_9 + x_{10}=\Tr[(\rho^{T_B})^3]+\Tr[(\rho^{T_A})^3]
    &
\end{align}

Finally, one can compute $\tilde{\bm{M}}^{-1}$ to express the local invariants in $\tilde{\vec{x}}^{(2)}_3$ in terms of measured quantities in $\vec{y}^{(2)}_3$ to detect entanglement.
\subsection{Entanglement criteria from third--order correlations in bipartite systems via local unitaries}

Let us recall that the reduction map~\cite{Horodecki1999}
\begin{equation}
    R(\rho) = \mathrm{Tr}(\rho) I - \rho
\end{equation}
is a positive, but not completely positive map, hence positivity of its amplification provides a separability criterion:
\begin{equation}
    (I \otimes R) \rho = \rho_A \otimes I_B - \rho \ge 0
\end{equation}
Positivity of $\rho_A \otimes I_B - \rho$ means that $\forall \ X\ge 0 \ \mathrm{Tr}(X(\rho_A \otimes I_B - \rho)) \ge 0$. Choosing $X = \rho$ one obtains:
\begin{equation}
    \mathrm{Tr}(\rho(\rho_A \otimes I_B - \rho)) = \mathrm{Tr}(\rho_A^2) - \mathrm{Tr}(\rho) \ge 0,
\end{equation}
reproducing criterion (\ref{sep_from_purities}). The other (linear in $\rho$) choices of $X$ would be $I \otimes \rho_B$ and $\rho_A \otimes I$, but they do not produce non-trivial criteria. The criterion (\ref{sep_from_purities}) is the only bipartite separability criterion using second-order invariants.

To obtain a third-order correlation criterion one should take $X$ being $2$-nd order expression in $\rho, \rho_A, \rho_B$. 
Choosing $X = \rho^2$ 
one obtains
\begin{equation}\label{sep_from_3rd_order}
    \mathrm{Tr}(\rho^3) \le \mathrm{Tr} ( \rho_A \mathrm{Tr}_B \rho^2 )
\end{equation}
Choosing $X = (\rho_A\otimes I - \rho)^2$ one obtain a condition for positivity of the third moment of $(I \otimes R) \rho$, but this condition is strictly weaker than (\ref{sep_from_3rd_order}).

There is one more issue concerning criterion (\ref{sep_from_3rd_order}). The term $\mathrm{Tr} \rho^3$ is not measurable in the randomised measurement scenario, but the term $\mathrm{Tr} \rho^3 + \mathrm{Tr} (\rho^\Gamma)^3$ is. 
As the partial transposition maps a separable state to a separable state, let's add two criteria (\ref{sep_from_3rd_order}) for $\rho$ and $\rho^\Gamma$ to obtain:
\begin{equation} \label{sep_from_3rd_order_corr}
    \mathrm{Tr}(\rho^3) + \mathrm{Tr} (\rho^\Gamma)^3 \le 2 \mathrm{Tr} ( \rho_A \mathrm{Tr}_B \rho^2 ) 
\end{equation}
In terms of (\ref{x_for_3rd_order}) it reads as:
$x_S \le 2 x_8$.
For Bell-diagonal states of two qubits the criterion (\ref{sep_from_3rd_order_corr}) detects the entanglement of the same states as the 2nd order criterion, but for Werner states in arbitrary dimension: $p P_+ + (1-p)/d^2 I_d \otimes I_d$, the criteria take the form:
\begin{align}
    -(d+1) p^2 + 1 \ge 0 & \quad \text{(purity criterion)} \label{purity}\\
    -(d^2-4)(d+1)p^3 + 2(d+1)(d-3)p^2 + 2 \ge 0  & \quad \text{(3rd-order criterion)}\label{3rdcriterion}
\end{align}
with the critical point obtained with Cardano-Ferrari method,
\begin{align}
p^*(d) &= \sqrt[3]{-\frac{r}{2} + \sqrt{\Delta}} + \sqrt[3]{-\frac{r}{2} - \sqrt{\Delta}} - \frac{a}{3} 
\end{align}
\begin{align}
a= -\frac{2(d-3)}{(d-2)(d+2)},\quad
b= -\frac{2}{(d-2)(d+2)(d+1)}, \quad
q = -\frac{4(d-3)^2}{3(d-2)^2(d+2)^2} \nonumber\\
r = -\frac{16(d-3)^3}{27(d-2)^3(d+2)^3} -\frac{2}{(d-2)(d+2)(d+1)}, \quad
\Delta = \left(\frac{r}{2}\right)^2 + \left(\frac{q}{3}\right)^3.
\end{align}
\begin{figure}
    \centering
    \includegraphics[width=0.5\linewidth]{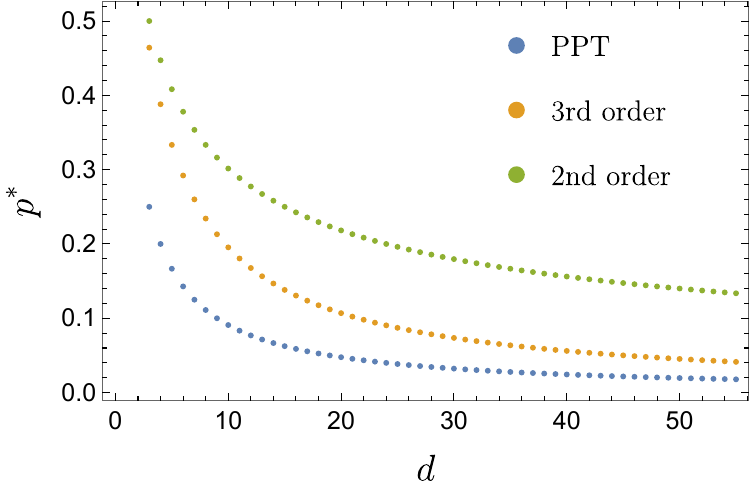}
    \caption{Comparison for $d>2$ of the purity criteria in Eq.\eqref{purity} obtained at the second order correlations and the one in Eq. \eqref{3rdcriterion} at the third-order correlations. Both are compared with the PPT criterion that for Werner state is equivalent to separability.
    }
    \label{fig:plot}
\end{figure}
and for $d=3$ the 2nd order criterion detects the states with $p > \frac 12$, while the criterion (\ref{sep_from_3rd_order_corr}) detects the states with $p > \frac 1{\sqrt[3]{10}}$.
\section{Conclusions}
Based on the framework of randomised measurements~\cite{cieslinski2024analysing}, we present a method for detecting entanglement in an unknown quantum state $\rho$ without requiring full quantum state tomography. Instead, our approach relies on estimating specific invariants of the state through experimentally accessible correlation functions. To this end, we propose an experimental scheme alongside two theoretical criteria: one based on second-order correlations, yielding the so-called \textit{purity criterion} in Eq.~\eqref{purity}, and another based on third-order correlations, leading to the entanglement criterion in Eq.~\eqref{3rdcriterion}. We benchmark both criteria on Werner states with equally dimensioned subsystems. For Werner states, the PPT property is equivalent to separability. 
This work opens several promising avenues for future research. A natural extension is to investigate
how the entanglement detection power improves with higher-order correlation measurements. Another direction lies in the development of quantum communication protocols exploiting these correlation-based criteria~\cite{Gigena2025,tyc2015quantum}.
Notably, the third-order criterion exploits the positivity of the reduction map, which has the property, that is invariant w.r. to local unitaries. All the higher-order criteria using reduction won't be able to detect entangled states. To go beyond, one has to construct out of the invariants a localy-invariant criterion, which won't be implied by positivity of a single localy-invariant map, but rather an localy-invarian family of non-invariant maps.   
Establishing such connections could significantly expand the toolbox for 
entanglement detection~\cite{arnault2015quantum,tran2016correlations,ketterer2019characterizing} and measures of entanglement~\cite{Grassl1998,Rains2000,Makhlin2002,Jarvis2014}.

\section*{Acknowledgements}
We thank D. Chru\'{s}ci\'{n}ski, C. Lupo, M. Grassl, R. Maggi, and A. Ac\'in for the fruitful scientific discussion. GSc thanks the Institute of Physics of the Nicolaus Copernicus University of Toru\'n for the hospitality. GSc is supported by Istituto Nazionale di Fisica
Nucleare (INFN) through the project ''QUANTUM'' and the European Union’s Horizon Europe research and innovation program under the project "Quantum Secure Networks Partnership" (QSNP, grant agreement No 101114043). GSa was supported by the National Science Centre project 2018/30/A/ST2/00837.

  \bibliographystyle{apsrev4-2}
  \bibliography{Entanglement_rand.bib}

\appendix

\section{Local invarants from 3-point correlations}\label{sec:3rd}

Let us write down explicitly the equation (\ref{eq:big}) 

\begin{align*}
\left[ \begin{array}{l} y_0 \\ \vdots \\ y_9 \end{array} \right] 
= \frac{1}{D_AD_B}
\mathrm{diag} \left[ \begin{array}{c}
1 \\
 d_B - 2 \\
 (d_B - 2) \cdot (d_B - 1) \\
 d_A - 2 \\
 (d_A - 2) \cdot (d_B - 2) \\
 (d_A - 2) \cdot (d_B - 2) \\
 (d_A - 2) \cdot (d_B - 2) \cdot (d_B - 1) \\
 (d_A - 2) \cdot (d_A - 1) \\
 (d_A - 2) \cdot (d_A - 1) \cdot (d_B - 2) \\
 (d_A - 2) \cdot (d_A - 1) \cdot (d_B - 2) \cdot (d_B - 1)
\end{array}\right]
\cdot
\left[\begin{matrix}\left(d_A^{2} - 2\right) \left(d_B^{2} - 2\right) & 3 d_B \cdot \left(2 - d_A^{2}\right)\\\left(d_A^{2} - 2\right) \left(d_B + 1\right) & \left(d_A^{2} - 2\right) \left(d_B - 1\right)\\d_A^{2} - 2 & 3 d_A^{2} - 6\\\left(d_A + 1\right) \left(d_B^{2} - 2\right) & - 3 d_B \left(d_A + 1\right)\\d_A d_B + d_A + d_B + 1 & d_A d_B - d_A + d_B - 1\\d_A d_B + d_A + d_B + 1 & d_A d_B - d_A + d_B - 1\\d_A + 1 & 3 d_A + 3\\d_B^{2} - 2 & - 3 d_B\\d_B + 1 & d_B - 1\\1 & 3\end{matrix}\right. \\
\left.\begin{matrix}4 d_A^{2} - 8 & 3 d_A \cdot \left(2 - d_B^{2}\right) & 6 d_A d_B & 3 d_A d_B & - 12 d_A & 4 d_B^{2} - 8 & - 12 d_B & 8 & 8\\4 - 2 d_A^{2} & - 3 d_A \left(d_B + 1\right) & 2 d_A \cdot \left(1 - d_B\right) & d_A \cdot \left(1 - d_B\right) & 6 d_A & 4 d_B + 4 & 4 d_B - 4 & -4 & -4\\2 d_A^{2} - 4 & - 3 d_A & - 6 d_A & - 3 d_A & - 6 d_A & 4 & 12 & 4 & 4\\4 d_A + 4 & \left(d_A - 1\right) \left(d_B^{2} - 2\right) & 2 d_B \cdot \left(1 - d_A\right) & d_B \cdot \left(1 - d_A\right) & 4 d_A - 4 & 4 - 2 d_B^{2} & 6 d_B & -4 & -4\\- 2 d_A - 2 & d_A d_B + d_A - d_B - 1 & d_A d_B + 2 & - d_A - d_B - 1 & 2 - 2 d_A & - 2 d_B - 2 & 2 - 2 d_B & 2 & 2\\- 2 d_A - 2 & d_A d_B + d_A - d_B - 1 & - 2 d_A - 2 d_B - 2 & d_A d_B + d_A + d_B + 3 & 2 - 2 d_A & - 2 d_B - 2 & 2 - 2 d_B & 2 & 2\\2 d_A + 2 & d_A - 1 & 2 d_A - 2 & d_A - 1 & 2 d_A - 2 & -2 & -6 & -2 & -2\\4 & 3 d_B^{2} - 6 & - 6 d_B & - 3 d_B & 12 & 2 d_B^{2} - 4 & - 6 d_B & 4 & 4\\-2 & 3 d_B + 3 & 2 d_B - 2 & d_B - 1 & -6 & 2 d_B + 2 & 2 d_B - 2 & -2 & -2\\2 & 3 & 6 & 3 & 6 & 2 & 6 & 2 & 2\end{matrix}\right]
\left[ \begin{array}{l} x_0 \\ \vdots \\ x_{10} \end{array} \right]
\end{align*}
where $D_K = \frac{1}{d_K\left(d_K^{2}-4\right)\left(d_K^{2}-1\right)}$, with $K=A;B$. Observe that the last two columns of $A$ are equal, hence one can treat $(x_9 + x_{10})/2$ as one variable and $A$ becomes a $10\times 10$ matrix after summing its two last columns.
Next, the difference of $y_4 - y_5$ is:
\begin{equation}
y_4-y_5 = \frac{(d_A^2 - 4)(d_B^2 - 4)}{D_1D_2} (x_4 - x_5)
\end{equation}
solving it we obtain:
\begin{equation}
x_4 = x_5 + 
\Delta, \quad \text{where} \quad \Delta = 
(y_4-y_5) d_A(d_A^2-1)d_B(d_B^2-1)
\end{equation}
and substituting the solution we obtain:
\begin{align*}
D_1 D_2 \cdot \mathrm{diag} \left[ \begin{array}{c}
1 \\
 d_B - 2 \\
 (d_B - 2) \cdot (d_B - 1) \\
 d_A - 2 \\
 (d_A - 2) \cdot (d_B - 2) \\
 (d_A - 2) \cdot (d_B - 2) \cdot (d_B - 1) \\
 (d_A - 2) \cdot (d_A - 1) \\
 (d_A - 2) \cdot (d_A - 1) \cdot (d_B - 2) \\
 (d_A - 2) \cdot (d_A - 1) \cdot (d_B - 2) \cdot (d_B - 1)
\end{array}\right]^{-1} 
\left[ \begin{array}{l} y_0 \\ y_1 \\ y_2 \\ y_3 \\ y_5 \\y_6 \\ y_7 \\ y_8 \\ y_9 \end{array} \right]
- \Delta \left[\begin{matrix}3 d_A d_B\\d_A \cdot \left(1 - d_B\right)\\- 3 d_A\\d_B \cdot \left(1 - d_A\right)\\d_Ad_B+d_A+d_B+3\\d_A - 1\\- 3 d_B\\d_B - 1\\3\end{matrix}\right]
=
\left[\begin{matrix}\left(d_A^{2} - 2\right) \left(d_B^{2} - 2\right)\\\left(d_A^{2} - 2\right) \left(d_B + 1\right)\\d_A^{2} - 2\\\left(d_A + 1\right) \left(d_B^{2} - 2\right)\\d_A d_B + d_A + d_B + 1\\d_A + 1\\d_B^{2} - 2\\d_B + 1\\1\end{matrix}\right. \\
\left.\begin{matrix}3 d_B \cdot \left(2 - d_A^{2}\right) & 4 d_A^{2} - 8 & 3 d_A \cdot \left(2 - d_B^{2}\right) & 9 d_A d_B & - 12 d_A & 4 d_B^{2} - 8 & - 12 d_B & 16\\\left(d_A^{2} - 2\right) \left(d_B - 1\right) & 4 - 2 d_A^{2} & - 3 d_A \left(d_B + 1\right) & 3 d_A \cdot \left(1 - d_B\right) & 6 d_A & 4 d_B + 4 & 4 d_B - 4 & -8\\3 d_A^{2} - 6 & 2 d_A^{2} - 4 & - 3 d_A & - 9 d_A & - 6 d_A & 4 & 12 & 8\\- 3 d_B \left(d_A + 1\right) & 4 d_A + 4 & \left(d_A - 1\right) \left(d_B^{2} - 2\right) & 3 d_B \cdot \left(1 - d_A\right) & 4 d_A - 4 & 4 - 2 d_B^{2} & 6 d_B & -8\\d_A d_B - d_A + d_B - 1 & - 2 d_A - 2 & d_A d_B + d_A - d_B - 1 & d_A d_B - d_A - d_B + 1 & 2 - 2 d_A & - 2 d_B - 2 & 2 - 2 d_B & 4\\3 d_A + 3 & 2 d_A + 2 & d_A - 1 & 3 d_A - 3 & 2 d_A - 2 & -2 & -6 & -4\\- 3 d_B & 4 & 3 d_B^{2} - 6 & - 9 d_B & 12 & 2 d_B^{2} - 4 & - 6 d_B & 8\\d_B - 1 & -2 & 3 d_B + 3 & 3 d_B - 3 & -6 & 2 d_B + 2 & 2 d_B - 2 & -4\\3 & 2 & 3 & 9 & 6 & 2 & 6 & 4\end{matrix}\right]
\left[ \begin{array}{c} x_0 \\ x_1 \\ x_2 \\ x_3 \\ x_5 \\ x_6 \\ x_7 \\ x_8 \\ (x_9 + x_{10})/2 \end{array} \right]
\end{align*}

\begin{align*}
 D_1 \left[\begin{array}{ccc} 1 & 0 & 0 \\ 0 & d_A-2 & 0 \\ 0 & 0 & (d_A-1)(d_A-2) \end{array} \right]^{-1}
 \otimes
 D_2 \left[\begin{array}{ccc} 1 & 0 & 0 \\ 0 & d_B-2 & 0 \\ 0 & 0 & (d_B-1)(d_B-2) \end{array} \right]^{-1}
\left[ \begin{array}{l} y_0 \\ y_1 \\ y_2 \\ y_3 \\ y_5 \\y_6 \\ y_7 \\ y_8 \\ y_9 \end{array} \right]
- \Delta
\left[\begin{matrix}3 d_A d_B\\d_A \cdot \left(1 - d_B\right)\\- 3 d_A\\d_B \cdot \left(1 - d_A\right)\\d_Ad_B+d_A+d_B+3\\d_A - 1\\- 3 d_B\\d_B - 1\\3\end{matrix}\right]
\\
= \left[\begin{matrix}d_A^{2} - 2 & - 3 d_A & 4\\d_A + 1 & d_A - 1 & -2\\1 & 3 & 2\end{matrix}\right]
\otimes
\left[\begin{matrix}d_B^{2} - 2 & - 3 d_B & 4\\d_B + 1 & d_B - 1 & -2\\1 & 3 & 2\end{matrix}\right]
\left[ \begin{array}{c} x_0 \\ x_1 \\ x_2 \\ x_3 \\ x_5 \\ x_6 \\ x_7 \\ x_8 \\ (x_9 + x_{10})/2 \end{array} \right]
\end{align*}
and finally
\begin{align*}
\left[ \begin{array}{c} x_0 \\ x_1 \\ x_2 \\ x_3 \\ x_5 \\ x_6 \\ x_7 \\ x_8 \\ (x_9 + x_{10})/2 \end{array} \right]
= 
\frac{d_A(d_A+1)}{d_A+2} \cdot \frac{d_B(d_B+1)}{d_B+2}
\left[\begin{matrix}\left(d_A - 2\right) \left(d_A - 1\right) & 3 d_A - 3 & 1\\- \left(d_A - 2\right) \left(d_A - 1\right) & \left(d_A - 2\right) \left(d_A - 1\right) & d_A\\\left(d_A - 2\right) \left(d_A - 1\right) & - \frac 32 \left(d_A - 1\right)^{2} & \frac 12 (d_A^{2} + 1) \end{matrix}\right]
\otimes \\
\left[\begin{matrix}\left(d_A - 2\right) \left(d_A - 1\right) & 3 d_A - 3 & 1\\- \left(d_A - 2\right) \left(d_A - 1\right) & \left(d_A - 2\right) \left(d_A - 1\right) & d_A\\\left(d_A - 2\right) \left(d_A - 1\right) & - \frac 32 \left(d_A - 1\right)^{2} & \frac 12 (d_A^{2} + 1) \end{matrix}\right]
\left[ \begin{array}{l} y_0 \\ y_1 \\ y_2 \\ y_3 \\ y_5 \\y_6 \\ y_7 \\ y_8 \\ y_9 \end{array} \right]
-
(y_4-y_5) d_A(d_A^2-1)d_B(d_B^2-1)
\left[\begin{matrix}6\\2 \left(d_B - 2\right)\\- 3 \left(d_B - 1\right)\\2 \left(d_A - 2\right)\\d_A d_B - d_A - d_B + 3\\- \left(d_A - 2\right) \left(d_B - 1\right)\\- 3 \left(d_A - 1\right)\\- \left(d_A - 1\right) \left(d_B - 2\right)\\\frac 32 \left(d_A - 1\right) \left(d_B - 1\right)\end{matrix}\right]
\end{align*}


\end{document}